%% file: JPCM-118505.R2.tex
\documentclass[12pt]{iopart}
\usepackage{graphicx}
\usepackage[utf8]{inputenc}
\usepackage[T1]{fontenc}
\usepackage{cite}
\DeclareUnicodeCharacter{0119}{\k{e}}

\begin{document}

\title{Coexistence of $s$- and $d$-wave gaps due to pair-hopping and exchange interactions}
\ead{shigeami@secondlab.co.jp}
\author{Shigeru Koikegami}
\address{Second Lab, LLC, 19-27 Inarimae, Tsukuba 305-0061, Japan}
\vspace{10pt}
\begin{indented}
\item[]Received 12 April 2021, revised 26 June 2021
\end{indented}

\begin{abstract}
I investigate the superconductivity of the three-band $t$--$J$--$U$ model derived from the three-band Hubbard model using the Schrieffer--Wolff transformation. 
My model is designed considering the hole-doped high-$T_{\mathrm{c}}$ superconducting cuprate. 
The model does not exclude the double occupancy of Cu sites by $d$ electrons, 
and there is a pair-hopping interaction between the $d$ and $p$ bands together with the exchange interaction. 
I analyse the superconducting transition temperature, electronic state, and superconducting gap function based on strong coupling theory 
and find that the superconductivity emerges due to the pair-hopping and exchange interactions via the Suhl-Kondo mechanism. 
In the superconducting state, the extended $s$- and $d_{x^2-y^2}$-wave superconducting gaps coexist, 
where both charge fluctuations and $d$--$p$ band hybridization are key ingredients.
\end{abstract}

\section{Introduction}
The $t$--$J$ model is one of the model Hamiltonians that form the basis of many theoretical studies of strongly correlated electron systems~\cite{Chao77,Chao78}. 
The $t$--$J$ model can also be derived as the low-energy effective Hamiltonian of the two-dimensional (2D) multiband Hubbard model~\cite{FCZhang88,Zaanen88,Matsukawa89}, 
regarded as the fundamental model Hamiltonian for the high-$T_{\mathrm{c}}$ superconducting cuprate (HTSC). 
Many theoretical studies of HTSC to date use the $t$--$J$ model as the model Hamiltonian~\cite{Mori02,Plakida02,Rosch04,Tohyama04,Ishihara04,Mishchenko04,Mishchenko06,Mallik20}. 
These studies often exclude the double occupancy of Cu sites by $d$ electrons, 
considering that the on-site Coulomb repulsion between $d$ orbitals is much larger than the transfer energy between the $d$ and $p$ orbitals. 
As a result, the $t$--$J$ model contains only one electron (or hole) band and a localized spin. 

However, the double occupancy of Cu sites need not necessarily be excluded 
when the on-site Coulomb repulsion $U$ is comparable to the transfer energy. 
Relaxing the single occupancy constraint and explicitly considering $U$ instead results in 
the $t$--$J$--$U$ model that includes both the $t$--$J$ model and the single-band Hubbard model as one of its limits~\cite{FCZhang03,Yuan05,Wang06,Abram13,Spalek17,Zegrodnik17,Fidrysiak20}. 
Thus, the $t$--$J$--$U$ model serves as an interpolation between the $t$--$J$ model and the single-band Hubbard model 
and is able to account for more properties caused by strong correlation. 
However, the charge transfer gap should be comparable to $t$ in the charge transfer regime. 
In this case, $p$ electron scattering by $d$ electrons cannot be negligible, and both $p$ and $d$ electrons must be considered.

In this paper, I derive the three-band $t$--$J$--$U$ model from the 2D three-band Hubbard model as its effective Hamiltonian 
by using the Schrieffer--Wolff (SW) transformation~\cite{Schrieffer66} and assume that double occupancy is not excluded. 
In my model, the pair-hopping interaction between the $d$ and $p$ bands exists separately from the exchange interaction. 
Treating these interactions using iterative perturbation theory (IPT) approximation, I investigate the superconductivity of the model in a strong coupling framework. 
The results show that the multicomponent superconductivity emerges with the hole doping, which introduces the $d$-$p$ band hybridization through 
exchange and pair-hopping interactions. This emergence of the superconductivity is due to the pair-hopping and exchange interactions via the Suhl-Kondo (SK) mechanism~\cite{Suhl59,Kondo63,Kondo02}, which stabilizes the superconducting gaps with different signs in a multiband system. 
In the superconducting state, the extended $s$- and $d_{x^2-y^2}$-wave superconducting gaps coexist,  
%%% 5
and the $s$- and $d$-wave gaps emerge due to the pair-hopping and exchange interactions, respectively. 
%%% 5

\section{Formulation}

Consider the three-band Hubbard model~\cite{Emery87} that expresses the Hamiltonian as ${\mathcal{H}} = {\mathcal{H}}_{0}+\sum_{\alpha}{\mathcal{H}}_{1}^{\alpha}$, where 
\begin{equation}
{\mathcal{H}}_{0} = \varepsilon_d^{}\sum_{j \sigma}d_{j \sigma}^{\dagger} d_{j \sigma}^{}
+ \varepsilon_p^{}\sum_{\alpha}\sum_{{\mathbf{k}} \sigma}p_{{\mathbf{k}} \sigma}^{\alpha\dagger} p_{{\mathbf{k}} \sigma}^{\alpha}
+ U \sum_j d_{j \uparrow}^{\dagger} d_{j \uparrow}^{} d_{j \downarrow}^{\dagger} d_{j \downarrow}^{}
\label{eq:01}
\end{equation} 
and
\begin{equation}
{\mathcal{H}}_{1}^{\alpha}  = \frac{1}{\sqrt{N}} \sum_j  \sum_{{\mathbf{k}} \sigma} 
\left(V_{\alpha{\mathbf{k}}}^{} e^{-{\mathrm{i}}{\mathbf{k}}\cdot{\mathbf{R}}_j} p_{{\mathbf{k}} \sigma}^{\alpha\dagger} d_{j \sigma}^{} + {\mathrm{H.c.}}
\right).
\label{eq:02}
\end{equation}
Here, $\alpha \in \{x,y\}$; $d_{j \sigma}^{}(d_{j \sigma}^{\dagger})$ is the annihilation (creation) operator for the $d$ electron of spin $\sigma$ at Cu site $j$; 
$p_{{\mathbf{k}} \sigma}^{\alpha}(p_{{\mathbf{k}} \sigma}^{\alpha\dagger})$ 
is the annihilation (creation) operator for $p^{\alpha}$ electrons of spin $\sigma$ with momentum ${\mathbf{k}}$, based on oxygen sites in real space; 
$\varepsilon_d^{}$ and $\varepsilon_p^{}$ are the $d$ and $p$ electron site energies, respectively; 
$U$ is the on-site Coulomb repulsion between $d$ orbitals; and $N$ is the number of k-space points in the first Brillouin zone (FBZ).
The lattice constant of the square lattice of Cu sites is the length unit. 
Thus, $V_{x{\mathbf{k}}}^{}=2{\mathrm{i}}t_{pd}^{} \sin \frac{k_x}{2}$ and $V_{y{\mathbf{k}}}^{}=-2{\mathrm{i}}t_{pd}^{} \sin \frac{k_y}{2}$, 
where $t_{pd}^{}$ is the transfer energy between the $d$ orbital and the neighbouring $p^{\alpha}$ orbital.

In order to derive the effective Hamiltonian for $\mathcal{H}$, I adopt the SW transformation as follows:
\begin{eqnarray}
\hspace{-6em}
e^{\sum_{\alpha}{\mathcal{S}}^{\alpha}} {\mathcal{H}} e^{-\sum_{\beta}{\mathcal{S}}^{\beta}}
& = & {\mathcal{H}}_{0}+ \sum_{\alpha}{\mathcal{H}}_{1}^{\alpha} 
+ \sum_{\alpha}\left[{\mathcal{S}}^{\alpha}, {\mathcal{H}}_{0}^{}\right]
+ \sum_{\alpha\beta}\left[{\mathcal{S}}^{\alpha}, {\mathcal{H}}_{1}^{\beta} \right]
+ \frac{1}{2}\sum_{\alpha\beta}\left[{\mathcal{S}}^{\alpha}, \left[{\mathcal{S}}^{\beta}, {\mathcal{H}}_0^{} \right] \right] + \ldots  \nonumber \\
& = & {\mathcal{H}}_{0}+ \frac{1}{2}\sum_{\alpha\beta}\left[{\mathcal{S}}^{\alpha}, {\mathcal{H}}_{1}^{\beta} \right] + \ldots,
\label{eq:03}
\end{eqnarray}
using ${\mathcal{H}}_{1}^{\alpha}+\left[{\mathcal{S}}^{\alpha},{\mathcal{H}}_{0}\right]=0$ and 
\begin{equation}
\hspace{-6em}
{\mathcal{S}}^{\alpha} = \frac{1}{\sqrt{N}} \sum_j  \sum_{{\mathbf{k}} \sigma} 
\left(
\frac{V_{\alpha{\mathbf{k}}}^{} e^{-{\mathrm{i}}{\mathbf{k}}\cdot{\mathbf{R}}_j}}{\mathit{\Delta}_{pd}^{}-U} 
 \,n_{d\,j-\sigma}^{} p_{{\mathbf{k}} \sigma}^{\alpha\dagger} d_{j \sigma}^{}  
+\frac{V_{\alpha{\mathbf{k}}}^{} e^{-{\mathrm{i}}{\mathbf{k}}\cdot{\mathbf{R}}_j}}{\mathit{\Delta}_{pd}^{}} 
(1-n_{d\,j-\sigma}^{}) p_{{\mathbf{k}} \sigma}^{\alpha\dagger} d_{j \sigma}^{}
\right) -{\mathrm{H.c.}}
\label{eq:04}
\end{equation}
Here, ${\mathit{\Delta}}_{pd}^{} \equiv \varepsilon_p^{}-\varepsilon_d^{}$, $n_{d\,j\sigma}^{} \equiv d_{j\sigma}^{\dagger} d_{j\sigma}^{}$, 
and H.c. indicates the Hermitian conjugate of the terms already written. 
The observable $n_{d\,j\sigma}^{}$ has $0$ or $1$ as its eigenvalue for each $j$ and $\sigma$. Using Eqs.~(\ref{eq:02}) and (\ref{eq:04}), the following results:
\begin{eqnarray}
\hspace{-6em}
\left[{\mathcal{S}}^{\alpha}, {\mathcal{H}}_{1}^{\beta}\right] & & \hspace{-2em} = -\frac{\delta_{\alpha\beta}}{N} \sum_{jj^\prime} \sum_{{\mathbf{k}} \sigma} \left(
\frac{n_{d\,j-\sigma}^{}}{\mathit{\Delta}_{pd}^{}-U}+\frac{1-n_{d\,j-\sigma}^{}}{\mathit{\Delta}_{pd}^{}}
\right)V_{\alpha{\mathbf{k}}}^{}V_{\beta{\mathbf{k}}}^{*}e^{-{\mathrm{i}}{\mathbf{k}}\cdot{\mathbf{R}}_j} e^{{\mathrm{i}}{\mathbf{k}}\cdot{\mathbf{R}}_{j^\prime}}
d_{j^\prime \sigma}^{\dagger} d_{j \sigma}^{} \nonumber \\
& & \hspace{-1em} +\frac{1}{N} \sum_j \sum_{{\mathbf{k}} {\mathbf{k}}^\prime \sigma} \left(
\frac{n_{d\,j-\sigma}^{}}{\mathit{\Delta}_{pd}^{}-U}+\frac{1-n_{d\,j-\sigma}^{}}{\mathit{\Delta}_{pd}^{}}
\right)V_{\alpha{\mathbf{k}}}^{}V_{\beta{\mathbf{k}}^\prime}^{*}e^{-{\mathrm{i}}{\mathbf{k}}\cdot{\mathbf{R}}_j} e^{{\mathrm{i}}{\mathbf{k}}^\prime\cdot{\mathbf{R}}_j}
p_{{\mathbf{k}} \sigma}^{\alpha \dagger} p_{{\mathbf{k}}^\prime \sigma}^{\beta} \nonumber \\
& & \hspace{-1em} -\frac{1}{N} \sum_j \sum_{{\mathbf{k}} {\mathbf{k}}^\prime \sigma} \left(
\frac{1}{\mathit{\Delta}_{pd}^{}-U}-\frac{1}{\mathit{\Delta}_{pd}^{}}
\right)V_{\alpha{\mathbf{k}}}^{}V_{\beta{\mathbf{k}}^\prime}^{*}e^{-{\mathrm{i}}{\mathbf{k}}\cdot{\mathbf{R}}_j} e^{{\mathrm{i}}{\mathbf{k}}^\prime\cdot{\mathbf{R}}_j}
p_{{\mathbf{k}} \sigma}^{\alpha\dagger} p_{{\mathbf{k}}^\prime -\sigma}^{\beta} d_{j\,-\sigma}^{\dagger} d_{j \sigma}^{} \nonumber \\
& & \hspace{-1em} -\frac{1}{N} \sum_j \sum_{{\mathbf{k}} {\mathbf{k}}^\prime \sigma} \left(
\frac{1}{\mathit{\Delta}_{pd}^{}-U}-\frac{1}{\mathit{\Delta}_{pd}^{}}
\right)V_{\alpha{\mathbf{k}}}^{}V_{\beta{\mathbf{k}}^\prime}^{} e^{-{\mathrm{i}}{\mathbf{k}}\cdot{\mathbf{R}}_j} e^{-{\mathrm{i}}{\mathbf{k}}^\prime\cdot{\mathbf{R}}_j}
p_{{\mathbf{k}} \sigma}^{\alpha\dagger} p_{{\mathbf{k}}^\prime -\sigma}^{\beta\dagger} d_{j\,-\sigma}^{} d_{j \sigma}^{} + {\mathrm{H.c.}} \nonumber \\
\label{eq:06}
\end{eqnarray}

Hereafter, I consider only the first two terms of the right-hand side of Eq.~(\ref{eq:03}), i.e., up to the second order of $t_{pd}^{}$. 
Now, I assume that the distribution of the $d$ electron is spatially uniform in the ground state and that the ground state is paramagnetic.
Thus, $\langle n_{d\,j\uparrow}^{} +  n_{d\,j\downarrow}^{} \rangle_0 \equiv n_{d}^{}$ and 
$\langle n_{d\,j\uparrow}^{}\rangle_0 = \langle n_{d\,j\downarrow}^{}\rangle_0$ for any $j$ where $n_{d}^{}$ 
is a c-number equal to the number of $d$ electrons in the ground state, 
where $\langle \cdots \rangle_0$ indicates the average in the ground state. 
I apply this approximation to Eqs.~(\ref{eq:03}) and (\ref{eq:06}) and treat $n_{d}^{}$ as a parameter that should be determined self-consistently. 
When I set $\varepsilon_p^{}$ to zero, i.e., $\mathit{\Delta}_{pd}^{} = - \varepsilon_d^{}$, 
and omit the constant terms, I obtain the effective Hamiltonian:
\begin{equation}
{\mathcal{H}}_{\mathrm{eff}}^{} = {\mathcal{H}}_{\mathrm{HF}}^{} + {\mathcal{H}}_{\mathrm{ex}}^{} + {\mathcal{H}}_{\mathrm{pair}}^{} + {\mathcal{H}}_{U}^{\prime}. 
\label{eq:08}
\end{equation}
${\mathcal{H}}_{\mathrm{HF}}^{}$ is the Hartree-Fock approximation of ${\mathcal{H}}_{0}$:
\begin{eqnarray}
{\mathcal{H}}_{\mathrm{HF}}^{} & = &  \sum_{{\mathbf{k}} \sigma} \varepsilon_{d{\mathbf{k}}}^{} d_{{\mathbf{k}} \sigma}^{\dagger} d_{{\mathbf{k}} \sigma}^{} 
+ \sum_{\alpha\beta} \sum_{{\mathbf{k}} \sigma} \varepsilon_{\alpha\beta{\mathbf{k}}}^{} p_{{\mathbf{k}} \sigma}^{\alpha\dagger} p_{{\mathbf{k}} \sigma}^{\beta},
\label{eq:07}
\end{eqnarray}
where 
$d_{{\mathbf{k}} \sigma}^{\dagger} = \frac{1}{\sqrt{N}} \sum_j d_{j \sigma}^{\dagger} e^{{\mathrm{i}}{\mathbf{k}}\cdot{\mathbf{R}}_j}, 
d_{{\mathbf{k}} \sigma} = \frac{1}{\sqrt{N}} \sum_j d_{j \sigma}^{}  e^{-{\mathrm{i}}{\mathbf{k}}\cdot{\mathbf{R}}_j}$, 
\begin{equation}
\varepsilon_{d{\mathbf{k}}}^{} = \varepsilon_{d}^{} + \frac{U}{2}n_{d}^{} + t\left(v_{x{\mathbf{k}}}^{}v_{x{\mathbf{k}}}^{*}+v_{y{\mathbf{k}}}^{}v_{y{\mathbf{k}}}^{*}\right),
\label{eq:17}
\end{equation}
and
\begin{equation}
\varepsilon_{\alpha\beta{\mathbf{k}}}^{} =  \left(Jn_{d}^{}-t \right) v_{\alpha{\mathbf{k}}}^{}v_{\beta{\mathbf{k}}}^{*},
\label{eq:13}
\end{equation}
with 
$v_{x{\mathbf{k}}}^{} = {\mathrm{i}}\sin \frac{k_x}{2}$, $v_{y{\mathbf{k}}}^{} = -{\mathrm{i}}\sin \frac{k_y}{2}$,
\begin{equation}
t = 4\,t_{pd}^2 \left(\frac{n_{d}^{}}{\varepsilon_d^{}+U}+\frac{1-n_{d}^{}}{\varepsilon_d^{}}\right),
\label{eq:09}
\end{equation}
and 
\begin{equation}
J = 2\,t_{pd}^2 \left(\frac{1}{\varepsilon_d^{}+U}-\frac{1}{\varepsilon_d^{}}\right).
\label{eq:10}
\end{equation}
${\mathcal{H}}_{\mathrm{ex}}^{}$ is an exchange interaction term:
\begin{equation}
{\mathcal{H}}_{\mathrm{ex}}^{} = \frac{J}{N} \sum_{\alpha\beta} \sum_{{\mathbf{k}} {\mathbf{k}}^\prime \sigma} \sum_{{\mathbf{q}}} 
v_{\alpha{\mathbf{k}}}^{} v_{\beta{\mathbf{k}}^\prime}^{*}
p_{{\mathbf{k}} \sigma}^{\alpha\dagger} p_{{\mathbf{k}}^\prime -\sigma}^{\beta} d_{{\mathbf{k}}^\prime+{\mathbf{q}}\,-\sigma}^{\dagger} d_{{\mathbf{k}}+{\mathbf{q}}\,\sigma}^{} 
+ {\mathrm{H.c.}}
\label{eq:11}
\end{equation}
${\mathcal{H}}_{\mathrm{pair}}^{}$ is a pair-hopping term:
\begin{equation}
{\mathcal{H}}_{\mathrm{pair}}^{} = \frac{J}{N} \sum_{\alpha\beta} \sum_{{\mathbf{k}} {\mathbf{k}}^\prime \sigma} \sum_{{\mathbf{q}}} 
v_{\alpha{\mathbf{k}}}^{} v_{\beta{\mathbf{k}}^\prime}^{}
p_{{\mathbf{k}} \sigma}^{\alpha\dagger} p_{{\mathbf{k}}^\prime -\sigma}^{\beta\dagger} d_{{\mathbf{k}}^\prime-{\mathbf{q}}\,-\sigma}^{} d_{{\mathbf{k}}+{\mathbf{q}}\,\sigma}^{}
+ {\mathrm{H.c.}}
\label{eq:12}
\end{equation}
${\mathcal{H}}_{U}^{\prime}$ is the Coulomb interaction term excluding the component with ${\mathbf{q}} = {\mathbf{0}}$:
\begin{equation}
{\mathcal{H}}_{U}^{\prime} = \frac{U}{N} \sum_{{\mathbf{k}} {\mathbf{k}}^\prime} \sum_{{\mathbf{q}} \neq {\mathbf{0}}} 
d_{{\mathbf{k}}+{\mathbf{q}}\,\uparrow}^{\dagger} d_{{\mathbf{k}} \uparrow}^{} d_{{\mathbf{k}}^\prime-{\mathbf{q}}\,\downarrow}^{\dagger} d_{{\mathbf{k}}^\prime \downarrow}^{}.
\label{eq:24}
\end{equation}
As a consequence, ${\mathcal{H}}_{\mathrm{eff}}^{}$ [Eq.~(\ref{eq:08})] can be characterized by the three parameters 
$t$ [Eq.~(\ref{eq:09})], $J$ [Eq.~(\ref{eq:10})], and $U$, and it can be regarded as the three-band $t$--$J$--$U$ model.

Here, $t$ in Eq.~(\ref{eq:09}) is positive near the half-filling in the charge-transfer regime, i.e., $U > - \varepsilon_d^{} > 0$. 
For instance, in the case $\varepsilon_d^{} = -U/2$, $t > 0$ for $n_{d}^{} > 0.5$, and 
the $d$ electron band dispersion $\varepsilon_{d{\mathbf{k}}}^{}$ in Eq.~(\ref{eq:17}) 
is the same as that for the single-band Hubbard model on a square lattice. $J$ in Eq.~(\ref{eq:10}) is always positive in the charge-transfer regime. 
Thus, ${\mathcal{H}}_{\mathrm{ex}}^{}$ in Eq.~(\ref{eq:11}) describes the transverse component of the antiferromagnetic exchange interaction 
between the $d$ and $p$ electrons, 
while the longitudinal component of this interaction narrows the bandwidth of $\varepsilon_{\alpha\beta{\mathbf{k}}}^{}$ in Eq.~(\ref{eq:13}) from $t$ to $t-Jn_{d}^{}$. 
Further, ${\mathcal{H}}_{\mathrm{ex}}^{}$ indicates that the $p$ electron is affected by the spin fluctuation of the $d$ electron. 
As will be shown later, the $d$-wave superconducting gap composed of $d$ and $p$ electrons emerges from ${\mathcal{H}}_{\mathrm{ex}}^{}$. 
${\mathcal{H}}_{\mathrm{pair}}^{}$ in Eq.~(\ref{eq:12}) appears for the first time by considering the double occupancy of Cu sites. 
The pair-hopping term is not included in the single-band $t$--$J$ model if double occupancy is excluded. 
In the model that includes the pair-hopping interaction, electrons favour pair formation~\cite{Penson86}. 
This is also true in the presence of the on-site interaction~\cite{Belkasri96,Robaszkiewicz99} and in the zero-bandwidth limit~\cite{Kapcia14}. 
Thus, the pair-hopping term in my model is expected to provide superconductivity in another way.

I introduce another assumption according to the speculation about the ground state of the three-band Hubbard model~\cite{KYamada92}. 
In the normal ground state, the $d$ and $p$ electrons should be combined to construct coherent quasi-particles through hybridization. 
The matrix elements of the hybridization between the $d$ and $p$ electrons can be found in the components with ${\mathbf{q}} = {\mathbf{0}}$ 
in Eqs.~(\ref{eq:11}) and (\ref{eq:12}) as follows. Defining
\begin{eqnarray}
& & \hspace{-7em} h_{pd}^{}  =  - \frac{{\mathrm{i}}}{N} \sum_{\alpha} \sum_{{\mathbf{k}}} \left[ 
 v_{\alpha{\mathbf{k}}}^{} \langle p_{{\mathbf{k}} \uparrow}^{\alpha\dagger} d_{{\mathbf{k}} \uparrow}^{} \rangle_0 - v_{\alpha{\mathbf{k}}}^{*} \langle 
d_{{\mathbf{k}} \uparrow}^{\dagger} p_{{\mathbf{k}} \uparrow}^{\alpha}  \rangle_0 \right] = - \frac{{\mathrm{i}}}{N} \sum_{\alpha} \sum_{{\mathbf{k}}} \left[  
 v_{\alpha{\mathbf{k}}}^{} \langle p_{{\mathbf{k}} \downarrow}^{\alpha\dagger} d_{{\mathbf{k}} \downarrow}^{} \rangle_0 - v_{\alpha{\mathbf{k}}}^{*} \langle  
d_{{\mathbf{k}} \downarrow}^{\dagger} p_{{\mathbf{k}} \downarrow}^{\alpha} \rangle_0 \right], \nonumber \\
\label{eq:25}
\end{eqnarray}
${\mathcal{H}}_{\mathrm{eff}}^{}$ can be rewritten as
\begin{equation}
{\mathcal{H}}_{\mathrm{eff}}^{} = {\mathcal{H}}_{0}^{\prime} + {\mathcal{H}}_{\mathrm{ex}}^{\prime} + {\mathcal{H}}_{\mathrm{pair}}^{\prime} + {\mathcal{H}}_{U}^{\prime},
\label{eq:26}
\end{equation}
where
\begin{equation}
{\mathcal{H}}_{0}^{\prime} = {\mathcal{H}}_{\mathrm{HF}}^{} + {\mathrm{i}} J  h_{pd}^{} \sum_{\alpha} \sum_{{\mathbf{k}} \sigma} 
\left(v_{\alpha{\mathbf{k}}}^{} p_{{\mathbf{k}} \sigma}^{\alpha\dagger} d_{{\mathbf{k}} \sigma}^{} - 
v_{\alpha{\mathbf{k}}}^{*} d_{{\mathbf{k}} \sigma}^{\dagger} p_{{\mathbf{k}} \sigma}^{\alpha} \right).
\label{eq:28}
\end{equation}
%%% 2
Here, $\langle A \rangle_0$ in Eq.~(\ref{eq:25}) means the expectation value of $A$ in the ground state of ${\mathcal{H}}_{0}^{\prime}$. 
%%% 2
${\mathcal{H}}_{\mathrm{ex}}^{\prime}$ and ${\mathcal{H}}_{\mathrm{pair}}^{\prime}$ indicate the exchange interaction and pair-hopping terms 
excluding the component with ${\mathbf{q}} = {\mathbf{0}}$ from Eqs.~(\ref{eq:11}) and (\ref{eq:12}), respectively. 
Thus, in the ground state of ${\mathcal{H}}_{0}^{\prime}$, the $d$ and $p$ electrons are combined to construct the coherent quasi-particles when $h_{pd}^{} > 0$. 

Hereafter, I treat ${\mathcal{H}}_{0}^{\prime}$ as the unperturbed part of ${\mathcal{H}}_{\mathrm{eff}}^{}$ on the assumption that $h_{pd}^{} > 0$. 
I diagonalize ${\mathcal{H}}_{0}^{\prime}$ and derive the unperturbed Green functions as follows:
\begin{eqnarray}
\label{eq:16}
& & G_{dd}^{0}(\mathbf{k},\mathrm{i}\epsilon_n) = \frac{{\mathrm{i}}\epsilon_n+\mu-\varepsilon_{xx{\mathbf{k}}}^{}-\varepsilon_{yy{\mathbf{k}}}^{}}{
({\mathrm{i}}\epsilon_n+\mu-\varepsilon_{{\mathbf{k}}}^{+})({\mathrm{i}}\epsilon_n+\mu-\varepsilon_{{\mathbf{k}}}^{-})}, \\
& & G_{d\alpha}^{0}(\mathbf{k},\mathrm{i}\epsilon_n) = \frac{{\mathrm{i}} J h_{pd}^{}v_{\alpha\mathbf{k}}^{}}{
({\mathrm{i}}\epsilon_n+\mu-\varepsilon_{{\mathbf{k}}}^{+})({\mathrm{i}}\epsilon_n+\mu-\varepsilon_{{\mathbf{k}}}^{-})}, \\
& & G_{\alpha d}^{0}(\mathbf{k},\mathrm{i}\epsilon_n) 
= \frac{-{\mathrm{i}} J h_{pd}^{}v_{\alpha\mathbf{k}}^{*}}{
({\mathrm{i}}\epsilon_n+\mu-\varepsilon_{{\mathbf{k}}}^{+})({\mathrm{i}}\epsilon_n+\mu-\varepsilon_{{\mathbf{k}}}^{-})},
\end{eqnarray}
and
\begin{eqnarray}
&  & \hspace{-7em}
\left(
\begin{array}{cc} G_{xx}^{0}(\mathbf{k},\mathrm{i}\epsilon_n) & G_{xy}^{0}(\mathbf{k},\mathrm{i}\epsilon_n) \\ 
G_{yx}^{0}(\mathbf{k},\mathrm{i}\epsilon_n) & G_{yy}^{0}(\mathbf{k},\mathrm{i}\epsilon_n) \\
\end{array}
\right) = 
\frac{1}{({\mathrm{i}}\epsilon_n+\mu)({\mathrm{i}}\epsilon_n+\mu-\varepsilon_{{\mathbf{k}}}^{+})({\mathrm{i}}\epsilon_n+\mu-\varepsilon_{{\mathbf{k}}}^{-})} \nonumber \\
&  & \hspace{-7em} \times \left(
\begin{array}{cc} 
({\mathrm{i}}\epsilon_n+\mu-\varepsilon_{d{\mathbf{k}}}^{}) ({\mathrm{i}}\epsilon_n+\mu-\varepsilon_{yy{\mathbf{k}}}^{}) - J^2 h_{pd}^{2} v_{y{\mathbf{k}}}^{}v_{y{\mathbf{k}}}^{*} & 
({\mathrm{i}}\epsilon_n+\mu-\varepsilon_{d{\mathbf{k}}}^{}) \varepsilon_{yx{\mathbf{k}}}^{} + J^2 h_{pd}^{2} v_{y{\mathbf{k}}}^{}v_{x{\mathbf{k}}}^{*} \\ 
({\mathrm{i}}\epsilon_n+\mu-\varepsilon_{d{\mathbf{k}}}^{}) \varepsilon_{xy{\mathbf{k}}}^{} + J^2 h_{pd}^{2} v_{x{\mathbf{k}}}^{}v_{y{\mathbf{k}}}^{*}  & 
({\mathrm{i}}\epsilon_n+\mu-\varepsilon_{d{\mathbf{k}}}^{}) ({\mathrm{i}}\epsilon_n+\mu-\varepsilon_{xx{\mathbf{k}}}^{}) - J^2 h_{pd}^{2} v_{x{\mathbf{k}}}^{}v_{x{\mathbf{k}}}^{*} \\
\end{array} \right). \nonumber \\
\end{eqnarray}
Here, I use the fermion Matsubara frequencies, $\epsilon_n=\pi T(2n+1)$, with integer $n$ and temperature $T$. $\mu$ is the chemical potential and 
\begin{equation}
\hspace{-4em} 
\varepsilon_{{\mathbf{k}}}^{\pm} =  \frac{\varepsilon_{d{\mathbf{k}}}^{}+\varepsilon_{xx{\mathbf{k}}}^{}+\varepsilon_{yy{\mathbf{k}}}^{}}{2}\pm
\sqrt{\left(\frac{\varepsilon_{d{\mathbf{k}}}^{}-\varepsilon_{xx{\mathbf{k}}}^{}-\varepsilon_{yy{\mathbf{k}}}^{}}{2}\right)^2+J^2 h_{pd}^{2}
\left(v_{x{\mathbf{k}}}^{}v_{x{\mathbf{k}}}^{*}+v_{y{\mathbf{k}}}^{}v_{y{\mathbf{k}}}^{*} \right)}.
\label{eq:29}
\end{equation}
For $h_{pd}^{} > 0$, Eq.~(\ref{eq:25}) can be rewritten as 
\begin{equation}
1 = \frac{2J}{N} \sum_{{\mathbf{k}}} 
\frac{v_{x{\mathbf{k}}}^{}v_{x{\mathbf{k}}}^{*}+v_{y{\mathbf{k}}}^{}v_{y{\mathbf{k}}}^{*}}{\varepsilon_{{\mathbf{k}}}^{+}-\varepsilon_{{\mathbf{k}}}^{-}}\,
\left\{\theta(\varepsilon_{{\mathbf{k}}}^{+}-\mu) - \theta(\varepsilon_{{\mathbf{k}}}^{-}-\mu)\right\},
\label{eq:27}
\end{equation}
where $\theta(x)$ means the Heaviside step function. 

In order to investigate the superconductivity in a strong coupling framework, I start with the Dyson-Gor'kov equations:
\begin{eqnarray}
& & \hspace{-7em} G_{\mu \nu}^{}({\mathbf{k}},\mathrm{i}\epsilon_n) = G_{\mu \nu}^{0}({\mathbf{k}},\mathrm{i}\epsilon_n) 
+G_{\mu \kappa}^{0}({\mathbf{k}},\mathrm{i}\epsilon_n){\mathit{\Sigma}}_{\kappa \lambda}^{}({\mathbf{k}},\mathrm{i}\epsilon_n)G_{\lambda \nu}^{}({\mathbf{k}},\mathrm{i}\epsilon_n)
+G_{\mu \kappa}^{0}({\mathbf{k}},\mathrm{i}\epsilon_n){\mathit{\Phi}}_{\kappa \lambda}^{}({\mathbf{k}},\mathrm{i}\epsilon_n)
F_{\lambda \nu}^{\dagger}(-{\mathbf{k}},-\mathrm{i}\epsilon_n), \nonumber \\ \\
& & \hspace{-7em} F_{\mu \nu}^{\dagger}({\mathbf{k}},\mathrm{i}\epsilon_n) = 
G_{\mu \kappa}^{0}({\mathbf{k}},\mathrm{i}\epsilon_n){\mathit{\Sigma}}_{\kappa \lambda}^{}({\mathbf{k}},\mathrm{i}\epsilon_n)F_{\lambda \nu}^{\dagger}({\mathbf{k}},\mathrm{i}\epsilon_n)
+G_{\mu \kappa}^{0}({\mathbf{k}},\mathrm{i}\epsilon_n){\mathit{\Phi}}_{\kappa \lambda}^{*}({\mathbf{k}},\mathrm{i}\epsilon_n)
G_{\lambda \nu}^{}(-{\mathbf{k}},-\mathrm{i}\epsilon_n), \\
& & \hspace{-7em} F_{\mu \nu}^{}({\mathbf{k}},\mathrm{i}\epsilon_n) = 
G_{\mu \kappa}^{0}({\mathbf{k}},\mathrm{i}\epsilon_n){\mathit{\Sigma}}_{\kappa \lambda}^{}({\mathbf{k}},\mathrm{i}\epsilon_n)F_{\lambda \nu}^{}({\mathbf{k}},\mathrm{i}\epsilon_n)
+G_{\mu \kappa}^{0}({\mathbf{k}},\mathrm{i}\epsilon_n){\mathit{\Phi}}_{\kappa \lambda}^{}({\mathbf{k}},\mathrm{i}\epsilon_n)
G_{\lambda \nu}^{}(-{\mathbf{k}},-\mathrm{i}\epsilon_n).
\end{eqnarray}
The orbital indices $\mu$, $\nu$, $\kappa$, and $\lambda$ run over $d$, $x$, and $y$, 
and I adopt the Einstein summation convention. $G_{\mu \nu}^{}({\mathbf{k}},\mathrm{i}\epsilon_n)$ and $F_{\mu \nu}^{}({\mathbf{k}},\mathrm{i}\epsilon_n)$ 
represent the normal and anomalous Green functions, respectively, and ${\mathit{\Sigma}}_{\mu \nu}^{}({\mathbf{k}},\mathrm{i}\epsilon_n)$ and 
${\mathit{\Phi}}_{\mu \nu}^{}({\mathbf{k}},\mathrm{i}\epsilon_n)$ correspond to the normal and anomalous self-energies, respectively. 
When ${\mathcal{H}}_{\mathrm{ex}}^{\prime} + {\mathcal{H}}_{\mathrm{pair}}^{\prime} + {\mathcal{H}}_{U}^{\prime}$ in Eq.~(\ref{eq:26}) is treated as a perturbation, 
the normal self-energies up to the second order of $J$ and $U$ are evaluated by the IPT approximation as follows:
\begin{eqnarray}
\label{eq:20}
& & \hspace{-6em} 
 {\mathit{\Sigma}}_{dd}^{}(\mathbf{k},\mathrm{i}\epsilon_n) = \frac{T}{N}\sum_{{\mathbf{k}}^\prime n^\prime} \left[ 
J^2\chi_J^G({\mathbf{k}}-{\mathbf{k}}^\prime,\mathrm{i}\epsilon_n-\mathrm{i}\epsilon_{n^\prime})G_{pp}^{0}({\mathbf{k}}^\prime,\mathrm{i}\epsilon_{n^\prime})
+ U^2\chi_U^{}({\mathbf{k}}-{\mathbf{k}}^\prime,\mathrm{i}\epsilon_n-\mathrm{i}\epsilon_{n^\prime})G_{dd}^{0}({\mathbf{k}}^\prime,\mathrm{i}\epsilon_{n^\prime})
\right], \nonumber \\ \\
& & 
{\mathit{\Sigma}}_{d \alpha}^{}(\mathbf{k},\mathrm{i}\epsilon_n) 
= - v_{\alpha\mathbf{k}}^{*}\frac{T}{N}\sum_{{\mathbf{k}}^\prime n^\prime}  \left[
J^2\chi_J^G({\mathbf{k}}-{\mathbf{k}}^\prime,\mathrm{i}\epsilon_n-\mathrm{i}\epsilon_{n^\prime})G_{pd}^{0}({\mathbf{k}}^\prime,\mathrm{i}\epsilon_{n^\prime})
\right], \\
& & 
{\mathit{\Sigma}}_{\alpha d}^{}(\mathbf{k},\mathrm{i}\epsilon_n) 
= - v_{\alpha\mathbf{k}}^{}\frac{T}{N}\sum_{{\mathbf{k}}^\prime n^\prime}  \left[
J^2\chi_J^G({\mathbf{k}}-{\mathbf{k}}^\prime,\mathrm{i}\epsilon_n-\mathrm{i}\epsilon_{n^\prime})G_{dp}^{0}({\mathbf{k}}^\prime,\mathrm{i}\epsilon_{n^\prime})
\right], \\
\label{eq:21}
& & {\mathit{\Sigma}}_{\alpha\beta}^{}(\mathbf{k},\mathrm{i}\epsilon_n) = v_{\alpha\mathbf{k}}^{} v_{\beta\mathbf{k}}^{*}
\frac{T}{N}\sum_{{\mathbf{k}}^\prime n^\prime} \left[
J^2\chi_J^G({\mathbf{k}}-{\mathbf{k}}^\prime,\mathrm{i}\epsilon_n-\mathrm{i}\epsilon_{n^\prime})G_{dd}^{0}({\mathbf{k}}^\prime,\mathrm{i}\epsilon_{n^\prime})
\right].
\end{eqnarray}
%
%%% 1
The IPT approximation was first applied in the study of the half-filled single-impurity Anderson model~\cite{KYamada75,KYosida75}, 
and it was adopted to solve the effective impurity model in the study of the $d=\infty$ Hubbard model~\cite{Georges92,XYZhang93}. 
In these works, it was shown that the second order perturbation theory in large energy scale $U$ could reproduce not only the coherent band 
but also the lower and upper incoherent bands. In a later section, it will be shown that my approach can reproduce similar band structure 
to be justified as the theory for the 2D three-band $t$--$J$--$U$ model. 
%%% 1

The anomalous self-energies up to the second order of $J$ and $U$ are evaluated as follows:
\begin{eqnarray}
\label{eq:18}
& & \hspace{-10em} {\mathit{\Phi}}_{dd}^{}(\mathbf{k},\mathrm{i}\epsilon_n) = - \frac{T}{N}\sum_{{\mathbf{k}}^\prime n^\prime} \left[\left\{J+
J^2\chi_{J}^{F}({\mathbf{k}}-{\mathbf{k}}^\prime,\mathrm{i}\epsilon_n-\mathrm{i}\epsilon_{n^\prime})\right\}\!F_{pp}^{}({\mathbf{k}}^\prime,\mathrm{i}\epsilon_{n^\prime}) 
+ \left\{U+U^2\chi_U^{}({\mathbf{k}}-{\mathbf{k}}^\prime,\mathrm{i}\epsilon_n-\mathrm{i}\epsilon_{n^\prime})\right\}\! F_{dd}^{}({\mathbf{k}}^\prime,\mathrm{i}\epsilon_{n^\prime})
\right], \nonumber \\ \\
\label{eq:30}
& & {\mathit{\Phi}}_{d\alpha}^{}(\mathbf{k},\mathrm{i}\epsilon_n) = v_{\alpha -\mathbf{k}}^{}\frac{T}{N}\sum_{{\mathbf{k}}^\prime n^\prime} \left[\left\{J+
J^2\chi_{J}^{F}({\mathbf{k}}-{\mathbf{k}}^\prime,\mathrm{i}\epsilon_n-\mathrm{i}\epsilon_{n^\prime})\right\}\! F_{pd}^{}({\mathbf{k}}^\prime,\mathrm{i}\epsilon_{n^\prime})
\right], \\
\label{eq:31}
& & {\mathit{\Phi}}_{\alpha d}^{}(\mathbf{k},\mathrm{i}\epsilon_n) = v_{\alpha \mathbf{k}}^{}\frac{T}{N}\sum_{{\mathbf{k}}^\prime n^\prime} \left[\left\{J+
J^2\chi_{J}^{F}({\mathbf{k}}-{\mathbf{k}}^\prime,\mathrm{i}\epsilon_n-\mathrm{i}\epsilon_{n^\prime})\right\}\! F_{dp}^{}({\mathbf{k}}^\prime,\mathrm{i}\epsilon_{n^\prime})
\right], \\
\label{eq:32}
& & {\mathit{\Phi}}_{\alpha \beta}^{}(\mathbf{k},\mathrm{i}\epsilon_n) = 
- v_{\alpha \mathbf{k}}^{} v_{\beta -\mathbf{k}}^{}
\frac{T}{N}\sum_{{\mathbf{k}}^\prime n^\prime} \left[\left\{J+
J^2\chi_J^F({\mathbf{k}}-{\mathbf{k}}^\prime,\mathrm{i}\epsilon_n-\mathrm{i}\epsilon_{n^\prime})\right\}\!F_{dd}^{}({\mathbf{k}}^\prime,\mathrm{i}\epsilon_{n^\prime})\right].
\end{eqnarray}
Here, the orbital indices $\alpha$ and $\beta$ run over $x$ and $y$, and 
\begin{eqnarray}
&  & \hspace{-4em} \chi_J^G({\mathbf{q}},\mathrm{i}\omega_m) = \chi_{dd,pp}^G({\mathbf{q}},\mathrm{i}\omega_m) - \chi_{dp,dp}^G({\mathbf{q}},\mathrm{i}\omega_m) 
- \chi_{pd,pd}^G({\mathbf{q}},\mathrm{i}\omega_m) + \chi_{pp,dd}^G({\mathbf{q}},\mathrm{i}\omega_m), \\
&  & \hspace{-4em} \chi_J^F({\mathbf{q}},\mathrm{i}\omega_m) = \chi_{dd,pp}^F({\mathbf{q}},\mathrm{i}\omega_m) - \chi_{dp,pd}^F({\mathbf{q}},\mathrm{i}\omega_m) 
- \chi_{pd,dp}^F({\mathbf{q}},\mathrm{i}\omega_m) + \chi_{pp,dd}^F({\mathbf{q}},\mathrm{i}\omega_m), \\
&  & \hspace{-4em} \chi_U^{}({\mathbf{q}},\mathrm{i}\omega_m) = \chi_{dd,dd}^G({\mathbf{q}},\mathrm{i}\omega_m) +\chi_{dd,dd}^F({\mathbf{q}},\mathrm{i}\omega_m), \\
\label{eq:23}
& & \chi_{\mu \nu,\kappa \lambda}^G({\mathbf{q}},\mathrm{i}\omega_m) = -\frac{T}{N}\sum_{{\mathbf{k}} n}
G_{\mu \nu}^{0}({\mathbf{q}}+{\mathbf{k}}, \mathrm{i}\omega_m+\mathrm{i}\epsilon_n)G_{\kappa \lambda}^{0}({\mathbf{k}}, \mathrm{i}\epsilon_n), \\
\label{eq:33}
& & \chi_{\mu \nu,\kappa \lambda}^F({\mathbf{q}},\mathrm{i}\omega_m) = -\frac{T}{N}\sum_{{\mathbf{k}} n}
F_{\mu \nu}^{}({\mathbf{q}}+{\mathbf{k}}, \mathrm{i}\omega_m+\mathrm{i}\epsilon_n)F_{\kappa \lambda}^{\dagger}({\mathbf{k}}, \mathrm{i}\epsilon_n), \\
& & G_{pp}^{0}({\mathbf{k}},\mathrm{i}\epsilon_n) = \sum_{\alpha \beta} v_{\alpha \mathbf{k}}^{*} v_{\beta \mathbf{k}}^{} G_{\alpha \beta}^{0}(\mathbf{k},\mathrm{i}\epsilon_n), \\
& & G_{dp}^{0}({\mathbf{k}},\mathrm{i}\epsilon_n) = \sum_{\alpha}  v_{\alpha \mathbf{k}}^{} G_{d \alpha}^{0}(\mathbf{k},\mathrm{i}\epsilon_n), \\
& & G_{pd}^{0}({\mathbf{k}},\mathrm{i}\epsilon_n) = \sum_{\alpha}  v_{\alpha \mathbf{k}}^{*} G_{\alpha d}^{0}(\mathbf{k},\mathrm{i}\epsilon_n), \\
& & F_{pp}^{}({\mathbf{k}},\mathrm{i}\epsilon_n) = \sum_{\alpha \beta} v_{\alpha \mathbf{k}}^{*} v_{\beta -\mathbf{k}}^{*} F_{\alpha  \beta}^{}(\mathbf{k},\mathrm{i}\epsilon_n), \\
& & F_{dp}^{}({\mathbf{k}},\mathrm{i}\epsilon_n) = \sum_{\alpha} v_{\alpha -\mathbf{k}}^{*} F_{d \alpha}^{}(\mathbf{k},\mathrm{i}\epsilon_n), \\
& & F_{pd}^{}({\mathbf{k}},\mathrm{i}\epsilon_n) = \sum_{\alpha} v_{\alpha \mathbf{k}}^{*} F_{\alpha d}^{}(\mathbf{k},\mathrm{i}\epsilon_n), \\
& & F_{pp}^{\dagger}({\mathbf{k}},\mathrm{i}\epsilon_n) = \sum_{\alpha \beta} v_{\alpha \mathbf{k}}^{} v_{\beta -\mathbf{k}}^{} F_{\alpha  \beta}^{\dagger}(\mathbf{k},\mathrm{i}\epsilon_n), \\
& & F_{dp}^{\dagger}({\mathbf{k}},\mathrm{i}\epsilon_n) = \sum_{\alpha} v_{\alpha -\mathbf{k}}^{} F_{d \alpha}^{\dagger}(\mathbf{k},\mathrm{i}\epsilon_n),{\mathrm{ and}} \\
& & F_{pd}^{\dagger}({\mathbf{k}},\mathrm{i}\epsilon_n) = \sum_{\alpha} v_{\alpha \mathbf{k}}^{} F_{\alpha  d}^{\dagger}(\mathbf{k},\mathrm{i}\epsilon_n),
\label{eq:14}
\end{eqnarray}
using the boson Matsubara frequencies, $\omega_m=2m \pi T$ with integer $m$. In Eqs.~(\ref{eq:23}) and (\ref{eq:33}), 
$\mu$, $\nu$, $\kappa$, and $\lambda$ denote $d$ or $p$, respectively. 
Note that $n_{d}^{}$, $t$, $h_{pd}^{}$, and the chemical potential $\mu$ must be determined self-consistently in the ground state of ${\mathcal{H}}_{0}^{\prime}$ 
through Eqs.~(\ref{eq:17})--(\ref{eq:09}), (\ref{eq:29}), and (\ref{eq:27}). 
To this end, I approximate $n_{d}^{}$ by the number of $d$ electrons in the ground state of ${\mathcal{H}}_{0}^{\prime}$:
\begin{equation}
\hspace{-6em} 
n_{d}^{} = 2 - \frac{2}{N}\sum_{\mathbf{k}} 
\frac{1}{\varepsilon_{{\mathbf{k}}}^{+}-\varepsilon_{{\mathbf{k}}}^{-}}
\left\{
(\varepsilon_{{\mathbf{k}}}^{+}-\varepsilon_{xx{\mathbf{k}}}^{}-\varepsilon_{yy{\mathbf{k}}}^{})\theta(\varepsilon_{{\mathbf{k}}}^{+}-\mu) -
(\varepsilon_{{\mathbf{k}}}^{-}-\varepsilon_{xx{\mathbf{k}}}^{}-\varepsilon_{yy{\mathbf{k}}}^{})\theta(\varepsilon_{{\mathbf{k}}}^{-}-\mu) 
\right\}.
\label{eq:15}
\end{equation}
Specifically, I regard $n_{d}^{}$ as a given parameter and solve Eqs.~(\ref{eq:17})--(\ref{eq:09}), (\ref{eq:29}), (\ref{eq:27}), and (\ref{eq:15}) 
to determine $t$, $h_{pd}^{}$, and the number of doped holes $\delta_{\mathrm{h}}^{0}$ for the ground state of ${\mathcal{H}}_{0}^{\prime}$, where 
\begin{equation}
\delta_{\mathrm{h}}^{0} =  \frac{2}{N}\sum_{\mathbf{k}} \left[
\theta(\varepsilon_{{\mathbf{k}}}^{+}-\mu) + \theta(\varepsilon_{{\mathbf{k}}}^{-}-\mu) + \theta(-\mu)
\right] - 1. \nonumber
\end{equation}
Once $t$, $h_{pd}^{}$, and $\delta_{\mathrm{h}}^{0}$ are determined for the ground state of ${\mathcal{H}}_{0}^{\prime}$, 
%%% 2
I treat $t$, $h_{pd}^{}$, and $\delta_{\mathrm{h}}^{0}$ as temperature independent parameters, whose values do not change from those at $T=0$. Then, 
%%% 2
Eqs.~(\ref{eq:17})--(\ref{eq:14}) are solved in a fully self-consistent manner 
to obtain ${\mathit{\Sigma}}_{\mu \nu}^{}({\mathbf{k}},\mathrm{i}\epsilon_n)$ and ${\mathit{\Phi}}_{\mu \nu}^{}({\mathbf{k}},\mathrm{i}\epsilon_n)$. 
To determine the transition temperature $T_{\mathrm{c}}$, 
I perform these calculations in two steps. First, ${\mathit{\Sigma}}_{\mu \nu}^{}({\mathbf{k}},\mathrm{i}\epsilon_n)$ 
is calculated with ${\mathit{\Phi}}_{\mu \nu}^{}({\mathbf{k}},\mathrm{i}\epsilon_n)=0$, and $\mu$ is self-consistently determined so that $\delta_{\mathrm{h}}^{}$ 
obtained from $G_{\mu \nu}^{}({\mathbf{k}},\mathrm{i}\epsilon_n)$ becomes equal to $\delta_{\mathrm{h}}^{0}$. 
%%% 3
In the first step, $\mu$ is correctly adjusted to compensate the temperature-dependent shift by ${\mathit{\Sigma}}_{\mu \nu}^{}({\mathbf{k}},\mathrm{i}\epsilon_n)$ with ${\mathit{\Phi}}_{\mu \nu}^{}({\mathbf{k}},\mathrm{i}\epsilon_n)=0$. 
%%% 3
Here, $\delta_{\mathrm{h}}^{} = n_{d\mathrm{h}}^{} + n_{p\mathrm{h}}^{} - 1$, where
\begin{eqnarray}
n_{d\mathrm{h}}^{} & = & 2 - 2\,\frac{T}{N} \sum_{{\mathbf{k}} n}G_{dd}^{}({\mathbf{k}},\mathrm{i}\epsilon_n) e^{\mathrm{i}\epsilon_n{0^+}},\\
n_{p\mathrm{h}}^{} & = & 4 - 2\,\frac{T}{N} \sum_{{\mathbf{k}} n}\left[G_{xx}^{}({\mathbf{k}},\mathrm{i}\epsilon_n)+G_{yy}^{}({\mathbf{k}},\mathrm{i}\epsilon_n)\right] e^{\mathrm{i}\epsilon_n{0^+}},
\end{eqnarray}
and $n_{d\mathrm{h}}^{}$ and $n_{p\mathrm{h}}^{}$ are the number of $d$ and $p$ holes, respectively. 
Next, using the determined $\mu$, fully self-consistent calculations are performed to obtain ${\mathit{\Sigma}}_{\mu \nu}^{}({\mathbf{k}},\mathrm{i}\epsilon_n)$ 
and ${\mathit{\Phi}}_{\mu \nu}^{}({\mathbf{k}},\mathrm{i}\epsilon_n)$. At this time, 
%%% 3
only the temperature-dependent shift by ${\mathit{\Phi}}_{\mu \nu}^{}({\mathbf{k}},\mathrm{i}\epsilon_n)$ is reflected in $\delta_{\mathrm{h}}^{}$ 
obtained from $G_{\mu \nu}^{}({\mathbf{k}},\mathrm{i}\epsilon_n)$. 
%%% 3
That is, if $\delta_{\mathrm{h}}^{}$ deviates from $\delta_{\mathrm{h}}^{0}$, ${\mathit{\Phi}}_{\mu \nu}^{}({\mathbf{k}},\mathrm{i}\epsilon_n) \neq 0$. 
Therefore, the temperature at which $\delta_{\mathrm{h}}^{}$ deviates from $\delta_{\mathrm{h}}^{0}$ is $T_{\mathrm{c}}$. 
%%% 3
Also in the second step, $\mu$ can be self-consistently determined so that $\delta_{\mathrm{h}}^{}$ 
obtained from $G_{\mu \nu}^{}({\mathbf{k}},\mathrm{i}\epsilon_n)$ becomes equal to $\delta_{\mathrm{h}}^{0}$ 
with ${\mathit{\Phi}}_{\mu \nu}^{}({\mathbf{k}},\mathrm{i}\epsilon_n) \neq 0$. In this case, the temperature at which $\mu$ deviates from the value with 
${\mathit{\Phi}}_{\mu \nu}^{}({\mathbf{k}},\mathrm{i}\epsilon_n) = 0$ is $T_{\mathrm{c}}$, which is consistent with the temperature at which $\delta_{\mathrm{h}}^{}$ deviates from $\delta_{\mathrm{h}}^{0}$ with fixed $\mu$.
%%% 3

\section{Results and discussion}
To perform the numerical calculations, I divide the FBZ into a $64 \times 64$ meshes and prepare $2048$ or $4096$ Matsubara frequencies. 
I commonly use $t_{pd} = 10000\,{\mathrm{K}}$ for my calculations, and here, I only consider the case $\varepsilon_d^{} = -U/2$. 
For this case, we have $J=t_{pd}$ when $U=8\,t_{pd}$. I find fully self-consistent solutions with $h_{pd}^{} > 0$ in $\delta_{\mathrm{h}}^{} \geq 0.117$.  
%%% 4
The ones in $0.117 \leq \delta_{\mathrm{h}}^{} \leq 0.139$ have ${\mathit{\Phi}}_{\mu \nu}^{}({\mathbf{k}},\mathrm{i}\epsilon_n) = 0$ and 
the others in $0.166 \leq \delta_{\mathrm{h}}^{} \leq 0.285$ have ${\mathit{\Phi}}_{\mu \nu}^{}({\mathbf{k}},\mathrm{i}\epsilon_n) \neq 0$. 
The former solutions correspond to metallic phase and the latter to superconducting phase. 
Although I find other fully self-consistent solutions with $h_{pd}^{} = 0$ in $\delta_{\mathrm{h}}^{} \leq 0.031$, which correspond to insulating phase, 
I cannot find any solutions in $0.031 < \delta_{\mathrm{h}}^{} < 0.117$. The absence of solutions in this doping range indicates that some of my assumptions break down. In particular, it is difficult to achieve the spatially uniform distribution of the $d$ electron in this range. 
For instance, the chemical potential shift suppression is observed in La$_{2-x}$Sr$_x$CuO$_4$ ($0 < x < 0.12$) by photoemission spectroscopy~\cite{Ino1997,Fujimori1998}. 
This suppression suggests the possibility of electronic phase separation between the insulating phase and the superconducting phase~\cite{Fujimori2001}, 
where the electrons are inhomogeneously distributed due to the strong electron correlation. 
Therefore, the theory in $0.031 < \delta_{\mathrm{h}}^{} < 0.117$ should consider the possibility of the spatially non-uniform distribution of the $d$ electron.

\begin{figure}[ht]
\begin{center}
\resizebox{88mm}{!}{\input{fig1.tex}}
\end{center}
\vspace{-4ex}
\caption{Doping dependences of $T_{\mathrm{c}}$ and $h_{pd}^{}$. "I", "M" and "S" indicate insulating, metallic and superconducting phases, respectively. 
The shaded region indicates $0.031 < \delta_{\mathrm{h}}^{} < 0.117$ in which any solutions cannot be found. 
$T_{\mathrm{c}}^{0}$ is the temperature at which the divergence of the Cooper susceptibility occurs. 
$\delta_{\mathrm{h}}^{}$ for $T_{\mathrm{c}}^{0}$ and $h_{pd}^{}$ are evaluated at $T=170\,\mathrm{K}$.}
\label{figure:1}
\end{figure}
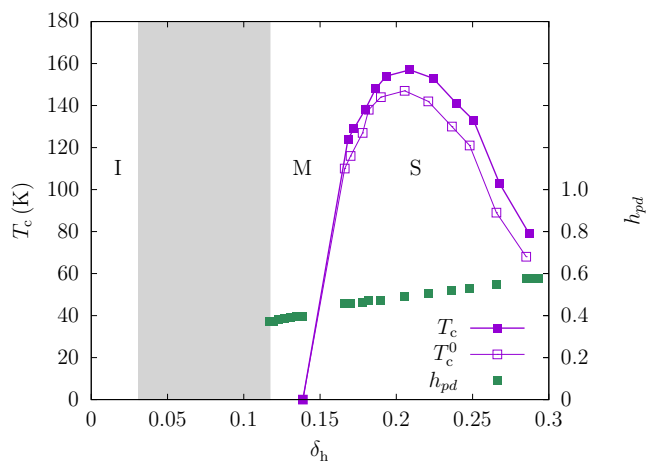
Figure~\ref{figure:1} summarizes these results with the doping dependences of $T_{\mathrm{c}}$ and $h_{pd}^{}$. 
Comparing $T_{\mathrm{c}}$ with $T_{\mathrm{c}}^{0}$, at which the divergence of the Cooper susceptibility occurs, 
$T_{\mathrm{c}}$ is higher than $T_{\mathrm{c}}^{0}$ by $10\sim14\,\mathrm{K}$ since $T_{\mathrm{c}}$ reflects the fluctuation of 
${\mathit{\Phi}}_{\mu \nu}^{}({\mathbf{k}},\mathrm{i}\epsilon_n)$. 
%%% 4
While $h_{pd}^{}$ increases monotonically with $\delta_{\mathrm{h}}^{}$, 
$T_{\mathrm{c}}$ reaches its maximum, $157\,\mathrm{K}$, at $\delta_{\mathrm{h}}^{} = 0.209$ and then decreases. 
This doping dependence of $T_{\mathrm{c}}$ reproduces the dome-shaped superconducting phase 
that is typical for the hole-doped HTSC~\cite{Tsukada06,TYoshida06}. 
%%% 4
This behavior is related to the doping dependence of the density of states, and it will be explained later. 

\begin{figure}[ht]
\begin{center}
\resizebox{88mm}{!}{\input{fig2.tex}}
\end{center}
\vspace{-11ex}
\caption{Temperature dependences of $\delta_{\mathrm{h}}^{}-\delta_{\mathrm{h}}^{0}$: (a) $\delta_{\mathrm{h}}^{0} \leq 0.190$ and (b) $\delta_{\mathrm{h}}^{0} \geq 0.205$.}
\label{figure:2}
\end{figure}
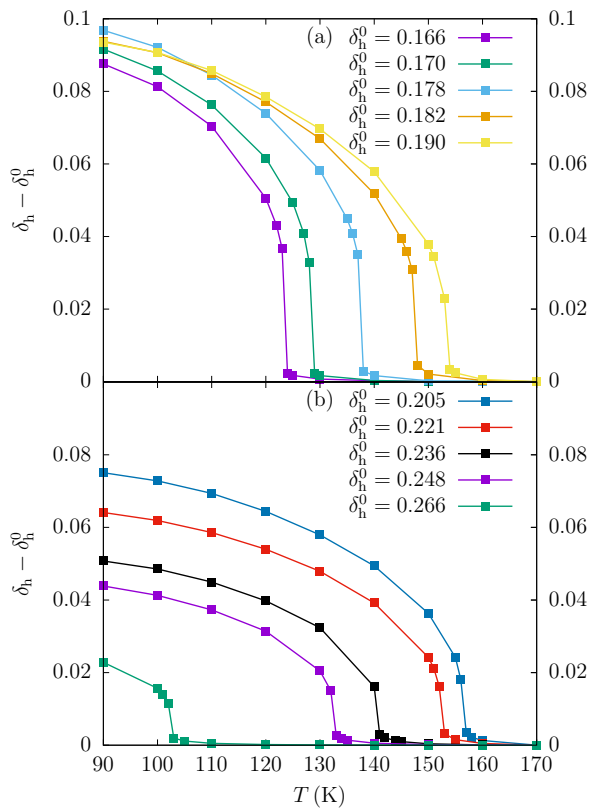

Figure~\ref{figure:2} shows the temperature dependences of $\delta_{\mathrm{h}}^{}-\delta_{\mathrm{h}}^{0}$ for every $\delta_{\mathrm{h}}^{0}$, 
which are used to determine $T_{\mathrm{c}}$. 
Here, I define the temperature at which $\delta_{\mathrm{h}}^{}-\delta_{\mathrm{h}}^{0}$ jumps as $T_{\mathrm{c}}$. 
The jumps of $\delta_{\mathrm{h}}^{}-\delta_{\mathrm{h}}^{0}$ at $T_{\mathrm{c}}$ in the underdoped regime, $\delta_{\mathrm{h}}^{0} \leq 0.190$ [Fig.~\ref{figure:2}(a)], 
are larger than those in the overdoped regime, $\delta_{\mathrm{h}}^{0} \geq 0.205$ [Fig.~\ref{figure:2}(b)]. In other words, 
while strong coupling superconductivity is established in the underdoped regime, the superconductivity in the overdoped regime remains with weak coupling. 
This tendency must be reflected in the superconducting gap magnitude, which has been shown to decrease with doping 
by the low-temperature specific heats of La$_{2-x}$Sr$_x$CuO$_4$~\cite{Wen05,Wang07}.

\begin{figure}[ht]
\begin{center}
\resizebox{88mm}{!}{\input{fig3.tex}}
\end{center}
\vspace{-4ex}
\caption{Doping dependences of $n_{d\mathrm{h}}^{}$, $n_{p\mathrm{h}}^{}$, $n_{d\mathrm{h}}^{0}$, and $n_{p\mathrm{h}}^{0}$ at $T=170\,\mathrm{K}$.}
\label{figure:3}
\end{figure}
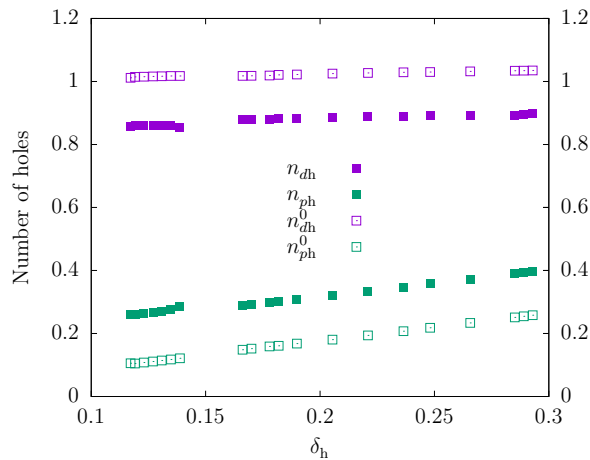

The electronic states of the obtained solutions are reconstructed from the unperturbed ground state. 
Figure~\ref{figure:3} shows the doping dependences of $n_{d\mathrm{h}}^{}$, $n_{p\mathrm{h}}^{}$, $n_{d\mathrm{h}}^{0}$, and $n_{p\mathrm{h}}^{0}$ 
at $T=170\,\mathrm{K}$. 
Here, $n_{d\mathrm{h}}^{0} = 2 - n_{d}^{}$ and $n_{p\mathrm{h}}^{0} = \delta_{\mathrm{h}}^{0} + n_{d}^{} - 1$, 
and $n_{d\mathrm{h}}^{0}$ and $n_{p\mathrm{h}}^{0}$ are the numbers of $d$ and $p$ holes in the unperturbed ground state, respectively.
As shown in Fig.~\ref{figure:3}, holes are transferred from the $d$ band to the $p$ band due to the charge fluctuations $\chi_J^G({\mathbf{q}},\mathrm{i}\omega_m)$
via the normal self-energies ${\mathit{\Sigma}}_{\mu \nu}^{}({\mathbf{k}},\mathrm{i}\epsilon_n)$ in Eqs.~(\ref{eq:20})--(\ref{eq:21}).
As a consequence, while $n_{p\mathrm{h}}^{}$ mainly increases with $\delta_{\mathrm{h}}^{}$, 
$n_{d\mathrm{h}}^{} < 1$, which means that the $d$ band is always electron doped. Since the $d$ band deviates from the half-filling due to the charge fluctuations, 
there is room for the pair-hopping interaction in Eq.~(\ref{eq:12}) to work effectively between the $p$ and $d$ electrons despite the strong correlations among $d$ electrons. 
Later, I show how the pair-hopping interaction works for the superconductivity in the analysis of the superconducting gap function.  

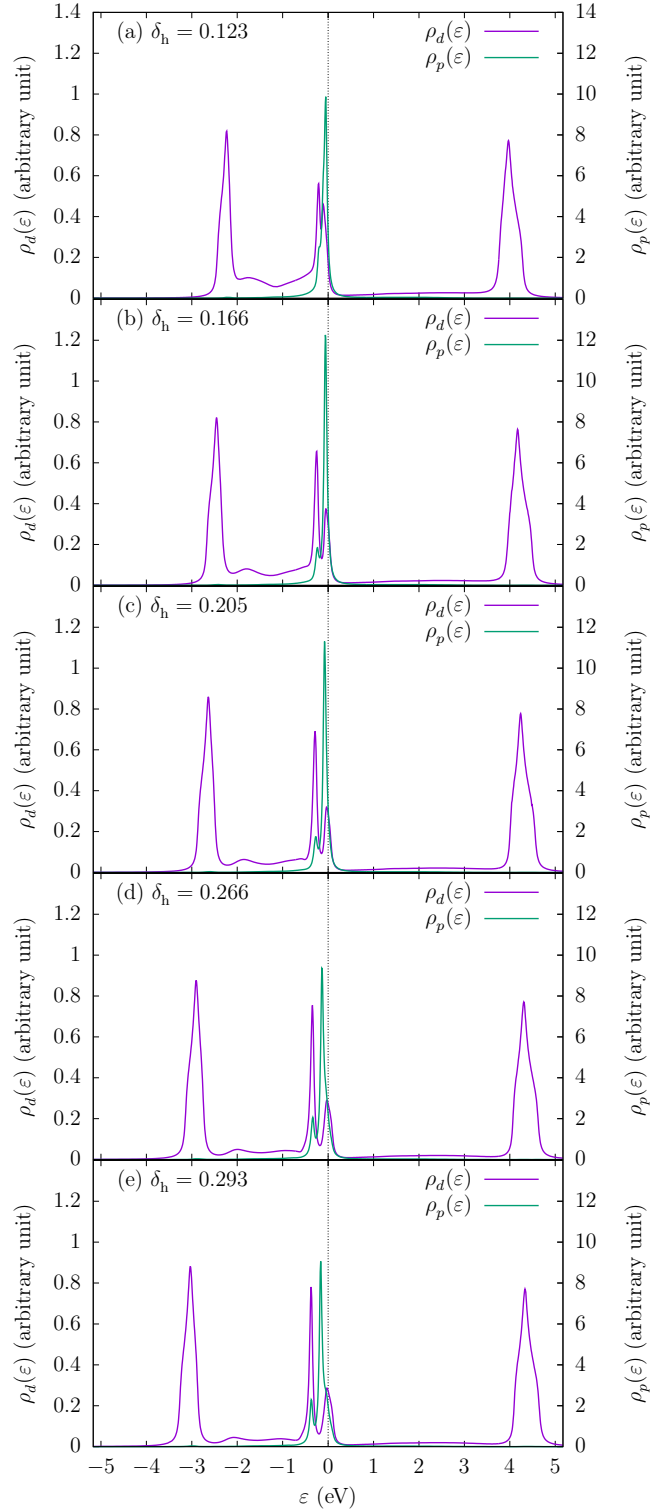
\begin{figure}[ht]
\begin{center}
\resizebox{88mm}{!}{\input{fig4.tex}}
\end{center}
\vspace{-58ex}
\caption{$\rho_{d}^{}(\varepsilon)$ and $\rho_{p}^{}(\varepsilon)$ at $T=170\,\mathrm{K}$: (a) $\delta_{\mathrm{h}}^{} = 0.123$, (b) $\delta_{\mathrm{h}}^{} = 0.166$, (c) $\delta_{\mathrm{h}}^{} = 0.205$, (d) $\delta_{\mathrm{h}}^{} = 0.266$, and (e) $\delta_{\mathrm{h}}^{} = 0.293$.}
\label{figure:4}
\end{figure}

\begin{figure}[ht]
\begin{center}
\resizebox{88mm}{!}{\input{fig5.tex}}
\end{center}
\vspace{-9ex}
\caption{Doping dependences of $\rho_{d}^{}(\varepsilon)$ and $\rho_{p}^{}(\varepsilon)$ at $T=170\,\mathrm{K}$ around the coherent band: (a) $\rho_{d}^{}(\varepsilon)$ and (b) $\rho_{p}^{}(\varepsilon)$.}
\label{figure:5}
\end{figure}
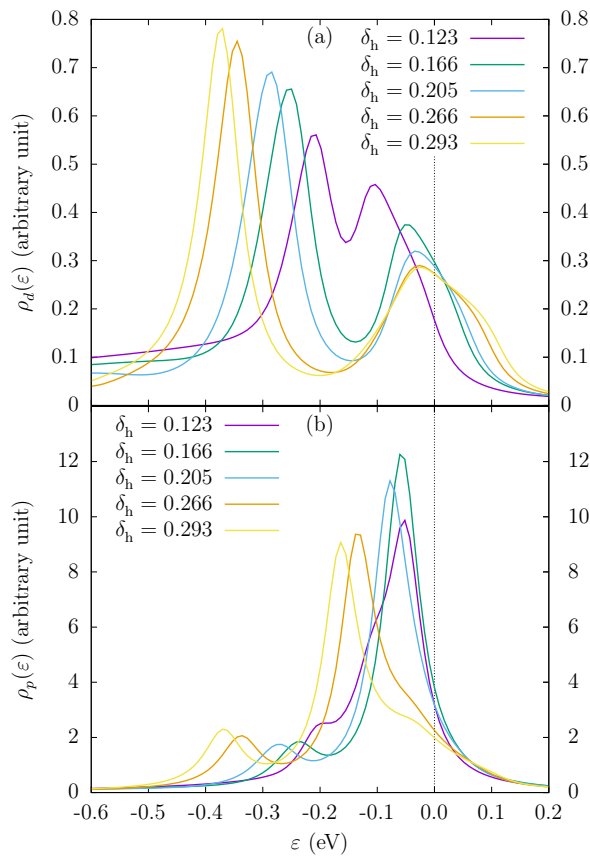

Figures~\ref{figure:4} and \ref{figure:5} show the doping dependences of $\rho_{d}^{}(\varepsilon)$ and $\rho_{p}^{}(\varepsilon)$ at $T = 170\,\mathrm{K}$, 
which elucidate how the dome-shaped superconducting phase develops.
Here, 
\begin{eqnarray}
\rho_{d}^{}(\varepsilon) & = & - \,\frac{1}{\pi N} \sum_{{\mathbf{k}}} \left.{\mathrm{Im}}\,
G_{dd}^{}({\mathbf{k}},\mathrm{i}\epsilon_n)\right|_{\mathrm{i}\epsilon_n \rightarrow \varepsilon + i\eta},\\
\rho_{p}^{}(\varepsilon) & = & - \,\frac{1}{\pi N} \sum_{{\mathbf{k}}} \left[\left.{\mathrm{Im}}\,G_{xx}^{}({\mathbf{k}},\mathrm{i}\epsilon_n)\right|_{\mathrm{i}\epsilon_n \rightarrow \varepsilon + i\eta}+\left.{\mathrm{Im}}\,G_{yy}^{}({\mathbf{k}},\mathrm{i}\epsilon_n)\right|_{\mathrm{i}\epsilon_n \rightarrow \varepsilon + i\eta}\right].
\end{eqnarray}
$\mathrm{i}\epsilon_n \rightarrow \varepsilon + i\eta$ indicates the performance of analytic continuation, for which I use the Pad$\acute{\mathrm{e}}$ approximation~\cite{Vidberg77} and $\eta=0.04\,t_{pd}$. 
$\rho_{d}^{}(\varepsilon)$ and $\rho_{p}^{}(\varepsilon)$ represent the density of states (DOS) of the $d$ and $p$ bands, respectively. 
It has been confirmed that the peak positions of $\rho_{d}^{}(\varepsilon)$ and $\rho_{p}^{}(\varepsilon)$ hardly change 
even if $\eta$ is changed to $0.02\,t_{pd}$. 
The three blocks appearing in $\rho_{d}^{}(\varepsilon)$ correspond to the lower Hubbard band, coherent band, and upper Hubbard band. 
The coherent band is split due to the hybridization with the $p$ band, and the higher peak energy approaches the Fermi level with the hole doping [Fig.~\ref{figure:5}(a)]. 
In contrast, $\rho_{p}^{}(\varepsilon)$ is large in the coherent band only. 
Reflecting that the holes are mainly doped into the $p$ band, as shown in Fig.~\ref{figure:3}, 
the peak energy moves away from the Fermi level with the hole doping [Fig.~\ref{figure:5}(b)]. 
Due to the competitive effect of these changes in DOS in the coherent band, there is a dome-shaped superconducting phase.

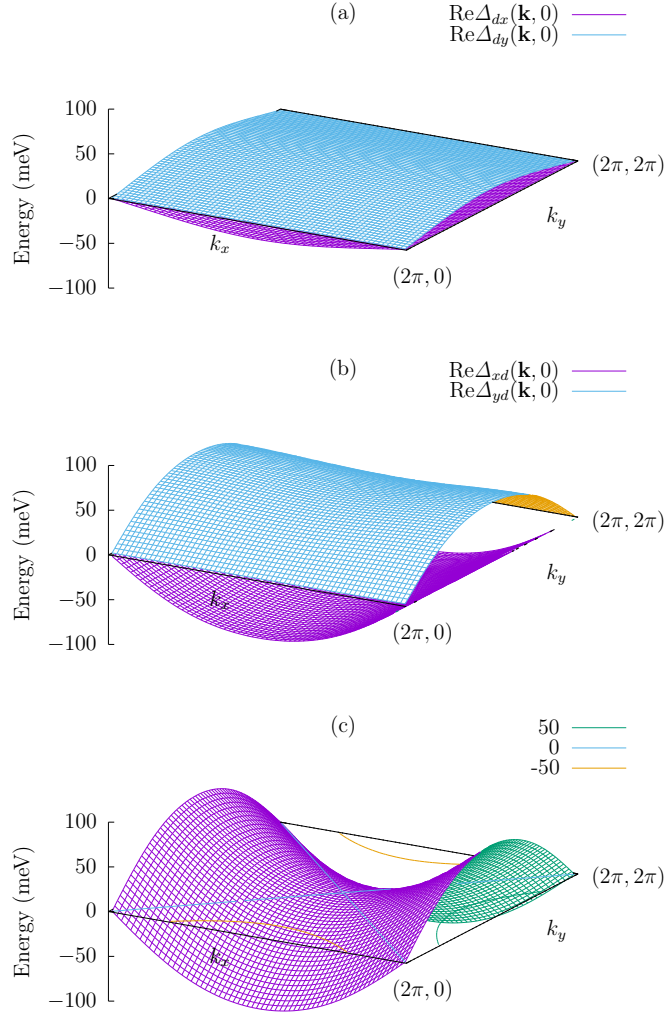
\begin{figure}[ht]
\begin{center}
\resizebox{88mm}{!}{\input{fig6a.tex}}\vspace{-8ex}
\resizebox{88mm}{!}{\input{fig6b.tex}}\vspace{-8ex}
\resizebox{88mm}{!}{\input{fig6c.tex}}
\end{center}
\vspace{-8ex}
\caption{${\mathit \Delta}_{d\alpha}({\mathbf{k}},0)$ and ${\mathit \Delta}_{\alpha d}({\mathbf{k}},0)$ at $\delta_{\mathrm{h}}^{} = 0.237$ and $T=110\,{\mathrm{K}}$: (a) ${\mathrm{Re}}{\mathit \Delta}_{dx}({\mathbf k},0)$ and ${\mathrm{Re}}{\mathit \Delta}_{dy}({\mathbf k},0)$, (b) ${\mathrm{Re}}{\mathit \Delta}_{xd}({\mathbf k},0)$ and ${\mathrm{Re}}{\mathit \Delta}_{yd}({\mathbf k},0)$, and (c) ${\mathrm{Re}}\{{\mathit{\Delta}}_{dx}^{}({\mathbf{k}},0)+{\mathit{\Delta}}_{dy}^{}({\mathbf{k}},0)+{\mathit{\Delta}}_{xd}^{}({\mathbf{k}},0)+{\mathit{\Delta}}_{yd}^{}({\mathbf{k}},0)\}$.}
\label{figure:6}
\end{figure}

The superconducting gap function, given in matrix form by 
$[\hat{\mathit{\Delta}}({\mathbf{k}},\varepsilon)]_{\mu\nu}^{}  \equiv {\mathit{\Delta}}_{\mu\nu}^{}({\mathbf{k}},\varepsilon)$, 
is defined as follows:
\begin{equation}
\label{eq:22}
\hat{\mathit{\Delta}}({\mathbf{k}},\varepsilon) =  \left.\mathrm{i}\epsilon_n\,{\mathrm{Im}}\,\hat{G}({\mathbf{k}},\mathrm{i}\epsilon_n) \cdot
\hat{\mathit{\Phi}}(\mathbf{k},\mathrm{i}\epsilon_n)\right|_{\mathrm{i}\epsilon_n \rightarrow \varepsilon + i\eta},
\end{equation}
where $[\hat{G}({\mathbf{k}},\mathrm{i}\epsilon_n)]_{\mu\nu}^{} \equiv G_{\mu\nu}^{}({\mathbf{k}},\mathrm{i}\epsilon_n)$ 
and $[\hat{\mathit{\Phi}}({\mathbf{k}},\mathrm{i}\epsilon_n)]_{\mu\nu}^{} \equiv {\mathit{\Phi}}_{\mu\nu}^{}({\mathbf{k}},\mathrm{i}\epsilon_n)$. 
Here, I use the Pad$\acute{\mathrm{e}}$ approximation for analytic continuation and $\eta=0.04\,t_{pd}$. 
It has been confirmed that ${\mathit{\Delta}}_{\mu\nu}^{}({\mathbf{k}},0)$ hardly changes even if $\eta$ is changed to $0.02\,t_{pd}$. 
The components of the superconducting gap function are classified into two classes. 
The first class is composed of ${\mathit{\Delta}}_{d\alpha}({\mathbf{k}},\varepsilon)$ and ${\mathit{\Delta}}_{\alpha d}({\mathbf{k}},\varepsilon)$, 
where $\alpha$ runs over $x$ and $y$. The real parts of these components with $\varepsilon=0$ are shown in Fig.~\ref{figure:6}. 
The imaginary parts of these components with $\varepsilon=0$ are all zero. One can see that 
${\mathrm{Re}}{\mathit{\Delta}}_{d\alpha}({\mathbf{k}},0)$ [Fig.~\ref{figure:6}(a)] and ${\mathrm{Re}}{\mathit{\Delta}}_{\alpha d}({\mathbf{k}},0)$ [Fig.~\ref{figure:6}(b)] are roughly proportional to $\sin \frac{k_\alpha}{2}$. These momentum dependences are derived from the first-order terms of $J$ 
in Eqs.~(\ref{eq:30}) and (\ref{eq:31}), which originate from the exchange interaction in Eq.~(\ref{eq:11}). Thus, ${\mathit{\Delta}}_{d\alpha}({\mathbf{k}},\varepsilon)$ and ${\mathit{\Delta}}_{\alpha d}({\mathbf{k}},\varepsilon)$ emerge due to the exchange interaction via the SK mechanism. 
It can be verified that the SK mechanism can work effectively with the exchange interaction only if $h_{pd}^{} > 0$. 
Moreover, the signs of ${\mathrm{Re}}{\mathit{\Delta}}_{dy}^{}({\mathbf{k}},0)$ and ${\mathrm{Re}}{\mathit{\Delta}}_{yd}^{}({\mathbf{k}},0)$
differ from the signs of ${\mathrm{Re}}{\mathit{\Delta}}_{dx}^{}({\mathbf{k}},0)$ and ${\mathrm{Re}}{\mathit{\Delta}}_{xd}^{}({\mathbf{k}},0)$, respectively. 
Therefore, as shown in Fig.~\ref{figure:6}(c), the linear combination ${\mathrm{Re}}\{{\mathit{\Delta}}_{dx}^{}({\mathbf{k}},0)+{\mathit{\Delta}}_{dy}^{}({\mathbf{k}},0)+{\mathit{\Delta}}_{xd}^{}({\mathbf{k}},0)+{\mathit{\Delta}}_{yd}^{}({\mathbf{k}},0)\}$ has line nodes at $k_x=k_y$ and $k_x=-k_y$ 
and behaves like a nodal $d_{x^2-y^2}$-wave superconducting gap.

The second class is composed of ${\mathit{\Delta}}_{dd}({\mathbf{k}},\varepsilon)$ and ${\mathit{\Delta}}_{\alpha \beta}({\mathbf{k}},\varepsilon)$, 
where $\alpha$ and $\beta$ run over $x$ and $y$. The real part of these components with $\varepsilon=0$ are shown in Fig.~\ref{figure:7}. 
The imaginary part of these components with $\varepsilon=0$ are all zero. 
${\mathrm{Re}}{\mathit{\Delta}}_{\alpha \beta}({\mathbf{k}},0)$ [Fig.~\ref{figure:7}(b) and (c)] is roughly proportional to $\sin \frac{k_\alpha}{2}\sin \frac{k_\beta}{2}$. 
This momentum dependence is derived from the first-order term of $J$ in Eq.~(\ref{eq:32}), which originates from the pair-hopping interaction in Eq.~(\ref{eq:12}). 
${\mathrm{Re}}{\mathit{\Delta}}_{dd}({\mathbf{k}},0)$ [Fig.~\ref{figure:7}(a)] has the momentum dependence of an extended $s$-wave, 
and its sign differs from the signs of ${\mathrm{Re}}{\mathit{\Delta}}_{xx}^{}({\mathbf{k}},0)$ and ${\mathrm{Re}}{\mathit{\Delta}}_{yy}^{}({\mathbf{k}},0)$. 
Thus, ${\mathit{\Delta}}_{dd}({\mathbf{k}},\varepsilon)$--as well as ${\mathit{\Delta}}_{\alpha \beta}({\mathbf{k}},\varepsilon)$--emerges due to the pair-hopping interaction via the SK mechanism, although it is affected by the terms of $U$ and $U^2$ 
in Eq.~(\ref{eq:18}). It can be verified that the SK mechanism can work with the pair-hopping interaction even if $h_{pd}^{} = 0$. 
Moreover, the absolute values of ${\mathrm{Re}}{\mathit{\Delta}}_{dd}({\mathbf{k}},0)$ are larger than those of ${\mathrm{Re}}{\mathit{\Delta}}_{\alpha \beta}({\mathbf{k}},0)$ 
for all $\alpha$, $\beta$, and ${\mathbf{k}}$. Therefore, as shown in Fig.~\ref{figure:7}(d), the linear combination ${\mathrm{Re}}\{{\mathit{\Delta}}_{dd}^{}({\mathbf{k}},0)+{\mathit{\Delta}}_{xx}^{}({\mathbf{k}},0)+{\mathit{\Delta}}_{xy}^{}({\mathbf{k}},0)+{\mathit{\Delta}}_{yx}^{}({\mathbf{k}},0)+{\mathit{\Delta}}_{yy}^{}({\mathbf{k}},0)\}$ 
behaves like an extended $s$-wave superconducting gap.
\begin{figure}[ht]
\begin{center}
\resizebox{88mm}{!}{\input{fig7a.tex}}\vspace{-8ex}
\resizebox{88mm}{!}{\input{fig7b.tex}}\vspace{-8ex}
\resizebox{88mm}{!}{\input{fig7c.tex}}\vspace{-8ex}
\resizebox{88mm}{!}{\input{fig7d.tex}}
\end{center}
\vspace{-8ex}
\caption{${\mathit \Delta}_{dd}({\mathbf{k}},0)$ and ${\mathit \Delta}_{\alpha \beta}({\mathbf{k}},0)$ at $\delta_{\mathrm{h}}^{} = 0.237$ and $T=110\,{\mathrm{K}}$: (a) ${\mathrm{Re}}{\mathit \Delta}_{dd}({\mathbf k},0)$, (b) ${\mathrm{Re}}{\mathit \Delta}_{xx}({\mathbf k},0)$ and ${\mathrm{Re}}{\mathit \Delta}_{xy}({\mathbf k},0)$, (b) ${\mathrm{Re}}{\mathit \Delta}_{yx}({\mathbf k},0)$ and ${\mathrm{Re}}{\mathit \Delta}_{yy}({\mathbf k},0)$, and (d) ${\mathrm{Re}}\{{\mathit{\Delta}}_{dd}^{}({\mathbf{k}},0)+{\mathit{\Delta}}_{xx}^{}({\mathbf{k}},0)+{\mathit{\Delta}}_{xy}^{}({\mathbf{k}},0)+{\mathit{\Delta}}_{yx}^{}({\mathbf{k}},0)+{\mathit{\Delta}}_{yy}^{}({\mathbf{k}},0)\}$.}
\label{figure:7}
\end{figure}

I have shown that the coexistence of extended $s$- and $d$-wave gaps is theoretically possible in the three-band $t$--$J$--$U$ model. 
The coexistence of $s$- and $d$-wave gaps was originally proposed to explain the apparently conflicting results of 
scanning tunnelling spectroscopy in HTSC~\cite{Muller95}. 
%%%
So far, the experiments on Bi$_2$Sr$_2$CaCu$_2$O$_{8+\delta}$ (Bi2212) utilizing tunneling effect in the superconducting phase, which include $c$-axis twist Josephson 
experiments~\cite{Li99,Takano02,Takano03, Takano04, Latyshev04, Klemm05, Zhu21}, $c$-axis scanning tunnelling microscopy~\cite{Misra02, Hoogenboom03, Zhong16}, 
and intrinsic Josephson junction terahertz emission~\cite{Kashiwagi15_1, Kashiwagi15_2}, provide clear evidences that 
the superconducting gap has $s$-wave symmetry. These experiments can directly observe the superconducting gap without breaking the gap into quasiparticles, 
and this result is also reasonable for the coexistence of $s$- and $d$-wave gaps. In the superconducting phase, where $s$- and $d$-wave gaps coexist, 
the $s$-wave gap is dominant over the $d$-wave gap in the energy $|\varepsilon| < {\mathit \Delta}_s$, where ${\mathit \Delta}_s$ indicates the $s$-wave gap magnitude. 
On the other hand, only when the $d$-wave gap magnitude ${\mathit \Delta}_d$ satisfies ${\mathit \Delta}_d > {\mathit \Delta}_s$, 
the $d$-wave gap becomes dominant over the $s$-wave gap in the energy $|\varepsilon| > {\mathit \Delta}_s$.
%%%

%%% low-temperature specific heat
In contrast, the quasiparticles from the $d$-wave gap can be observed in the energy $|\varepsilon| < {\mathit \Delta}_s$, 
where their excitation energies are always smaller than those of the quasiparticles from the $s$-wave gap. 
Thus, the experimental method breaking the gap into quasiparticles does mainly observe the $d$-wave gap. 
For example, both temperature and magnetic field dependences of low-temperature specific heat indicate that the $d$-wave superconducting gap exists 
in near optimally doped Bi$_2$Sr$_{2-x}$La$_x$CuO$_{6+\delta}\,(x \sim 0.4)$~\cite{Wang11}. 
%%% low-temperature specific heat

%%% ARPES
The above discussion holds even if the $d$-wave gap is not a superconducting gap. The angle-resolved photoemission spectroscopy (ARPES) experiment on Bi2212 
shows the marked change of temperature dependence of spectral intensity across critical value $p_c \sim 0.19$ with hole doping~\cite{Chen19}. 
This change with hole doping $p$ can be interpreted as a result of the coexistence of $s$- and $d$-wave gaps 
when we replace energy with temperature in the above discussion. 
For $p < p_c$, the $d$-wave gap affects the electronic structure above $T_c$ if ${\mathit \Delta}_d > {\mathit \Delta}_s$. 
The electronic structure affected by the $d$-wave gap is called pseudogap. 
However, for $p > p_c$, both the $s$- and $d$-wave gaps do not affect the electronic structure above $T_c$ if ${\mathit \Delta}_d < {\mathit \Delta}_s$. 
Therefore, the pseudogap disappears across $p_c$ with hole doping, which has also been observed by the ARPES experiment~\cite{Chen19}. 
%%% ARPES

Furthermore, Raman spectroscopy~\cite{Masui03} and the 
magnetic field penetration depth measurement by muon-spin rotation~\cite{Khasanov07_01,Khasanov07_02,Khasanov08} have provided 
evidence that supports the coexistence of $s$- and $d$-wave gaps in hole-doped HTSC. 
In theoretical work, the possibility of the coexistence of an 
extended $s$- and $d$-wave superconducting state has been shown with the analysis of the 2D $t$--$J$ model considering fluctuation effects~\cite{Mallik20}, 
and further experimental and theoretical research that assumes such coexistence is desired in the future. 

I conclude by comparing the obtained superconducting state to that found in other theoretical work. 
The $d_{x^2-y^2}$-wave superconducting gap composed of ${\mathit{\Delta}}_{d\alpha}({\mathbf{k}},\varepsilon)$ and ${\mathit{\Delta}}_{\alpha d}({\mathbf{k}},\varepsilon)$, 
which emerges due to the exchange interaction via the SK mechanism, corresponds to the one mediated by antiferromagnetic spin fluctuations (AFSF)~\cite{Moriya00}. 
This is clear because the superexchange interaction among $d$ electrons, which is responsible for the AFSF, 
can be derived from the exchange interaction between $d$ and $p$ electrons. 
In general, once the superexchange interaction acts between charge carriers, the $d_{x^2-y^2}$-wave superconductivity can emerge~\cite{Que87}.
Moreover, the $d_{x^2-y^2}$-wave superconductivity in my model can emerge only with the $d$-$p$ band hybridization. 
Therefore, it must be important that the $d$ electron is implicitly hybridized with the $p$ electron in the AFSF-mediated superconductivity. 
This speculation is supported by the studies of Kondo lattice models proposed for copper oxide~\cite{Prelovsek88,Ramsak89,Castellani88_1,Castellani88_2,Cancrini91,Kamimura88,Andrei89,Hatsugai89}. 
The Kondo interaction between localized $d$ spin and $p$ electron in Kondo lattice models corresponds to the exchange interaction between $d$ and $p$ electrons 
in the large-$U$ limit of my model. The studies of Kondo lattice models indicate that superconductivity emerges due to the Kondo effect, 
the compensation for the localized $d$ spin by the $p$ electrons via the Kondo interaction. 
As the Kondo effect corresponds to the formation of a Fermi liquid state through the $d$-$p$ band hybridization~\cite{KYamada92}, 
the superconductivity in Kondo lattice models is consistent with the $d_{x^2-y^2}$-wave superconductivity in my model.

The extended $s$-wave superconducting gap composed 
of ${\mathit{\Delta}}_{dd}({\mathbf{k}},\varepsilon)$ and ${\mathit{\Delta}}_{\alpha \beta}({\mathbf{k}},\varepsilon)$, 
which emerges due to the pair-hopping interaction via the SK mechanism, corresponds to the kinetic-energy-driven superconductivity 
of the single-band $t$--$J$ model~\cite{Hirsch92,Tsunetsugu98,Tsunetsugu99,Imada01,Sarker09,Sarker12,Feng12,Feng15,Gao18}. 
In the kinetic-energy-driven superconductivity, the charge carriers form the superconducting pairs to gain kinetic energy. 
This energy gain can be derived from the pair-hopping interaction between $p$ and $d$ electrons, 
which works to form the extended $s$-wave superconducting gap in my model. 

\section{Summary}
In summary, the three-band $t$--$J$--$U$ model is derived assuming that the double occupancy by $d$ electrons is not excluded. 
When the $d$ electron is hybridized with the $p$ electron through exchange and pair-hopping interactions, 
the dome-shaped superconducting phase can be reproduced despite the strong correlations among $d$ electrons. 
In the superconducting phase, the extended $s$- and $d_{x^2-y^2}$-wave superconducting gaps coexist. 
The extended $s$-wave gap emerges due to the pair-hopping interaction via the SK mechanism, which works effectively due to the charge fluctuations. 
In contrast, the $d_{x^2-y^2}$-wave gap emerges due to the exchange interaction via the SK mechanism, 
which can effectively work only with the $d$-$p$ band hybridization. The obtained superconducting state is consistent with those in other theoretical work, 
which include AFSF-mediated superconductivity and kinetic-energy-driven superconductivity.  

\section*{Acknowledgements}
The author would like to thank Prof. T. Tohyama and Prof. H. Yamase for their invaluable comments. 
The author is also grateful to anonymous reviewers for providing informations on many important references and insightful comments.

\section*{References}

\end{document}

%% file: fig1.tex
% GNUPLOT: LaTeX picture with Postscript
\begingroup
  \makeatletter
  \providecommand\color[2][]{%
    \GenericError{(gnuplot) \space\space\space\@spaces}{%
      Package color not loaded in conjunction with
      terminal option `colourtext'%
    }{See the gnuplot documentation for explanation.%
    }{Either use 'blacktext' in gnuplot or load the package
      color.sty in LaTeX.}%
    \renewcommand\color[2][]{}%
  }%
  \providecommand\includegraphics[2][]{%
    \GenericError{(gnuplot) \space\space\space\@spaces}{%
      Package graphicx or graphics not loaded%
    }{See the gnuplot documentation for explanation.%
    }{The gnuplot epslatex terminal needs graphicx.sty or graphics.sty.}%
    \renewcommand\includegraphics[2][]{}%
  }%
  \providecommand\rotatebox[2]{#2}%
  \@ifundefined{ifGPcolor}{%
    \newif\ifGPcolor
    \GPcolortrue
  }{}%
  \@ifundefined{ifGPblacktext}{%
    \newif\ifGPblacktext
    \GPblacktexttrue
  }{}%
  % define a \g@addto@macro without @ in the name:
  \let\gplgaddtomacro\g@addto@macro
  % define empty templates for all commands taking text:
  \gdef\gplbacktext{}%
  \gdef\gplfronttext{}%
  \makeatother
  \ifGPblacktext
    % no textcolor at all
    \def\colorrgb#1{}%
    \def\colorgray#1{}%
  \else
    % gray or color?
    \ifGPcolor
      \def\colorrgb#1{\color[rgb]{#1}}%
      \def\colorgray#1{\color[gray]{#1}}%
      \expandafter\def\csname LTw\endcsname{\color{white}}%
      \expandafter\def\csname LTb\endcsname{\color{black}}%
      \expandafter\def\csname LTa\endcsname{\color{black}}%
      \expandafter\def\csname LT0\endcsname{\color[rgb]{1,0,0}}%
      \expandafter\def\csname LT1\endcsname{\color[rgb]{0,1,0}}%
      \expandafter\def\csname LT2\endcsname{\color[rgb]{0,0,1}}%
      \expandafter\def\csname LT3\endcsname{\color[rgb]{1,0,1}}%
      \expandafter\def\csname LT4\endcsname{\color[rgb]{0,1,1}}%
      \expandafter\def\csname LT5\endcsname{\color[rgb]{1,1,0}}%
      \expandafter\def\csname LT6\endcsname{\color[rgb]{0,0,0}}%
      \expandafter\def\csname LT7\endcsname{\color[rgb]{1,0.3,0}}%
      \expandafter\def\csname LT8\endcsname{\color[rgb]{0.5,0.5,0.5}}%
    \else
      % gray
      \def\colorrgb#1{\color{black}}%
      \def\colorgray#1{\color[gray]{#1}}%
      \expandafter\def\csname LTw\endcsname{\color{white}}%
      \expandafter\def\csname LTb\endcsname{\color{black}}%
      \expandafter\def\csname LTa\endcsname{\color{black}}%
      \expandafter\def\csname LT0\endcsname{\color{black}}%
      \expandafter\def\csname LT1\endcsname{\color{black}}%
      \expandafter\def\csname LT2\endcsname{\color{black}}%
      \expandafter\def\csname LT3\endcsname{\color{black}}%
      \expandafter\def\csname LT4\endcsname{\color{black}}%
      \expandafter\def\csname LT5\endcsname{\color{black}}%
      \expandafter\def\csname LT6\endcsname{\color{black}}%
      \expandafter\def\csname LT7\endcsname{\color{black}}%
      \expandafter\def\csname LT8\endcsname{\color{black}}%
    \fi
  \fi
    \setlength{\unitlength}{0.0500bp}%
    \ifx\gptboxheight\undefined%
      \newlength{\gptboxheight}%
      \newlength{\gptboxwidth}%
      \newsavebox{\gptboxtext}%
    \fi%
    \setlength{\fboxrule}{0.5pt}%
    \setlength{\fboxsep}{1pt}%
\begin{picture}(7200.00,5040.00)%
    \gplgaddtomacro\gplbacktext{%
      \csname LTb\endcsname%%
      \put(814,704){\makebox(0,0)[r]{\strut{}0}}%
      \put(814,1161){\makebox(0,0)[r]{\strut{}20}}%
      \put(814,1618){\makebox(0,0)[r]{\strut{}40}}%
      \put(814,2076){\makebox(0,0)[r]{\strut{}60}}%
      \put(814,2533){\makebox(0,0)[r]{\strut{}80}}%
      \put(814,2990){\makebox(0,0)[r]{\strut{}100}}%
      \put(814,3447){\makebox(0,0)[r]{\strut{}120}}%
      \put(814,3905){\makebox(0,0)[r]{\strut{}140}}%
      \put(814,4362){\makebox(0,0)[r]{\strut{}160}}%
      \put(814,4819){\makebox(0,0)[r]{\strut{}180}}%
      \put(946,484){\makebox(0,0){\strut{}$0$}}%
      \put(1776,484){\makebox(0,0){\strut{}$0.05$}}%
      \put(2605,484){\makebox(0,0){\strut{}$0.1$}}%
      \put(3435,484){\makebox(0,0){\strut{}$0.15$}}%
      \put(4264,484){\makebox(0,0){\strut{}$0.2$}}%
      \put(5094,484){\makebox(0,0){\strut{}$0.25$}}%
      \put(5923,484){\makebox(0,0){\strut{}$0.3$}}%
      \put(6055,704){\makebox(0,0)[l]{\strut{}0}}%
      \put(6055,1161){\makebox(0,0)[l]{\strut{}0.2}}%
      \put(6055,1618){\makebox(0,0)[l]{\strut{}0.4}}%
      \put(6055,2076){\makebox(0,0)[l]{\strut{}0.6}}%
      \put(6055,2533){\makebox(0,0)[l]{\strut{}0.8}}%
      \put(6055,2990){\makebox(0,0)[l]{\strut{}1.0}}%
      \put(1195,3219){\makebox(0,0)[l]{\strut{}I}}%
      \put(3136,3219){\makebox(0,0)[l]{\strut{}M}}%
      \put(4413,3219){\makebox(0,0)[l]{\strut{}S}}%
    }%
    \gplgaddtomacro\gplfronttext{%
      \csname LTb\endcsname%%
      \put(198,2761){\rotatebox{-270}{\makebox(0,0){\strut{}$T_{\mathrm c}\,({\mathrm K})$}}}%
      \put(6847,2761){\rotatebox{-270}{\makebox(0,0){\strut{}$h_{pd}$}}}%
      \put(3434,154){\makebox(0,0){\strut{}$\delta_{\mathrm h}^{}$}}%
      \csname LTb\endcsname%%
      \put(4936,1482){\makebox(0,0)[r]{\strut{}$T_{\mathrm c}$}}%
      \csname LTb\endcsname%%
      \put(4936,1196){\makebox(0,0)[r]{\strut{}$T_{\mathrm c}^{0}$}}%
      \csname LTb\endcsname%%
      \put(4936,910){\makebox(0,0)[r]{\strut{}$h_{pd}$}}%
    }%
    \gplbacktext
    \put(0,0){\includegraphics{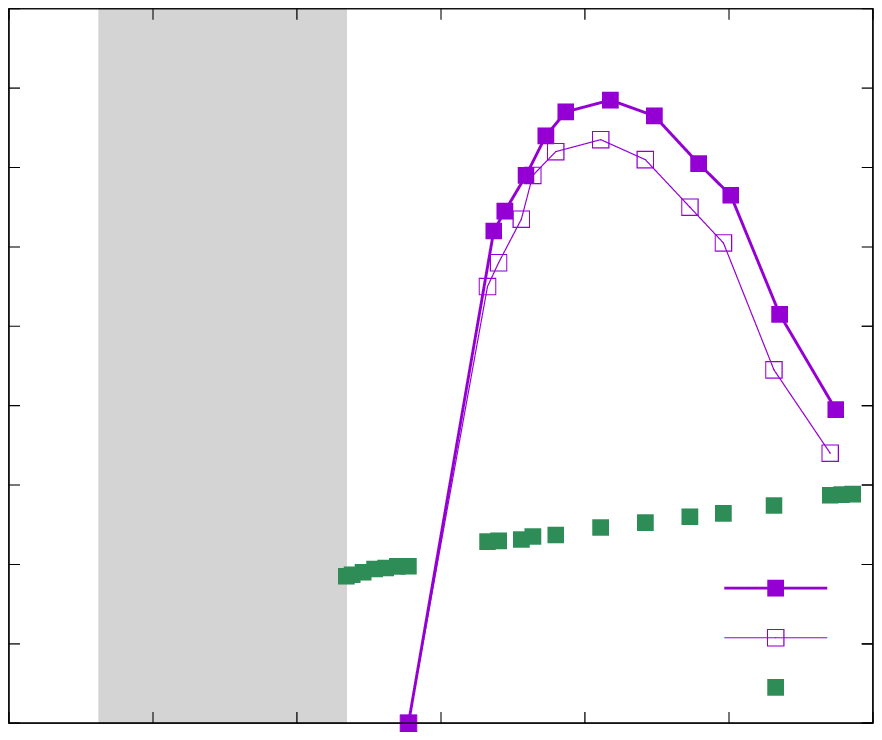}}%
    \gplfronttext
  \end{picture}%
\endgroup

%% file: fig2.tex
% GNUPLOT: LaTeX picture with Postscript
\begingroup
  \makeatletter
  \providecommand\color[2][]{%
    \GenericError{(gnuplot) \space\space\space\@spaces}{%
      Package color not loaded in conjunction with
      terminal option `colourtext'%
    }{See the gnuplot documentation for explanation.%
    }{Either use 'blacktext' in gnuplot or load the package
      color.sty in LaTeX.}%
    \renewcommand\color[2][]{}%
  }%
  \providecommand\includegraphics[2][]{%
    \GenericError{(gnuplot) \space\space\space\@spaces}{%
      Package graphicx or graphics not loaded%
    }{See the gnuplot documentation for explanation.%
    }{The gnuplot epslatex terminal needs graphicx.sty or graphics.sty.}%
    \renewcommand\includegraphics[2][]{}%
  }%
  \providecommand\rotatebox[2]{#2}%
  \@ifundefined{ifGPcolor}{%
    \newif\ifGPcolor
    \GPcolortrue
  }{}%
  \@ifundefined{ifGPblacktext}{%
    \newif\ifGPblacktext
    \GPblacktexttrue
  }{}%
  % define a \g@addto@macro without @ in the name:
  \let\gplgaddtomacro\g@addto@macro
  % define empty templates for all commands taking text:
  \gdef\gplbacktext{}%
  \gdef\gplfronttext{}%
  \makeatother
  \ifGPblacktext
    % no textcolor at all
    \def\colorrgb#1{}%
    \def\colorgray#1{}%
  \else
    % gray or color?
    \ifGPcolor
      \def\colorrgb#1{\color[rgb]{#1}}%
      \def\colorgray#1{\color[gray]{#1}}%
      \expandafter\def\csname LTw\endcsname{\color{white}}%
      \expandafter\def\csname LTb\endcsname{\color{black}}%
      \expandafter\def\csname LTa\endcsname{\color{black}}%
      \expandafter\def\csname LT0\endcsname{\color[rgb]{1,0,0}}%
      \expandafter\def\csname LT1\endcsname{\color[rgb]{0,1,0}}%
      \expandafter\def\csname LT2\endcsname{\color[rgb]{0,0,1}}%
      \expandafter\def\csname LT3\endcsname{\color[rgb]{1,0,1}}%
      \expandafter\def\csname LT4\endcsname{\color[rgb]{0,1,1}}%
      \expandafter\def\csname LT5\endcsname{\color[rgb]{1,1,0}}%
      \expandafter\def\csname LT6\endcsname{\color[rgb]{0,0,0}}%
      \expandafter\def\csname LT7\endcsname{\color[rgb]{1,0.3,0}}%
      \expandafter\def\csname LT8\endcsname{\color[rgb]{0.5,0.5,0.5}}%
    \else
      % gray
      \def\colorrgb#1{\color{black}}%
      \def\colorgray#1{\color[gray]{#1}}%
      \expandafter\def\csname LTw\endcsname{\color{white}}%
      \expandafter\def\csname LTb\endcsname{\color{black}}%
      \expandafter\def\csname LTa\endcsname{\color{black}}%
      \expandafter\def\csname LT0\endcsname{\color{black}}%
      \expandafter\def\csname LT1\endcsname{\color{black}}%
      \expandafter\def\csname LT2\endcsname{\color{black}}%
      \expandafter\def\csname LT3\endcsname{\color{black}}%
      \expandafter\def\csname LT4\endcsname{\color{black}}%
      \expandafter\def\csname LT5\endcsname{\color{black}}%
      \expandafter\def\csname LT6\endcsname{\color{black}}%
      \expandafter\def\csname LT7\endcsname{\color{black}}%
      \expandafter\def\csname LT8\endcsname{\color{black}}%
    \fi
  \fi
    \setlength{\unitlength}{0.0500bp}%
    \ifx\gptboxheight\undefined%
      \newlength{\gptboxheight}%
      \newlength{\gptboxwidth}%
      \newsavebox{\gptboxtext}%
    \fi%
    \setlength{\fboxrule}{0.5pt}%
    \setlength{\fboxsep}{1pt}%
\begin{picture}(7200.00,10080.00)%
    \gplgaddtomacro\gplbacktext{%
      \csname LTb\endcsname%%
      \put(946,5744){\makebox(0,0)[r]{\strut{}$0$}}%
      \put(946,6534){\makebox(0,0)[r]{\strut{}$0.02$}}%
      \put(946,7324){\makebox(0,0)[r]{\strut{}$0.04$}}%
      \put(946,8114){\makebox(0,0)[r]{\strut{}$0.06$}}%
      \put(946,8904){\makebox(0,0)[r]{\strut{}$0.08$}}%
      \put(946,9694){\makebox(0,0)[r]{\strut{}$0.1$}}%
      \put(1078,5524){\makebox(0,0){\strut{} }}%
      \put(1667,5524){\makebox(0,0){\strut{} }}%
      \put(2256,5524){\makebox(0,0){\strut{} }}%
      \put(2845,5524){\makebox(0,0){\strut{} }}%
      \put(3434,5524){\makebox(0,0){\strut{} }}%
      \put(4024,5524){\makebox(0,0){\strut{} }}%
      \put(4613,5524){\makebox(0,0){\strut{} }}%
      \put(5202,5524){\makebox(0,0){\strut{} }}%
      \put(5791,5524){\makebox(0,0){\strut{} }}%
      \put(5923,5744){\makebox(0,0)[l]{\strut{}$0$}}%
      \put(5923,6534){\makebox(0,0)[l]{\strut{}$0.02$}}%
      \put(5923,7324){\makebox(0,0)[l]{\strut{}$0.04$}}%
      \put(5923,8114){\makebox(0,0)[l]{\strut{}$0.06$}}%
      \put(5923,8904){\makebox(0,0)[l]{\strut{}$0.08$}}%
      \put(5923,9694){\makebox(0,0)[l]{\strut{}$0.1$}}%
    }%
    \gplgaddtomacro\gplfronttext{%
      \csname LTb\endcsname%%
      \put(198,7719){\rotatebox{-270}{\makebox(0,0){\strut{}$\delta_{\mathrm{h}}^{}-\delta_{\mathrm{h}}^{0}$}}}%
      \put(6847,7719){\rotatebox{-270}{\makebox(0,0){\strut{} }}}%
      \put(3434,5194){\makebox(0,0){\strut{} }}%
      \put(3434,9496){\makebox(0,0){\strut{}(a)}}%
      \csname LTb\endcsname%%
      \put(4804,9488){\makebox(0,0)[r]{\strut{}$\delta_{\mathrm{h}}^{0}=0.166$}}%
      \csname LTb\endcsname%%
      \put(4804,9202){\makebox(0,0)[r]{\strut{}$\delta_{\mathrm{h}}^{0}=0.170$}}%
      \csname LTb\endcsname%%
      \put(4804,8916){\makebox(0,0)[r]{\strut{}$\delta_{\mathrm{h}}^{0}=0.178$}}%
      \csname LTb\endcsname%%
      \put(4804,8630){\makebox(0,0)[r]{\strut{}$\delta_{\mathrm{h}}^{0}=0.182$}}%
      \csname LTb\endcsname%%
      \put(4804,8344){\makebox(0,0)[r]{\strut{}$\delta_{\mathrm{h}}^{0}=0.190$}}%
    }%
    \gplgaddtomacro\gplbacktext{%
      \csname LTb\endcsname%%
      \put(946,1787){\makebox(0,0)[r]{\strut{}$0$}}%
      \put(946,2577){\makebox(0,0)[r]{\strut{}$0.02$}}%
      \put(946,3367){\makebox(0,0)[r]{\strut{}$0.04$}}%
      \put(946,4158){\makebox(0,0)[r]{\strut{}$0.06$}}%
      \put(946,4948){\makebox(0,0)[r]{\strut{}$0.08$}}%
      \put(1078,1567){\makebox(0,0){\strut{}90}}%
      \put(1667,1567){\makebox(0,0){\strut{}100}}%
      \put(2256,1567){\makebox(0,0){\strut{}110}}%
      \put(2845,1567){\makebox(0,0){\strut{}120}}%
      \put(3434,1567){\makebox(0,0){\strut{}130}}%
      \put(4024,1567){\makebox(0,0){\strut{}140}}%
      \put(4613,1567){\makebox(0,0){\strut{}150}}%
      \put(5202,1567){\makebox(0,0){\strut{}160}}%
      \put(5791,1567){\makebox(0,0){\strut{}170}}%
      \put(5923,1787){\makebox(0,0)[l]{\strut{}$0$}}%
      \put(5923,2577){\makebox(0,0)[l]{\strut{}$0.02$}}%
      \put(5923,3367){\makebox(0,0)[l]{\strut{}$0.04$}}%
      \put(5923,4158){\makebox(0,0)[l]{\strut{}$0.06$}}%
      \put(5923,4948){\makebox(0,0)[l]{\strut{}$0.08$}}%
    }%
    \gplgaddtomacro\gplfronttext{%
      \csname LTb\endcsname%%
      \put(198,3762){\rotatebox{-270}{\makebox(0,0){\strut{}$\delta_{\mathrm{h}}-\delta_{\mathrm{h}}^{0}$}}}%
      \put(6847,3762){\rotatebox{-270}{\makebox(0,0){\strut{} }}}%
      \put(3434,1237){\makebox(0,0){\strut{}$T\,({\mathrm K})$}}%
      \put(3434,5540){\makebox(0,0){\strut{}(b)}}%
      \csname LTb\endcsname%%
      \put(4804,5532){\makebox(0,0)[r]{\strut{}$\delta_{\mathrm{h}}^{0}=0.205$}}%
      \csname LTb\endcsname%%
      \put(4804,5246){\makebox(0,0)[r]{\strut{}$\delta_{\mathrm{h}}^{0}=0.221$}}%
      \csname LTb\endcsname%%
      \put(4804,4960){\makebox(0,0)[r]{\strut{}$\delta_{\mathrm{h}}^{0}=0.236$}}%
      \csname LTb\endcsname%%
      \put(4804,4674){\makebox(0,0)[r]{\strut{}$\delta_{\mathrm{h}}^{0}=0.248$}}%
      \csname LTb\endcsname%%
      \put(4804,4388){\makebox(0,0)[r]{\strut{}$\delta_{\mathrm{h}}^{0}=0.266$}}%
    }%
    \gplbacktext
    \put(0,0){\includegraphics{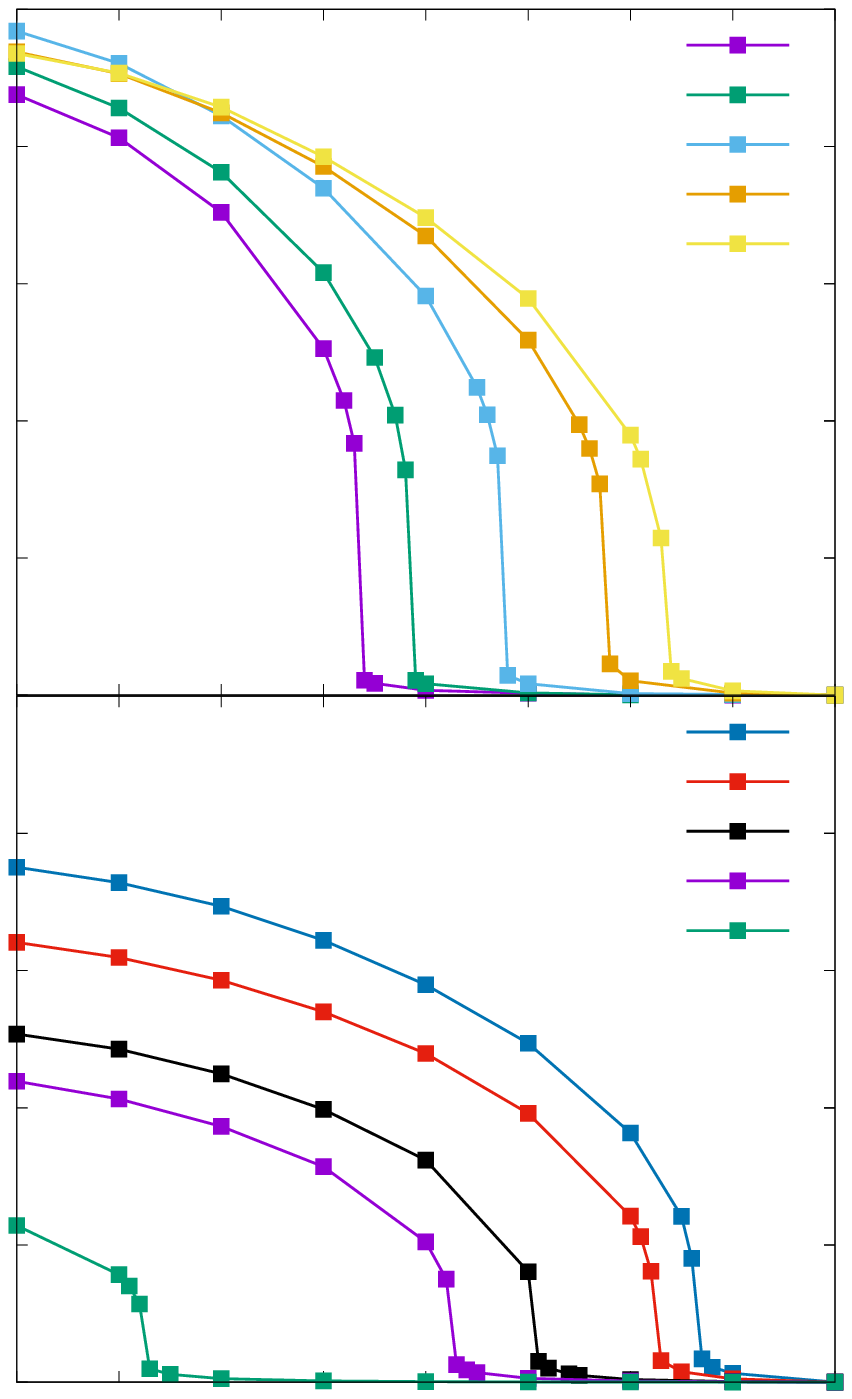}}%
    \gplfronttext
  \end{picture}%
\endgroup

%% file: fig3.tex
% GNUPLOT: LaTeX picture with Postscript
\begingroup
  \makeatletter
  \providecommand\color[2][]{%
    \GenericError{(gnuplot) \space\space\space\@spaces}{%
      Package color not loaded in conjunction with
      terminal option `colourtext'%
    }{See the gnuplot documentation for explanation.%
    }{Either use 'blacktext' in gnuplot or load the package
      color.sty in LaTeX.}%
    \renewcommand\color[2][]{}%
  }%
  \providecommand\includegraphics[2][]{%
    \GenericError{(gnuplot) \space\space\space\@spaces}{%
      Package graphicx or graphics not loaded%
    }{See the gnuplot documentation for explanation.%
    }{The gnuplot epslatex terminal needs graphicx.sty or graphics.sty.}%
    \renewcommand\includegraphics[2][]{}%
  }%
  \providecommand\rotatebox[2]{#2}%
  \@ifundefined{ifGPcolor}{%
    \newif\ifGPcolor
    \GPcolortrue
  }{}%
  \@ifundefined{ifGPblacktext}{%
    \newif\ifGPblacktext
    \GPblacktexttrue
  }{}%
  % define a \g@addto@macro without @ in the name:
  \let\gplgaddtomacro\g@addto@macro
  % define empty templates for all commands taking text:
  \gdef\gplbacktext{}%
  \gdef\gplfronttext{}%
  \makeatother
  \ifGPblacktext
    % no textcolor at all
    \def\colorrgb#1{}%
    \def\colorgray#1{}%
  \else
    % gray or color?
    \ifGPcolor
      \def\colorrgb#1{\color[rgb]{#1}}%
      \def\colorgray#1{\color[gray]{#1}}%
      \expandafter\def\csname LTw\endcsname{\color{white}}%
      \expandafter\def\csname LTb\endcsname{\color{black}}%
      \expandafter\def\csname LTa\endcsname{\color{black}}%
      \expandafter\def\csname LT0\endcsname{\color[rgb]{1,0,0}}%
      \expandafter\def\csname LT1\endcsname{\color[rgb]{0,1,0}}%
      \expandafter\def\csname LT2\endcsname{\color[rgb]{0,0,1}}%
      \expandafter\def\csname LT3\endcsname{\color[rgb]{1,0,1}}%
      \expandafter\def\csname LT4\endcsname{\color[rgb]{0,1,1}}%
      \expandafter\def\csname LT5\endcsname{\color[rgb]{1,1,0}}%
      \expandafter\def\csname LT6\endcsname{\color[rgb]{0,0,0}}%
      \expandafter\def\csname LT7\endcsname{\color[rgb]{1,0.3,0}}%
      \expandafter\def\csname LT8\endcsname{\color[rgb]{0.5,0.5,0.5}}%
    \else
      % gray
      \def\colorrgb#1{\color{black}}%
      \def\colorgray#1{\color[gray]{#1}}%
      \expandafter\def\csname LTw\endcsname{\color{white}}%
      \expandafter\def\csname LTb\endcsname{\color{black}}%
      \expandafter\def\csname LTa\endcsname{\color{black}}%
      \expandafter\def\csname LT0\endcsname{\color{black}}%
      \expandafter\def\csname LT1\endcsname{\color{black}}%
      \expandafter\def\csname LT2\endcsname{\color{black}}%
      \expandafter\def\csname LT3\endcsname{\color{black}}%
      \expandafter\def\csname LT4\endcsname{\color{black}}%
      \expandafter\def\csname LT5\endcsname{\color{black}}%
      \expandafter\def\csname LT6\endcsname{\color{black}}%
      \expandafter\def\csname LT7\endcsname{\color{black}}%
      \expandafter\def\csname LT8\endcsname{\color{black}}%
    \fi
  \fi
    \setlength{\unitlength}{0.0500bp}%
    \ifx\gptboxheight\undefined%
      \newlength{\gptboxheight}%
      \newlength{\gptboxwidth}%
      \newsavebox{\gptboxtext}%
    \fi%
    \setlength{\fboxrule}{0.5pt}%
    \setlength{\fboxsep}{1pt}%
\begin{picture}(7200.00,5040.00)%
    \gplgaddtomacro\gplbacktext{%
      \csname LTb\endcsname%%
      \put(814,704){\makebox(0,0)[r]{\strut{}$0$}}%
      \put(814,1390){\makebox(0,0)[r]{\strut{}$0.2$}}%
      \put(814,2076){\makebox(0,0)[r]{\strut{}$0.4$}}%
      \put(814,2762){\makebox(0,0)[r]{\strut{}$0.6$}}%
      \put(814,3447){\makebox(0,0)[r]{\strut{}$0.8$}}%
      \put(814,4133){\makebox(0,0)[r]{\strut{}$1$}}%
      \put(814,4819){\makebox(0,0)[r]{\strut{}$1.2$}}%
      \put(946,484){\makebox(0,0){\strut{}$0.1$}}%
      \put(2190,484){\makebox(0,0){\strut{}$0.15$}}%
      \put(3435,484){\makebox(0,0){\strut{}$0.2$}}%
      \put(4679,484){\makebox(0,0){\strut{}$0.25$}}%
      \put(5923,484){\makebox(0,0){\strut{}$0.3$}}%
      \put(6055,704){\makebox(0,0)[l]{\strut{}$0$}}%
      \put(6055,1390){\makebox(0,0)[l]{\strut{}$0.2$}}%
      \put(6055,2076){\makebox(0,0)[l]{\strut{}$0.4$}}%
      \put(6055,2762){\makebox(0,0)[l]{\strut{}$0.6$}}%
      \put(6055,3447){\makebox(0,0)[l]{\strut{}$0.8$}}%
      \put(6055,4133){\makebox(0,0)[l]{\strut{}$1$}}%
      \put(6055,4819){\makebox(0,0)[l]{\strut{}$1.2$}}%
    }%
    \gplgaddtomacro\gplfronttext{%
      \csname LTb\endcsname%%
      \put(198,2761){\rotatebox{-270}{\makebox(0,0){\strut{}Number of holes}}}%
      \put(6847,2761){\rotatebox{-270}{\makebox(0,0){\strut{} }}}%
      \put(3434,154){\makebox(0,0){\strut{}$\delta_{\mathrm h}^{}$}}%
      \csname LTb\endcsname%%
      \put(3403,3190){\makebox(0,0)[r]{\strut{}$n_{d{\mathrm h}}^{}$}}%
      \csname LTb\endcsname%%
      \put(3403,2904){\makebox(0,0)[r]{\strut{}$n_{p{\mathrm h}}^{}$}}%
      \csname LTb\endcsname%%
      \put(3403,2618){\makebox(0,0)[r]{\strut{}$n_{d{\mathrm h}}^{0}$}}%
      \csname LTb\endcsname%%
      \put(3403,2332){\makebox(0,0)[r]{\strut{}$n_{p{\mathrm h}}^{0}$}}%
    }%
    \gplbacktext
    \put(0,0){\includegraphics{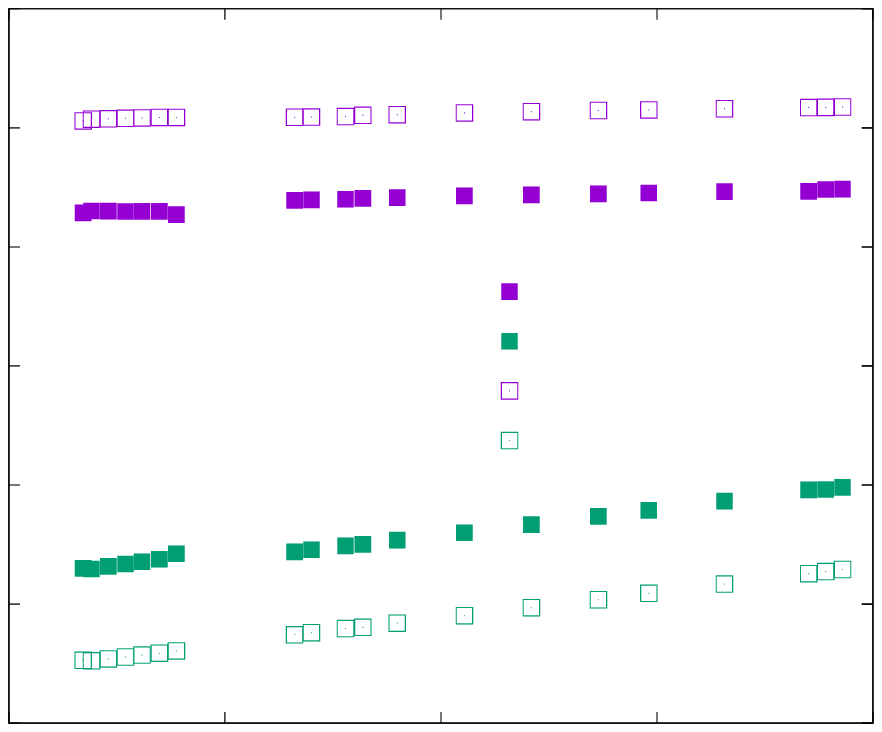}}%
    \gplfronttext
  \end{picture}%
\endgroup

%% file: fig4.tex
% GNUPLOT: LaTeX picture with Postscript
\begingroup
  \makeatletter
  \providecommand\color[2][]{%
    \GenericError{(gnuplot) \space\space\space\@spaces}{%
      Package color not loaded in conjunction with
      terminal option `colourtext'%
    }{See the gnuplot documentation for explanation.%
    }{Either use 'blacktext' in gnuplot or load the package
      color.sty in LaTeX.}%
    \renewcommand\color[2][]{}%
  }%
  \providecommand\includegraphics[2][]{%
    \GenericError{(gnuplot) \space\space\space\@spaces}{%
      Package graphicx or graphics not loaded%
    }{See the gnuplot documentation for explanation.%
    }{The gnuplot epslatex terminal needs graphicx.sty or graphics.sty.}%
    \renewcommand\includegraphics[2][]{}%
  }%
  \providecommand\rotatebox[2]{#2}%
  \@ifundefined{ifGPcolor}{%
    \newif\ifGPcolor
    \GPcolortrue
  }{}%
  \@ifundefined{ifGPblacktext}{%
    \newif\ifGPblacktext
    \GPblacktexttrue
  }{}%
  % define a \g@addto@macro without @ in the name:
  \let\gplgaddtomacro\g@addto@macro
  % define empty templates for all commands taking text:
  \gdef\gplbacktext{}%
  \gdef\gplfronttext{}%
  \makeatother
  \ifGPblacktext
    % no textcolor at all
    \def\colorrgb#1{}%
    \def\colorgray#1{}%
  \else
    % gray or color?
    \ifGPcolor
      \def\colorrgb#1{\color[rgb]{#1}}%
      \def\colorgray#1{\color[gray]{#1}}%
      \expandafter\def\csname LTw\endcsname{\color{white}}%
      \expandafter\def\csname LTb\endcsname{\color{black}}%
      \expandafter\def\csname LTa\endcsname{\color{black}}%
      \expandafter\def\csname LT0\endcsname{\color[rgb]{1,0,0}}%
      \expandafter\def\csname LT1\endcsname{\color[rgb]{0,1,0}}%
      \expandafter\def\csname LT2\endcsname{\color[rgb]{0,0,1}}%
      \expandafter\def\csname LT3\endcsname{\color[rgb]{1,0,1}}%
      \expandafter\def\csname LT4\endcsname{\color[rgb]{0,1,1}}%
      \expandafter\def\csname LT5\endcsname{\color[rgb]{1,1,0}}%
      \expandafter\def\csname LT6\endcsname{\color[rgb]{0,0,0}}%
      \expandafter\def\csname LT7\endcsname{\color[rgb]{1,0.3,0}}%
      \expandafter\def\csname LT8\endcsname{\color[rgb]{0.5,0.5,0.5}}%
    \else
      % gray
      \def\colorrgb#1{\color{black}}%
      \def\colorgray#1{\color[gray]{#1}}%
      \expandafter\def\csname LTw\endcsname{\color{white}}%
      \expandafter\def\csname LTb\endcsname{\color{black}}%
      \expandafter\def\csname LTa\endcsname{\color{black}}%
      \expandafter\def\csname LT0\endcsname{\color{black}}%
      \expandafter\def\csname LT1\endcsname{\color{black}}%
      \expandafter\def\csname LT2\endcsname{\color{black}}%
      \expandafter\def\csname LT3\endcsname{\color{black}}%
      \expandafter\def\csname LT4\endcsname{\color{black}}%
      \expandafter\def\csname LT5\endcsname{\color{black}}%
      \expandafter\def\csname LT6\endcsname{\color{black}}%
      \expandafter\def\csname LT7\endcsname{\color{black}}%
      \expandafter\def\csname LT8\endcsname{\color{black}}%
    \fi
  \fi
    \setlength{\unitlength}{0.0500bp}%
    \ifx\gptboxheight\undefined%
      \newlength{\gptboxheight}%
      \newlength{\gptboxwidth}%
      \newsavebox{\gptboxtext}%
    \fi%
    \setlength{\fboxrule}{0.5pt}%
    \setlength{\fboxsep}{1pt}%
\begin{picture}(7200.00,25200.00)%
    \gplgaddtomacro\gplbacktext{%
      \csname LTb\endcsname%%
      \put(814,21368){\makebox(0,0)[r]{\strut{}$0$}}%
      \put(814,21813){\makebox(0,0)[r]{\strut{}$0.2$}}%
      \put(814,22257){\makebox(0,0)[r]{\strut{}$0.4$}}%
      \put(814,22702){\makebox(0,0)[r]{\strut{}$0.6$}}%
      \put(814,23146){\makebox(0,0)[r]{\strut{}$0.8$}}%
      \put(814,23591){\makebox(0,0)[r]{\strut{}$1$}}%
      \put(814,24035){\makebox(0,0)[r]{\strut{}$1.2$}}%
      \put(814,24480){\makebox(0,0)[r]{\strut{}$1.4$}}%
      \put(1030,21148){\makebox(0,0){\strut{} }}%
      \put(1524,21148){\makebox(0,0){\strut{} }}%
      \put(2018,21148){\makebox(0,0){\strut{} }}%
      \put(2512,21148){\makebox(0,0){\strut{} }}%
      \put(3006,21148){\makebox(0,0){\strut{} }}%
      \put(3501,21148){\makebox(0,0){\strut{} }}%
      \put(3995,21148){\makebox(0,0){\strut{} }}%
      \put(4489,21148){\makebox(0,0){\strut{} }}%
      \put(4983,21148){\makebox(0,0){\strut{} }}%
      \put(5477,21148){\makebox(0,0){\strut{} }}%
      \put(5971,21148){\makebox(0,0){\strut{} }}%
      \put(6187,21368){\makebox(0,0)[l]{\strut{}0}}%
      \put(6187,21813){\makebox(0,0)[l]{\strut{}2}}%
      \put(6187,22257){\makebox(0,0)[l]{\strut{}4}}%
      \put(6187,22702){\makebox(0,0)[l]{\strut{}6}}%
      \put(6187,23146){\makebox(0,0)[l]{\strut{}8}}%
      \put(6187,23591){\makebox(0,0)[l]{\strut{}10}}%
      \put(6187,24035){\makebox(0,0)[l]{\strut{}12}}%
      \put(6187,24480){\makebox(0,0)[l]{\strut{}14}}%
      \csname LTb\endcsname%%
      \put(3501,24700){\makebox(0,0){\strut{} }}%
    }%
    \gplgaddtomacro\gplfronttext{%
      \csname LTb\endcsname%%
      \put(198,22924){\rotatebox{-270}{\makebox(0,0){\strut{}$\rho_{d}^{}(\varepsilon)$ (arbitrary unit)}}}%
      \put(6847,22924){\rotatebox{-270}{\makebox(0,0){\strut{}$\rho_{p}^{}(\varepsilon)$ (arbitrary unit)}}}%
      \put(3500,20818){\makebox(0,0){\strut{} }}%
      \put(1916,24260){\makebox(0,0){\strut{}(a) $\delta_{\mathrm{h}}^{}=0.123$}}%
      \csname LTb\endcsname%%
      \put(5068,24274){\makebox(0,0)[r]{\strut{}$\rho_{d}^{}(\varepsilon)$}}%
      \csname LTb\endcsname%%
      \put(5068,23988){\makebox(0,0)[r]{\strut{}$\rho_{p}^{}(\varepsilon)$}}%
    }%
    \gplgaddtomacro\gplbacktext{%
      \csname LTb\endcsname%%
      \put(814,18243){\makebox(0,0)[r]{\strut{}$0$}}%
      \put(814,18688){\makebox(0,0)[r]{\strut{}$0.2$}}%
      \put(814,19132){\makebox(0,0)[r]{\strut{}$0.4$}}%
      \put(814,19577){\makebox(0,0)[r]{\strut{}$0.6$}}%
      \put(814,20022){\makebox(0,0)[r]{\strut{}$0.8$}}%
      \put(814,20467){\makebox(0,0)[r]{\strut{}$1$}}%
      \put(814,20911){\makebox(0,0)[r]{\strut{}$1.2$}}%
      \put(1030,18023){\makebox(0,0){\strut{} }}%
      \put(1524,18023){\makebox(0,0){\strut{} }}%
      \put(2018,18023){\makebox(0,0){\strut{} }}%
      \put(2512,18023){\makebox(0,0){\strut{} }}%
      \put(3006,18023){\makebox(0,0){\strut{} }}%
      \put(3501,18023){\makebox(0,0){\strut{} }}%
      \put(3995,18023){\makebox(0,0){\strut{} }}%
      \put(4489,18023){\makebox(0,0){\strut{} }}%
      \put(4983,18023){\makebox(0,0){\strut{} }}%
      \put(5477,18023){\makebox(0,0){\strut{} }}%
      \put(5971,18023){\makebox(0,0){\strut{} }}%
      \put(6187,18243){\makebox(0,0)[l]{\strut{}0}}%
      \put(6187,18688){\makebox(0,0)[l]{\strut{}2}}%
      \put(6187,19132){\makebox(0,0)[l]{\strut{}4}}%
      \put(6187,19577){\makebox(0,0)[l]{\strut{}6}}%
      \put(6187,20022){\makebox(0,0)[l]{\strut{}8}}%
      \put(6187,20467){\makebox(0,0)[l]{\strut{}10}}%
      \put(6187,20911){\makebox(0,0)[l]{\strut{}12}}%
      \csname LTb\endcsname%%
      \put(3501,21576){\makebox(0,0){\strut{} }}%
    }%
    \gplgaddtomacro\gplfronttext{%
      \csname LTb\endcsname%%
      \put(198,19799){\rotatebox{-270}{\makebox(0,0){\strut{}$\rho_{d}^{}(\varepsilon)$ (arbitrary unit)}}}%
      \put(6847,19799){\rotatebox{-270}{\makebox(0,0){\strut{}$\rho_{p}^{}(\varepsilon)$ (arbitrary unit)}}}%
      \put(3500,17693){\makebox(0,0){\strut{} }}%
      \put(1916,21136){\makebox(0,0){\strut{}(b) $\delta_{\mathrm{h}}^{}=0.166$}}%
      \csname LTb\endcsname%%
      \put(5068,21150){\makebox(0,0)[r]{\strut{}$\rho_{d}^{}(\varepsilon)$}}%
      \csname LTb\endcsname%%
      \put(5068,20864){\makebox(0,0)[r]{\strut{}$\rho_{p}^{}(\varepsilon)$}}%
    }%
    \gplgaddtomacro\gplbacktext{%
      \csname LTb\endcsname%%
      \put(814,15118){\makebox(0,0)[r]{\strut{}$0$}}%
      \put(814,15563){\makebox(0,0)[r]{\strut{}$0.2$}}%
      \put(814,16008){\makebox(0,0)[r]{\strut{}$0.4$}}%
      \put(814,16453){\makebox(0,0)[r]{\strut{}$0.6$}}%
      \put(814,16897){\makebox(0,0)[r]{\strut{}$0.8$}}%
      \put(814,17342){\makebox(0,0)[r]{\strut{}$1$}}%
      \put(814,17787){\makebox(0,0)[r]{\strut{}$1.2$}}%
      \put(1030,14898){\makebox(0,0){\strut{} }}%
      \put(1524,14898){\makebox(0,0){\strut{} }}%
      \put(2018,14898){\makebox(0,0){\strut{} }}%
      \put(2512,14898){\makebox(0,0){\strut{} }}%
      \put(3006,14898){\makebox(0,0){\strut{} }}%
      \put(3501,14898){\makebox(0,0){\strut{} }}%
      \put(3995,14898){\makebox(0,0){\strut{} }}%
      \put(4489,14898){\makebox(0,0){\strut{} }}%
      \put(4983,14898){\makebox(0,0){\strut{} }}%
      \put(5477,14898){\makebox(0,0){\strut{} }}%
      \put(5971,14898){\makebox(0,0){\strut{} }}%
      \put(6187,15118){\makebox(0,0)[l]{\strut{}0}}%
      \put(6187,15563){\makebox(0,0)[l]{\strut{}2}}%
      \put(6187,16008){\makebox(0,0)[l]{\strut{}4}}%
      \put(6187,16453){\makebox(0,0)[l]{\strut{}6}}%
      \put(6187,16897){\makebox(0,0)[l]{\strut{}8}}%
      \put(6187,17342){\makebox(0,0)[l]{\strut{}10}}%
      \put(6187,17787){\makebox(0,0)[l]{\strut{}12}}%
      \csname LTb\endcsname%%
      \put(3501,18452){\makebox(0,0){\strut{} }}%
    }%
    \gplgaddtomacro\gplfronttext{%
      \csname LTb\endcsname%%
      \put(198,16675){\rotatebox{-270}{\makebox(0,0){\strut{}$\rho_{d}^{}(\varepsilon)$ (arbitrary unit)}}}%
      \put(6847,16675){\rotatebox{-270}{\makebox(0,0){\strut{}$\rho_{p}^{}(\varepsilon)$ (arbitrary unit)}}}%
      \put(3500,14568){\makebox(0,0){\strut{} }}%
      \put(1916,18012){\makebox(0,0){\strut{}(c) $\delta_{\mathrm{h}}^{}=0.205$}}%
      \csname LTb\endcsname%%
      \put(5068,18026){\makebox(0,0)[r]{\strut{}$\rho_{d}^{}(\varepsilon)$}}%
      \csname LTb\endcsname%%
      \put(5068,17740){\makebox(0,0)[r]{\strut{}$\rho_{p}^{}(\varepsilon)$}}%
    }%
    \gplgaddtomacro\gplbacktext{%
      \csname LTb\endcsname%%
      \put(814,11993){\makebox(0,0)[r]{\strut{}$0$}}%
      \put(814,12438){\makebox(0,0)[r]{\strut{}$0.2$}}%
      \put(814,12883){\makebox(0,0)[r]{\strut{}$0.4$}}%
      \put(814,13328){\makebox(0,0)[r]{\strut{}$0.6$}}%
      \put(814,13773){\makebox(0,0)[r]{\strut{}$0.8$}}%
      \put(814,14218){\makebox(0,0)[r]{\strut{}$1$}}%
      \put(814,14663){\makebox(0,0)[r]{\strut{}$1.2$}}%
      \put(1030,11773){\makebox(0,0){\strut{} }}%
      \put(1524,11773){\makebox(0,0){\strut{} }}%
      \put(2018,11773){\makebox(0,0){\strut{} }}%
      \put(2512,11773){\makebox(0,0){\strut{} }}%
      \put(3006,11773){\makebox(0,0){\strut{} }}%
      \put(3501,11773){\makebox(0,0){\strut{} }}%
      \put(3995,11773){\makebox(0,0){\strut{} }}%
      \put(4489,11773){\makebox(0,0){\strut{} }}%
      \put(4983,11773){\makebox(0,0){\strut{} }}%
      \put(5477,11773){\makebox(0,0){\strut{} }}%
      \put(5971,11773){\makebox(0,0){\strut{} }}%
      \put(6187,11993){\makebox(0,0)[l]{\strut{}0}}%
      \put(6187,12438){\makebox(0,0)[l]{\strut{}2}}%
      \put(6187,12883){\makebox(0,0)[l]{\strut{}4}}%
      \put(6187,13328){\makebox(0,0)[l]{\strut{}6}}%
      \put(6187,13773){\makebox(0,0)[l]{\strut{}8}}%
      \put(6187,14218){\makebox(0,0)[l]{\strut{}10}}%
      \put(6187,14663){\makebox(0,0)[l]{\strut{}12}}%
      \csname LTb\endcsname%%
      \put(3501,15328){\makebox(0,0){\strut{} }}%
    }%
    \gplgaddtomacro\gplfronttext{%
      \csname LTb\endcsname%%
      \put(198,13550){\rotatebox{-270}{\makebox(0,0){\strut{}$\rho_{d}^{}(\varepsilon)$ (arbitrary unit)}}}%
      \put(6847,13550){\rotatebox{-270}{\makebox(0,0){\strut{}$\rho_{p}^{}(\varepsilon)$ (arbitrary unit)}}}%
      \put(3500,11443){\makebox(0,0){\strut{} }}%
      \put(1916,14888){\makebox(0,0){\strut{}(d) $\delta_{\mathrm{h}}^{}=0.266$}}%
      \csname LTb\endcsname%%
      \put(5068,14902){\makebox(0,0)[r]{\strut{}$\rho_{d}^{}(\varepsilon)$}}%
      \csname LTb\endcsname%%
      \put(5068,14616){\makebox(0,0)[r]{\strut{}$\rho_{p}^{}(\varepsilon)$}}%
    }%
    \gplgaddtomacro\gplbacktext{%
      \csname LTb\endcsname%%
      \put(814,8868){\makebox(0,0)[r]{\strut{}$0$}}%
      \put(814,9313){\makebox(0,0)[r]{\strut{}$0.2$}}%
      \put(814,9758){\makebox(0,0)[r]{\strut{}$0.4$}}%
      \put(814,10203){\makebox(0,0)[r]{\strut{}$0.6$}}%
      \put(814,10649){\makebox(0,0)[r]{\strut{}$0.8$}}%
      \put(814,11094){\makebox(0,0)[r]{\strut{}$1$}}%
      \put(814,11539){\makebox(0,0)[r]{\strut{}$1.2$}}%
      \put(1030,8648){\makebox(0,0){\strut{}$-5$}}%
      \put(1524,8648){\makebox(0,0){\strut{}$-4$}}%
      \put(2018,8648){\makebox(0,0){\strut{}$-3$}}%
      \put(2512,8648){\makebox(0,0){\strut{}$-2$}}%
      \put(3006,8648){\makebox(0,0){\strut{}$-1$}}%
      \put(3501,8648){\makebox(0,0){\strut{}$0$}}%
      \put(3995,8648){\makebox(0,0){\strut{}$1$}}%
      \put(4489,8648){\makebox(0,0){\strut{}$2$}}%
      \put(4983,8648){\makebox(0,0){\strut{}$3$}}%
      \put(5477,8648){\makebox(0,0){\strut{}$4$}}%
      \put(5971,8648){\makebox(0,0){\strut{}$5$}}%
      \put(6187,8868){\makebox(0,0)[l]{\strut{}0}}%
      \put(6187,9313){\makebox(0,0)[l]{\strut{}2}}%
      \put(6187,9758){\makebox(0,0)[l]{\strut{}4}}%
      \put(6187,10203){\makebox(0,0)[l]{\strut{}6}}%
      \put(6187,10649){\makebox(0,0)[l]{\strut{}8}}%
      \put(6187,11094){\makebox(0,0)[l]{\strut{}10}}%
      \put(6187,11539){\makebox(0,0)[l]{\strut{}12}}%
      \csname LTb\endcsname%%
      \put(3501,12204){\makebox(0,0){\strut{} }}%
    }%
    \gplgaddtomacro\gplfronttext{%
      \csname LTb\endcsname%%
      \put(198,10426){\rotatebox{-270}{\makebox(0,0){\strut{}$\rho_{d}^{}(\varepsilon)$ (arbitrary unit)}}}%
      \put(6847,10426){\rotatebox{-270}{\makebox(0,0){\strut{}$\rho_{p}^{}(\varepsilon)$ (arbitrary unit)}}}%
      \put(3500,8318){\makebox(0,0){\strut{}$\varepsilon$ (eV)}}%
      \put(1916,11764){\makebox(0,0){\strut{}(e) $\delta_{\mathrm{h}}^{}=0.293$}}%
      \csname LTb\endcsname%%
      \put(5068,11778){\makebox(0,0)[r]{\strut{}$\rho_{d}^{}(\varepsilon)$}}%
      \csname LTb\endcsname%%
      \put(5068,11492){\makebox(0,0)[r]{\strut{}$\rho_{p}^{}(\varepsilon)$}}%
    }%
    \gplbacktext
    \put(0,0){\includegraphics{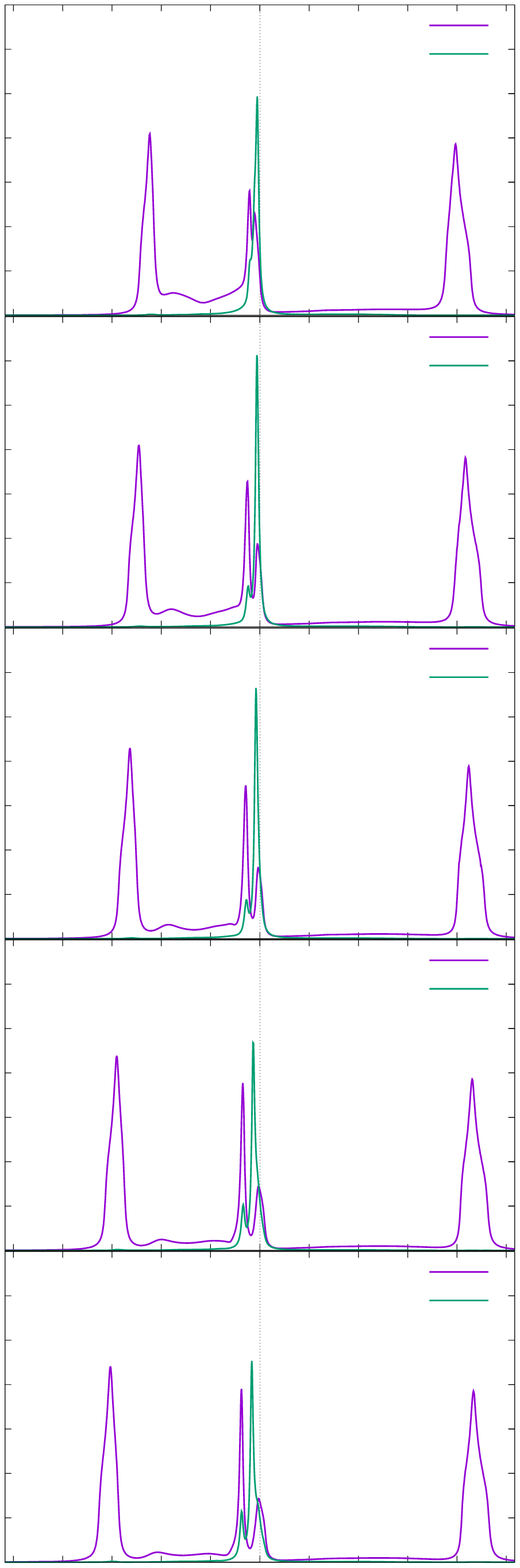}}%
    \gplfronttext
  \end{picture}%
\endgroup

%% file: fig5.tex
% GNUPLOT: LaTeX picture with Postscript
\begingroup
  \makeatletter
  \providecommand\color[2][]{%
    \GenericError{(gnuplot) \space\space\space\@spaces}{%
      Package color not loaded in conjunction with
      terminal option `colourtext'%
    }{See the gnuplot documentation for explanation.%
    }{Either use 'blacktext' in gnuplot or load the package
      color.sty in LaTeX.}%
    \renewcommand\color[2][]{}%
  }%
  \providecommand\includegraphics[2][]{%
    \GenericError{(gnuplot) \space\space\space\@spaces}{%
      Package graphicx or graphics not loaded%
    }{See the gnuplot documentation for explanation.%
    }{The gnuplot epslatex terminal needs graphicx.sty or graphics.sty.}%
    \renewcommand\includegraphics[2][]{}%
  }%
  \providecommand\rotatebox[2]{#2}%
  \@ifundefined{ifGPcolor}{%
    \newif\ifGPcolor
    \GPcolortrue
  }{}%
  \@ifundefined{ifGPblacktext}{%
    \newif\ifGPblacktext
    \GPblacktexttrue
  }{}%
  % define a \g@addto@macro without @ in the name:
  \let\gplgaddtomacro\g@addto@macro
  % define empty templates for all commands taking text:
  \gdef\gplbacktext{}%
  \gdef\gplfronttext{}%
  \makeatother
  \ifGPblacktext
    % no textcolor at all
    \def\colorrgb#1{}%
    \def\colorgray#1{}%
  \else
    % gray or color?
    \ifGPcolor
      \def\colorrgb#1{\color[rgb]{#1}}%
      \def\colorgray#1{\color[gray]{#1}}%
      \expandafter\def\csname LTw\endcsname{\color{white}}%
      \expandafter\def\csname LTb\endcsname{\color{black}}%
      \expandafter\def\csname LTa\endcsname{\color{black}}%
      \expandafter\def\csname LT0\endcsname{\color[rgb]{1,0,0}}%
      \expandafter\def\csname LT1\endcsname{\color[rgb]{0,1,0}}%
      \expandafter\def\csname LT2\endcsname{\color[rgb]{0,0,1}}%
      \expandafter\def\csname LT3\endcsname{\color[rgb]{1,0,1}}%
      \expandafter\def\csname LT4\endcsname{\color[rgb]{0,1,1}}%
      \expandafter\def\csname LT5\endcsname{\color[rgb]{1,1,0}}%
      \expandafter\def\csname LT6\endcsname{\color[rgb]{0,0,0}}%
      \expandafter\def\csname LT7\endcsname{\color[rgb]{1,0.3,0}}%
      \expandafter\def\csname LT8\endcsname{\color[rgb]{0.5,0.5,0.5}}%
    \else
      % gray
      \def\colorrgb#1{\color{black}}%
      \def\colorgray#1{\color[gray]{#1}}%
      \expandafter\def\csname LTw\endcsname{\color{white}}%
      \expandafter\def\csname LTb\endcsname{\color{black}}%
      \expandafter\def\csname LTa\endcsname{\color{black}}%
      \expandafter\def\csname LT0\endcsname{\color{black}}%
      \expandafter\def\csname LT1\endcsname{\color{black}}%
      \expandafter\def\csname LT2\endcsname{\color{black}}%
      \expandafter\def\csname LT3\endcsname{\color{black}}%
      \expandafter\def\csname LT4\endcsname{\color{black}}%
      \expandafter\def\csname LT5\endcsname{\color{black}}%
      \expandafter\def\csname LT6\endcsname{\color{black}}%
      \expandafter\def\csname LT7\endcsname{\color{black}}%
      \expandafter\def\csname LT8\endcsname{\color{black}}%
    \fi
  \fi
    \setlength{\unitlength}{0.0500bp}%
    \ifx\gptboxheight\undefined%
      \newlength{\gptboxheight}%
      \newlength{\gptboxwidth}%
      \newsavebox{\gptboxtext}%
    \fi%
    \setlength{\fboxrule}{0.5pt}%
    \setlength{\fboxsep}{1pt}%
\begin{picture}(7200.00,10080.00)%
    \gplgaddtomacro\gplbacktext{%
      \csname LTb\endcsname%%
      \put(814,5744){\makebox(0,0)[r]{\strut{}$0$}}%
      \put(814,6269){\makebox(0,0)[r]{\strut{}$0.1$}}%
      \put(814,6795){\makebox(0,0)[r]{\strut{}$0.2$}}%
      \put(814,7320){\makebox(0,0)[r]{\strut{}$0.3$}}%
      \put(814,7845){\makebox(0,0)[r]{\strut{}$0.4$}}%
      \put(814,8370){\makebox(0,0)[r]{\strut{}$0.5$}}%
      \put(814,8896){\makebox(0,0)[r]{\strut{}$0.6$}}%
      \put(814,9421){\makebox(0,0)[r]{\strut{}$0.7$}}%
      \put(814,9946){\makebox(0,0)[r]{\strut{}$0.8$}}%
      \put(946,5524){\makebox(0,0){\strut{} }}%
      \put(1568,5524){\makebox(0,0){\strut{} }}%
      \put(2190,5524){\makebox(0,0){\strut{} }}%
      \put(2812,5524){\makebox(0,0){\strut{} }}%
      \put(3435,5524){\makebox(0,0){\strut{} }}%
      \put(4057,5524){\makebox(0,0){\strut{} }}%
      \put(4679,5524){\makebox(0,0){\strut{} }}%
      \put(5301,5524){\makebox(0,0){\strut{} }}%
      \put(5923,5524){\makebox(0,0){\strut{} }}%
      \put(6055,5744){\makebox(0,0)[l]{\strut{}$0$}}%
      \put(6055,6269){\makebox(0,0)[l]{\strut{}$0.1$}}%
      \put(6055,6795){\makebox(0,0)[l]{\strut{}$0.2$}}%
      \put(6055,7320){\makebox(0,0)[l]{\strut{}$0.3$}}%
      \put(6055,7845){\makebox(0,0)[l]{\strut{}$0.4$}}%
      \put(6055,8370){\makebox(0,0)[l]{\strut{}$0.5$}}%
      \put(6055,8896){\makebox(0,0)[l]{\strut{}$0.6$}}%
      \put(6055,9421){\makebox(0,0)[l]{\strut{}$0.7$}}%
      \put(6055,9946){\makebox(0,0)[l]{\strut{}$0.8$}}%
      \csname LTb\endcsname%%
      \put(4679,10166){\makebox(0,0){\strut{} }}%
    }%
    \gplgaddtomacro\gplfronttext{%
      \csname LTb\endcsname%%
      \put(198,7845){\rotatebox{-270}{\makebox(0,0){\strut{}$\rho_{d}^{}(\varepsilon)$ (arbitrary unit)}}}%
      \put(6847,7845){\rotatebox{-270}{\makebox(0,0){\strut{} }}}%
      \put(3434,5194){\makebox(0,0){\strut{} }}%
      \put(3434,9748){\makebox(0,0){\strut{}(a)}}%
      \csname LTb\endcsname%%
      \put(4936,9740){\makebox(0,0)[r]{\strut{}$\delta_{\mathrm{h}}^{}=0.123$}}%
      \csname LTb\endcsname%%
      \put(4936,9454){\makebox(0,0)[r]{\strut{}$\delta_{\mathrm{h}}^{}=0.166$}}%
      \csname LTb\endcsname%%
      \put(4936,9168){\makebox(0,0)[r]{\strut{}$\delta_{\mathrm{h}}^{}=0.205$}}%
      \csname LTb\endcsname%%
      \put(4936,8882){\makebox(0,0)[r]{\strut{}$\delta_{\mathrm{h}}^{}=0.266$}}%
      \csname LTb\endcsname%%
      \put(4936,8596){\makebox(0,0)[r]{\strut{}$\delta_{\mathrm{h}}^{}=0.293$}}%
    }%
    \gplgaddtomacro\gplbacktext{%
      \csname LTb\endcsname%%
      \put(814,1530){\makebox(0,0)[r]{\strut{}  0}}%
      \put(814,2130){\makebox(0,0)[r]{\strut{}  2}}%
      \put(814,2731){\makebox(0,0)[r]{\strut{}  4}}%
      \put(814,3331){\makebox(0,0)[r]{\strut{}  6}}%
      \put(814,3932){\makebox(0,0)[r]{\strut{}  8}}%
      \put(814,4532){\makebox(0,0)[r]{\strut{} 10}}%
      \put(814,5133){\makebox(0,0)[r]{\strut{} 12}}%
      \put(946,1310){\makebox(0,0){\strut{}-0.6}}%
      \put(1568,1310){\makebox(0,0){\strut{}-0.5}}%
      \put(2190,1310){\makebox(0,0){\strut{}-0.4}}%
      \put(2812,1310){\makebox(0,0){\strut{}-0.3}}%
      \put(3435,1310){\makebox(0,0){\strut{}-0.2}}%
      \put(4057,1310){\makebox(0,0){\strut{}-0.1}}%
      \put(4679,1310){\makebox(0,0){\strut{}0.0}}%
      \put(5301,1310){\makebox(0,0){\strut{}0.1}}%
      \put(5923,1310){\makebox(0,0){\strut{}0.2}}%
      \put(6055,1530){\makebox(0,0)[l]{\strut{}  0}}%
      \put(6055,2130){\makebox(0,0)[l]{\strut{}  2}}%
      \put(6055,2731){\makebox(0,0)[l]{\strut{}  4}}%
      \put(6055,3331){\makebox(0,0)[l]{\strut{}  6}}%
      \put(6055,3932){\makebox(0,0)[l]{\strut{}  8}}%
      \put(6055,4532){\makebox(0,0)[l]{\strut{} 10}}%
      \put(6055,5133){\makebox(0,0)[l]{\strut{} 12}}%
      \csname LTb\endcsname%%
      \put(4679,5953){\makebox(0,0){\strut{} }}%
    }%
    \gplgaddtomacro\gplfronttext{%
      \csname LTb\endcsname%%
      \put(198,3631){\rotatebox{-270}{\makebox(0,0){\strut{}$\rho_{p}^{}(\varepsilon)$ (arbitrary unit)}}}%
      \put(6847,3631){\rotatebox{-270}{\makebox(0,0){\strut{} }}}%
      \put(3434,980){\makebox(0,0){\strut{}$\varepsilon$ (eV)}}%
      \put(3434,5535){\makebox(0,0){\strut{}(b)}}%
      \csname LTb\endcsname%%
      \put(2266,5527){\makebox(0,0)[r]{\strut{}$\delta_{\mathrm{h}}^{}=0.123$}}%
      \csname LTb\endcsname%%
      \put(2266,5241){\makebox(0,0)[r]{\strut{}$\delta_{\mathrm{h}}^{}=0.166$}}%
      \csname LTb\endcsname%%
      \put(2266,4955){\makebox(0,0)[r]{\strut{}$\delta_{\mathrm{h}}^{}=0.205$}}%
      \csname LTb\endcsname%%
      \put(2266,4669){\makebox(0,0)[r]{\strut{}$\delta_{\mathrm{h}}^{}=0.266$}}%
      \csname LTb\endcsname%%
      \put(2266,4383){\makebox(0,0)[r]{\strut{}$\delta_{\mathrm{h}}^{}=0.293$}}%
    }%
    \gplbacktext
    \put(0,0){\includegraphics{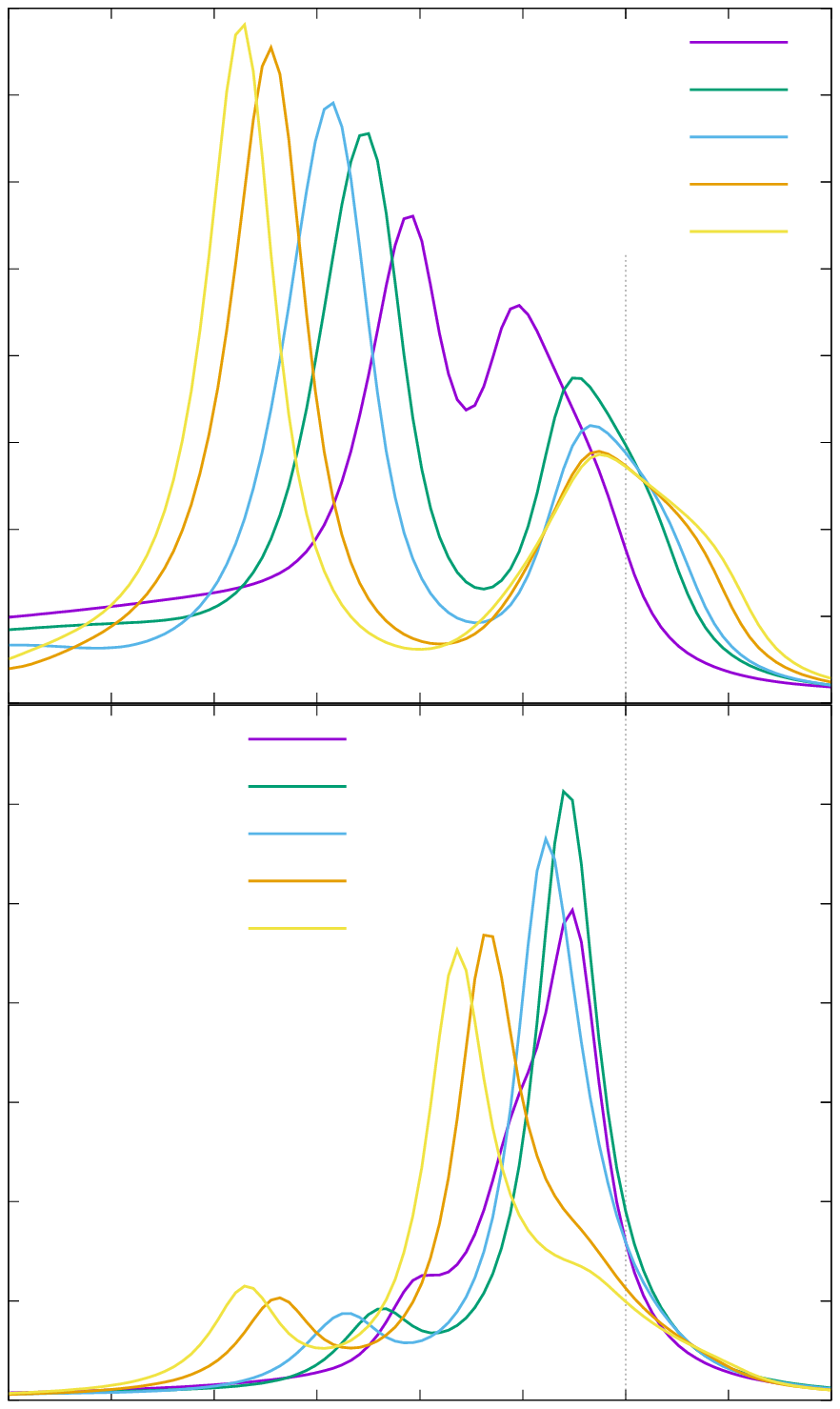}}%
    \gplfronttext
  \end{picture}%
\endgroup

%% file: fig6a.tex
% GNUPLOT: LaTeX picture with Postscript
\begingroup
  \makeatletter
  \providecommand\color[2][]{%
    \GenericError{(gnuplot) \space\space\space\@spaces}{%
      Package color not loaded in conjunction with
      terminal option `colourtext'%
    }{See the gnuplot documentation for explanation.%
    }{Either use 'blacktext' in gnuplot or load the package
      color.sty in LaTeX.}%
    \renewcommand\color[2][]{}%
  }%
  \providecommand\includegraphics[2][]{%
    \GenericError{(gnuplot) \space\space\space\@spaces}{%
      Package graphicx or graphics not loaded%
    }{See the gnuplot documentation for explanation.%
    }{The gnuplot epslatex terminal needs graphicx.sty or graphics.sty.}%
    \renewcommand\includegraphics[2][]{}%
  }%
  \providecommand\rotatebox[2]{#2}%
  \@ifundefined{ifGPcolor}{%
    \newif\ifGPcolor
    \GPcolortrue
  }{}%
  \@ifundefined{ifGPblacktext}{%
    \newif\ifGPblacktext
    \GPblacktexttrue
  }{}%
  % define a \g@addto@macro without @ in the name:
  \let\gplgaddtomacro\g@addto@macro
  % define empty templates for all commands taking text:
  \gdef\gplbacktext{}%
  \gdef\gplfronttext{}%
  \makeatother
  \ifGPblacktext
    % no textcolor at all
    \def\colorrgb#1{}%
    \def\colorgray#1{}%
  \else
    % gray or color?
    \ifGPcolor
      \def\colorrgb#1{\color[rgb]{#1}}%
      \def\colorgray#1{\color[gray]{#1}}%
      \expandafter\def\csname LTw\endcsname{\color{white}}%
      \expandafter\def\csname LTb\endcsname{\color{black}}%
      \expandafter\def\csname LTa\endcsname{\color{black}}%
      \expandafter\def\csname LT0\endcsname{\color[rgb]{1,0,0}}%
      \expandafter\def\csname LT1\endcsname{\color[rgb]{0,1,0}}%
      \expandafter\def\csname LT2\endcsname{\color[rgb]{0,0,1}}%
      \expandafter\def\csname LT3\endcsname{\color[rgb]{1,0,1}}%
      \expandafter\def\csname LT4\endcsname{\color[rgb]{0,1,1}}%
      \expandafter\def\csname LT5\endcsname{\color[rgb]{1,1,0}}%
      \expandafter\def\csname LT6\endcsname{\color[rgb]{0,0,0}}%
      \expandafter\def\csname LT7\endcsname{\color[rgb]{1,0.3,0}}%
      \expandafter\def\csname LT8\endcsname{\color[rgb]{0.5,0.5,0.5}}%
    \else
      % gray
      \def\colorrgb#1{\color{black}}%
      \def\colorgray#1{\color[gray]{#1}}%
      \expandafter\def\csname LTw\endcsname{\color{white}}%
      \expandafter\def\csname LTb\endcsname{\color{black}}%
      \expandafter\def\csname LTa\endcsname{\color{black}}%
      \expandafter\def\csname LT0\endcsname{\color{black}}%
      \expandafter\def\csname LT1\endcsname{\color{black}}%
      \expandafter\def\csname LT2\endcsname{\color{black}}%
      \expandafter\def\csname LT3\endcsname{\color{black}}%
      \expandafter\def\csname LT4\endcsname{\color{black}}%
      \expandafter\def\csname LT5\endcsname{\color{black}}%
      \expandafter\def\csname LT6\endcsname{\color{black}}%
      \expandafter\def\csname LT7\endcsname{\color{black}}%
      \expandafter\def\csname LT8\endcsname{\color{black}}%
    \fi
  \fi
    \setlength{\unitlength}{0.0500bp}%
    \ifx\gptboxheight\undefined%
      \newlength{\gptboxheight}%
      \newlength{\gptboxwidth}%
      \newsavebox{\gptboxtext}%
    \fi%
    \setlength{\fboxrule}{0.5pt}%
    \setlength{\fboxsep}{1pt}%
\begin{picture}(7200.00,5040.00)%
    \gplgaddtomacro\gplbacktext{%
      \csname LTb\endcsname%%
      \put(3600,4336){\makebox(0,0){\strut{}(a)}}%
    }%
    \gplgaddtomacro\gplfronttext{%
      \csname LTb\endcsname%%
      \put(5955,4316){\makebox(0,0)[r]{\strut{}${\mathrm{Re}}{\mathit \Delta}_{dx}({\mathbf k},0)$}}%
      \csname LTb\endcsname%%
      \put(5955,4096){\makebox(0,0)[r]{\strut{}${\mathrm{Re}}{\mathit \Delta}_{dy}({\mathbf k},0)$}}%
      \csname LTb\endcsname%%
      \put(4463,1459){\makebox(0,0){\strut{}$(2\pi,0)$}}%
      \put(2259,1822){\makebox(0,0){\strut{}$k_x$}}%
      \csname LTb\endcsname%%
      \put(6301,2683){\makebox(0,0)[l]{\strut{}$(2\pi,2\pi)$}}%
      \put(5922,2117){\makebox(0,0){\strut{}$k_y$}}%
      \put(920,1340){\makebox(0,0)[r]{\strut{}$-100$}}%
      \put(920,1827){\makebox(0,0)[r]{\strut{}$-50$}}%
      \put(920,2315){\makebox(0,0)[r]{\strut{}$0$}}%
      \put(920,2801){\makebox(0,0)[r]{\strut{}$50$}}%
      \put(920,3288){\makebox(0,0)[r]{\strut{}$100$}}%
      \put(122,2315){\rotatebox{90}{\makebox(0,0){\strut{}Energy (meV)}}}%
    }%
    \gplbacktext
    \put(0,0){\includegraphics{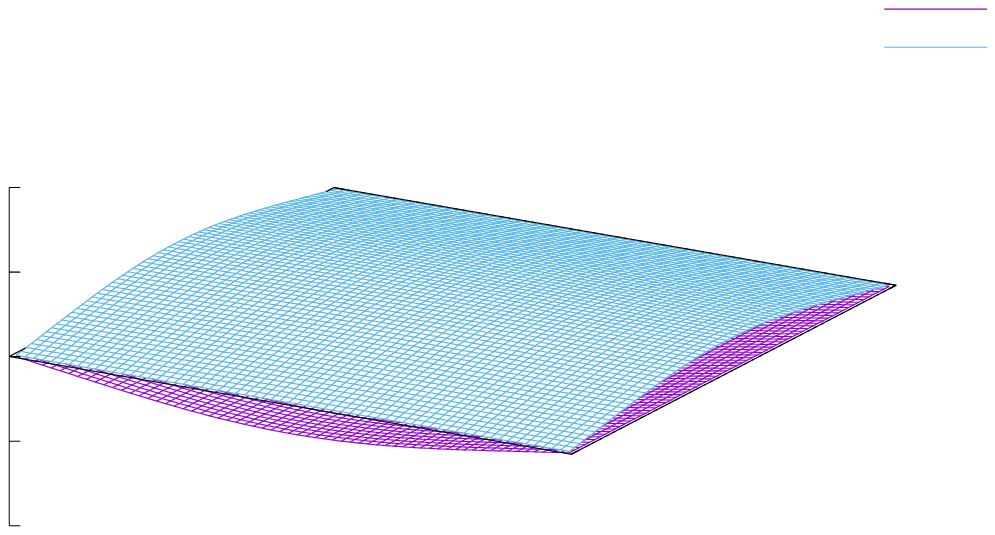}}%
    \gplfronttext
  \end{picture}%
\endgroup

%% file: fig6b.tex
% GNUPLOT: LaTeX picture with Postscript
\begingroup
  \makeatletter
  \providecommand\color[2][]{%
    \GenericError{(gnuplot) \space\space\space\@spaces}{%
      Package color not loaded in conjunction with
      terminal option `colourtext'%
    }{See the gnuplot documentation for explanation.%
    }{Either use 'blacktext' in gnuplot or load the package
      color.sty in LaTeX.}%
    \renewcommand\color[2][]{}%
  }%
  \providecommand\includegraphics[2][]{%
    \GenericError{(gnuplot) \space\space\space\@spaces}{%
      Package graphicx or graphics not loaded%
    }{See the gnuplot documentation for explanation.%
    }{The gnuplot epslatex terminal needs graphicx.sty or graphics.sty.}%
    \renewcommand\includegraphics[2][]{}%
  }%
  \providecommand\rotatebox[2]{#2}%
  \@ifundefined{ifGPcolor}{%
    \newif\ifGPcolor
    \GPcolortrue
  }{}%
  \@ifundefined{ifGPblacktext}{%
    \newif\ifGPblacktext
    \GPblacktexttrue
  }{}%
  % define a \g@addto@macro without @ in the name:
  \let\gplgaddtomacro\g@addto@macro
  % define empty templates for all commands taking text:
  \gdef\gplbacktext{}%
  \gdef\gplfronttext{}%
  \makeatother
  \ifGPblacktext
    % no textcolor at all
    \def\colorrgb#1{}%
    \def\colorgray#1{}%
  \else
    % gray or color?
    \ifGPcolor
      \def\colorrgb#1{\color[rgb]{#1}}%
      \def\colorgray#1{\color[gray]{#1}}%
      \expandafter\def\csname LTw\endcsname{\color{white}}%
      \expandafter\def\csname LTb\endcsname{\color{black}}%
      \expandafter\def\csname LTa\endcsname{\color{black}}%
      \expandafter\def\csname LT0\endcsname{\color[rgb]{1,0,0}}%
      \expandafter\def\csname LT1\endcsname{\color[rgb]{0,1,0}}%
      \expandafter\def\csname LT2\endcsname{\color[rgb]{0,0,1}}%
      \expandafter\def\csname LT3\endcsname{\color[rgb]{1,0,1}}%
      \expandafter\def\csname LT4\endcsname{\color[rgb]{0,1,1}}%
      \expandafter\def\csname LT5\endcsname{\color[rgb]{1,1,0}}%
      \expandafter\def\csname LT6\endcsname{\color[rgb]{0,0,0}}%
      \expandafter\def\csname LT7\endcsname{\color[rgb]{1,0.3,0}}%
      \expandafter\def\csname LT8\endcsname{\color[rgb]{0.5,0.5,0.5}}%
    \else
      % gray
      \def\colorrgb#1{\color{black}}%
      \def\colorgray#1{\color[gray]{#1}}%
      \expandafter\def\csname LTw\endcsname{\color{white}}%
      \expandafter\def\csname LTb\endcsname{\color{black}}%
      \expandafter\def\csname LTa\endcsname{\color{black}}%
      \expandafter\def\csname LT0\endcsname{\color{black}}%
      \expandafter\def\csname LT1\endcsname{\color{black}}%
      \expandafter\def\csname LT2\endcsname{\color{black}}%
      \expandafter\def\csname LT3\endcsname{\color{black}}%
      \expandafter\def\csname LT4\endcsname{\color{black}}%
      \expandafter\def\csname LT5\endcsname{\color{black}}%
      \expandafter\def\csname LT6\endcsname{\color{black}}%
      \expandafter\def\csname LT7\endcsname{\color{black}}%
      \expandafter\def\csname LT8\endcsname{\color{black}}%
    \fi
  \fi
    \setlength{\unitlength}{0.0500bp}%
    \ifx\gptboxheight\undefined%
      \newlength{\gptboxheight}%
      \newlength{\gptboxwidth}%
      \newsavebox{\gptboxtext}%
    \fi%
    \setlength{\fboxrule}{0.5pt}%
    \setlength{\fboxsep}{1pt}%
\begin{picture}(7200.00,5040.00)%
    \gplgaddtomacro\gplbacktext{%
      \csname LTb\endcsname%%
      \put(3600,4336){\makebox(0,0){\strut{}(b)}}%
    }%
    \gplgaddtomacro\gplfronttext{%
      \csname LTb\endcsname%%
      \put(5955,4316){\makebox(0,0)[r]{\strut{}${\mathrm{Re}}{\mathit \Delta}_{xd}({\mathbf k},0)$}}%
      \csname LTb\endcsname%%
      \put(5955,4096){\makebox(0,0)[r]{\strut{}${\mathrm{Re}}{\mathit \Delta}_{yd}({\mathbf k},0)$}}%
      \csname LTa\endcsname%%
      \put(4463,1459){\makebox(0,0){\strut{}$(2\pi,0)$}}%
      \csname LTb\endcsname%%
      \put(2259,1822){\makebox(0,0){\strut{}$k_x$}}%
      \csname LTb\endcsname%%
      \put(6301,2683){\makebox(0,0)[l]{\strut{}$(2\pi,2\pi)$}}%
      \put(5922,2117){\makebox(0,0){\strut{}$k_y$}}%
      \put(920,1340){\makebox(0,0)[r]{\strut{}$-100$}}%
      \put(920,1827){\makebox(0,0)[r]{\strut{}$-50$}}%
      \put(920,2315){\makebox(0,0)[r]{\strut{}$0$}}%
      \put(920,2801){\makebox(0,0)[r]{\strut{}$50$}}%
      \put(920,3288){\makebox(0,0)[r]{\strut{}$100$}}%
      \put(122,2315){\rotatebox{90}{\makebox(0,0){\strut{}Energy (meV)}}}%
    }%
    \gplbacktext
    \put(0,0){\includegraphics{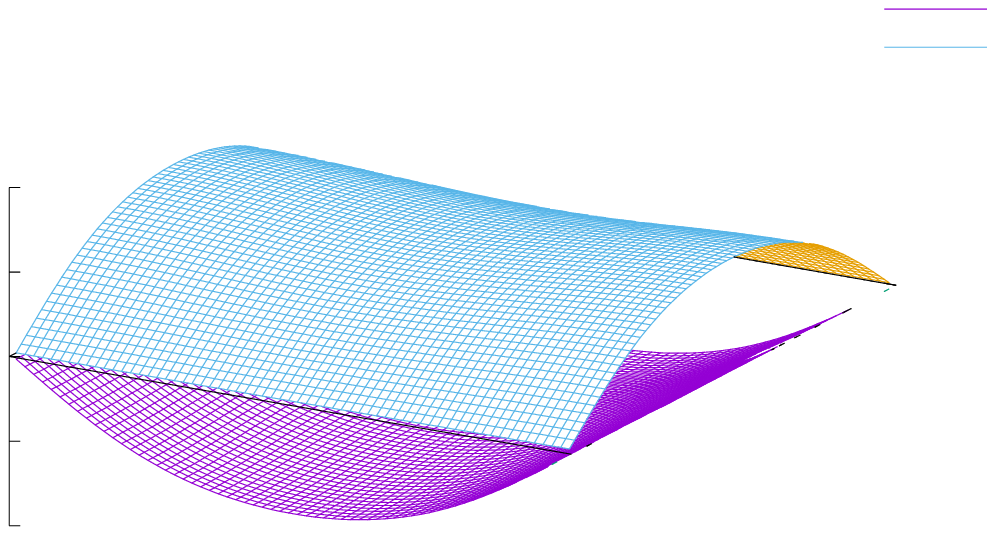}}%
    \gplfronttext
  \end{picture}%
\endgroup

%% file: fig6c.tex
% GNUPLOT: LaTeX picture with Postscript
\begingroup
  \makeatletter
  \providecommand\color[2][]{%
    \GenericError{(gnuplot) \space\space\space\@spaces}{%
      Package color not loaded in conjunction with
      terminal option `colourtext'%
    }{See the gnuplot documentation for explanation.%
    }{Either use 'blacktext' in gnuplot or load the package
      color.sty in LaTeX.}%
    \renewcommand\color[2][]{}%
  }%
  \providecommand\includegraphics[2][]{%
    \GenericError{(gnuplot) \space\space\space\@spaces}{%
      Package graphicx or graphics not loaded%
    }{See the gnuplot documentation for explanation.%
    }{The gnuplot epslatex terminal needs graphicx.sty or graphics.sty.}%
    \renewcommand\includegraphics[2][]{}%
  }%
  \providecommand\rotatebox[2]{#2}%
  \@ifundefined{ifGPcolor}{%
    \newif\ifGPcolor
    \GPcolortrue
  }{}%
  \@ifundefined{ifGPblacktext}{%
    \newif\ifGPblacktext
    \GPblacktexttrue
  }{}%
  % define a \g@addto@macro without @ in the name:
  \let\gplgaddtomacro\g@addto@macro
  % define empty templates for all commands taking text:
  \gdef\gplbacktext{}%
  \gdef\gplfronttext{}%
  \makeatother
  \ifGPblacktext
    % no textcolor at all
    \def\colorrgb#1{}%
    \def\colorgray#1{}%
  \else
    % gray or color?
    \ifGPcolor
      \def\colorrgb#1{\color[rgb]{#1}}%
      \def\colorgray#1{\color[gray]{#1}}%
      \expandafter\def\csname LTw\endcsname{\color{white}}%
      \expandafter\def\csname LTb\endcsname{\color{black}}%
      \expandafter\def\csname LTa\endcsname{\color{black}}%
      \expandafter\def\csname LT0\endcsname{\color[rgb]{1,0,0}}%
      \expandafter\def\csname LT1\endcsname{\color[rgb]{0,1,0}}%
      \expandafter\def\csname LT2\endcsname{\color[rgb]{0,0,1}}%
      \expandafter\def\csname LT3\endcsname{\color[rgb]{1,0,1}}%
      \expandafter\def\csname LT4\endcsname{\color[rgb]{0,1,1}}%
      \expandafter\def\csname LT5\endcsname{\color[rgb]{1,1,0}}%
      \expandafter\def\csname LT6\endcsname{\color[rgb]{0,0,0}}%
      \expandafter\def\csname LT7\endcsname{\color[rgb]{1,0.3,0}}%
      \expandafter\def\csname LT8\endcsname{\color[rgb]{0.5,0.5,0.5}}%
    \else
      % gray
      \def\colorrgb#1{\color{black}}%
      \def\colorgray#1{\color[gray]{#1}}%
      \expandafter\def\csname LTw\endcsname{\color{white}}%
      \expandafter\def\csname LTb\endcsname{\color{black}}%
      \expandafter\def\csname LTa\endcsname{\color{black}}%
      \expandafter\def\csname LT0\endcsname{\color{black}}%
      \expandafter\def\csname LT1\endcsname{\color{black}}%
      \expandafter\def\csname LT2\endcsname{\color{black}}%
      \expandafter\def\csname LT3\endcsname{\color{black}}%
      \expandafter\def\csname LT4\endcsname{\color{black}}%
      \expandafter\def\csname LT5\endcsname{\color{black}}%
      \expandafter\def\csname LT6\endcsname{\color{black}}%
      \expandafter\def\csname LT7\endcsname{\color{black}}%
      \expandafter\def\csname LT8\endcsname{\color{black}}%
    \fi
  \fi
    \setlength{\unitlength}{0.0500bp}%
    \ifx\gptboxheight\undefined%
      \newlength{\gptboxheight}%
      \newlength{\gptboxwidth}%
      \newsavebox{\gptboxtext}%
    \fi%
    \setlength{\fboxrule}{0.5pt}%
    \setlength{\fboxsep}{1pt}%
\begin{picture}(7200.00,5040.00)%
    \gplgaddtomacro\gplbacktext{%
      \csname LTb\endcsname%%
      \put(3600,4336){\makebox(0,0){\strut{}(c)}}%
    }%
    \gplgaddtomacro\gplfronttext{%
      \csname LTb\endcsname%%
      \put(5955,4316){\makebox(0,0)[r]{\strut{}      50}}%
      \csname LTb\endcsname%%
      \put(5955,4096){\makebox(0,0)[r]{\strut{}       0}}%
      \csname LTb\endcsname%%
      \put(5955,3876){\makebox(0,0)[r]{\strut{}     -50}}%
      \csname LTb\endcsname%%
      \put(4463,1459){\makebox(0,0){\strut{}$(2\pi,0)$}}%
      \put(2259,1822){\makebox(0,0){\strut{}$k_x$}}%
      \csname LTb\endcsname%%
      \put(6301,2683){\makebox(0,0)[l]{\strut{}$(2\pi,2\pi)$}}%
      \put(5922,2117){\makebox(0,0){\strut{}$k_y$}}%
      \put(920,1340){\makebox(0,0)[r]{\strut{}$-100$}}%
      \put(920,1827){\makebox(0,0)[r]{\strut{}$-50$}}%
      \put(920,2315){\makebox(0,0)[r]{\strut{}$0$}}%
      \put(920,2801){\makebox(0,0)[r]{\strut{}$50$}}%
      \put(920,3288){\makebox(0,0)[r]{\strut{}$100$}}%
      \put(122,2315){\rotatebox{90}{\makebox(0,0){\strut{}Energy (meV)}}}%
    }%
    \gplbacktext
    \put(0,0){\includegraphics{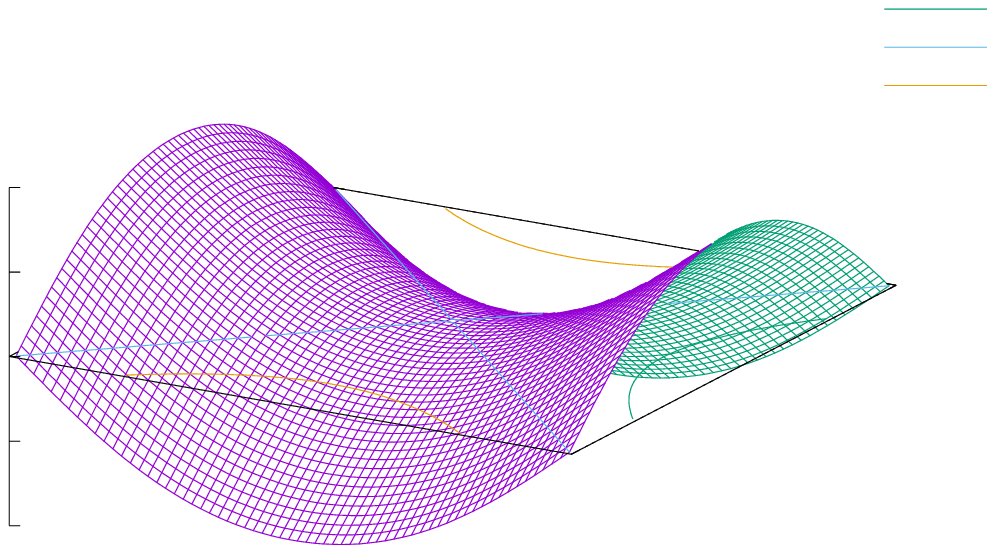}}%
    \gplfronttext
  \end{picture}%
\endgroup

%% file: fig7a.tex
% GNUPLOT: LaTeX picture with Postscript
\begingroup
  \makeatletter
  \providecommand\color[2][]{%
    \GenericError{(gnuplot) \space\space\space\@spaces}{%
      Package color not loaded in conjunction with
      terminal option `colourtext'%
    }{See the gnuplot documentation for explanation.%
    }{Either use 'blacktext' in gnuplot or load the package
      color.sty in LaTeX.}%
    \renewcommand\color[2][]{}%
  }%
  \providecommand\includegraphics[2][]{%
    \GenericError{(gnuplot) \space\space\space\@spaces}{%
      Package graphicx or graphics not loaded%
    }{See the gnuplot documentation for explanation.%
    }{The gnuplot epslatex terminal needs graphicx.sty or graphics.sty.}%
    \renewcommand\includegraphics[2][]{}%
  }%
  \providecommand\rotatebox[2]{#2}%
  \@ifundefined{ifGPcolor}{%
    \newif\ifGPcolor
    \GPcolortrue
  }{}%
  \@ifundefined{ifGPblacktext}{%
    \newif\ifGPblacktext
    \GPblacktexttrue
  }{}%
  % define a \g@addto@macro without @ in the name:
  \let\gplgaddtomacro\g@addto@macro
  % define empty templates for all commands taking text:
  \gdef\gplbacktext{}%
  \gdef\gplfronttext{}%
  \makeatother
  \ifGPblacktext
    % no textcolor at all
    \def\colorrgb#1{}%
    \def\colorgray#1{}%
  \else
    % gray or color?
    \ifGPcolor
      \def\colorrgb#1{\color[rgb]{#1}}%
      \def\colorgray#1{\color[gray]{#1}}%
      \expandafter\def\csname LTw\endcsname{\color{white}}%
      \expandafter\def\csname LTb\endcsname{\color{black}}%
      \expandafter\def\csname LTa\endcsname{\color{black}}%
      \expandafter\def\csname LT0\endcsname{\color[rgb]{1,0,0}}%
      \expandafter\def\csname LT1\endcsname{\color[rgb]{0,1,0}}%
      \expandafter\def\csname LT2\endcsname{\color[rgb]{0,0,1}}%
      \expandafter\def\csname LT3\endcsname{\color[rgb]{1,0,1}}%
      \expandafter\def\csname LT4\endcsname{\color[rgb]{0,1,1}}%
      \expandafter\def\csname LT5\endcsname{\color[rgb]{1,1,0}}%
      \expandafter\def\csname LT6\endcsname{\color[rgb]{0,0,0}}%
      \expandafter\def\csname LT7\endcsname{\color[rgb]{1,0.3,0}}%
      \expandafter\def\csname LT8\endcsname{\color[rgb]{0.5,0.5,0.5}}%
    \else
      % gray
      \def\colorrgb#1{\color{black}}%
      \def\colorgray#1{\color[gray]{#1}}%
      \expandafter\def\csname LTw\endcsname{\color{white}}%
      \expandafter\def\csname LTb\endcsname{\color{black}}%
      \expandafter\def\csname LTa\endcsname{\color{black}}%
      \expandafter\def\csname LT0\endcsname{\color{black}}%
      \expandafter\def\csname LT1\endcsname{\color{black}}%
      \expandafter\def\csname LT2\endcsname{\color{black}}%
      \expandafter\def\csname LT3\endcsname{\color{black}}%
      \expandafter\def\csname LT4\endcsname{\color{black}}%
      \expandafter\def\csname LT5\endcsname{\color{black}}%
      \expandafter\def\csname LT6\endcsname{\color{black}}%
      \expandafter\def\csname LT7\endcsname{\color{black}}%
      \expandafter\def\csname LT8\endcsname{\color{black}}%
    \fi
  \fi
    \setlength{\unitlength}{0.0500bp}%
    \ifx\gptboxheight\undefined%
      \newlength{\gptboxheight}%
      \newlength{\gptboxwidth}%
      \newsavebox{\gptboxtext}%
    \fi%
    \setlength{\fboxrule}{0.5pt}%
    \setlength{\fboxsep}{1pt}%
\begin{picture}(7200.00,5040.00)%
    \gplgaddtomacro\gplbacktext{%
      \csname LTb\endcsname%%
      \put(3600,4336){\makebox(0,0){\strut{}(a)}}%
    }%
    \gplgaddtomacro\gplfronttext{%
      \csname LTb\endcsname%%
      \put(5955,4316){\makebox(0,0)[r]{\strut{}${\mathrm{Re}}{\mathit \Delta}_{dd}({\mathbf k},0)$}}%
      \csname LTb\endcsname%%
      \put(4463,2433){\makebox(0,0){\strut{}$(2\pi,0)$}}%
      \put(2259,2795){\makebox(0,0){\strut{}$k_x$}}%
      \csname LTb\endcsname%%
      \put(6301,3657){\makebox(0,0)[l]{\strut{}$(2\pi,2\pi)$}}%
      \put(5922,3090){\makebox(0,0){\strut{}$k_y$}}%
      \put(920,1340){\makebox(0,0)[r]{\strut{}$-40$}}%
      \put(920,1827){\makebox(0,0)[r]{\strut{}$-30$}}%
      \put(920,2315){\makebox(0,0)[r]{\strut{}$-20$}}%
      \put(920,2801){\makebox(0,0)[r]{\strut{}$-10$}}%
      \put(920,3288){\makebox(0,0)[r]{\strut{}$0$}}%
      \put(122,2315){\rotatebox{90}{\makebox(0,0){\strut{}Energy (meV)}}}%
    }%
    \gplbacktext
    \put(0,0){\includegraphics{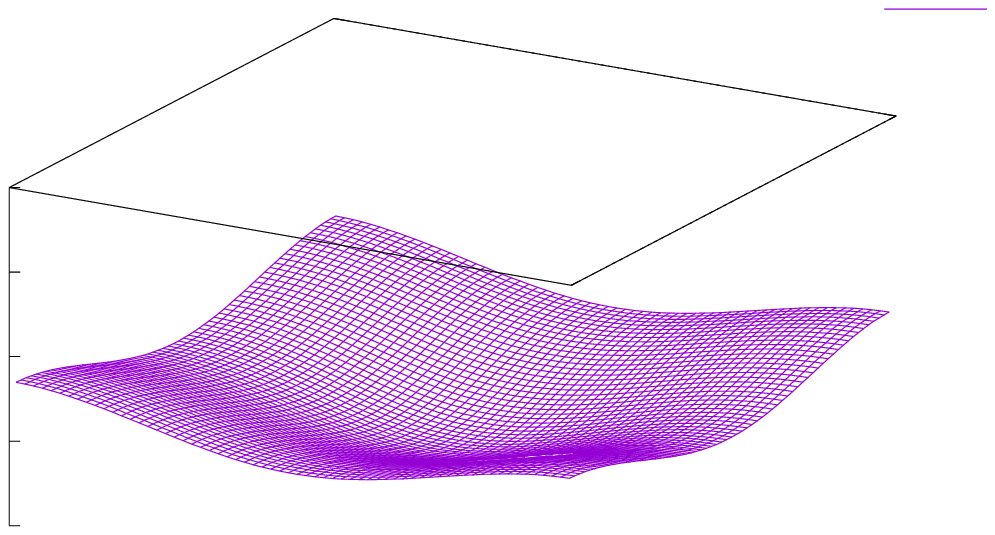}}%
    \gplfronttext
  \end{picture}%
\endgroup

%% file: fig7b.tex
% GNUPLOT: LaTeX picture with Postscript
\begingroup
  \makeatletter
  \providecommand\color[2][]{%
    \GenericError{(gnuplot) \space\space\space\@spaces}{%
      Package color not loaded in conjunction with
      terminal option `colourtext'%
    }{See the gnuplot documentation for explanation.%
    }{Either use 'blacktext' in gnuplot or load the package
      color.sty in LaTeX.}%
    \renewcommand\color[2][]{}%
  }%
  \providecommand\includegraphics[2][]{%
    \GenericError{(gnuplot) \space\space\space\@spaces}{%
      Package graphicx or graphics not loaded%
    }{See the gnuplot documentation for explanation.%
    }{The gnuplot epslatex terminal needs graphicx.sty or graphics.sty.}%
    \renewcommand\includegraphics[2][]{}%
  }%
  \providecommand\rotatebox[2]{#2}%
  \@ifundefined{ifGPcolor}{%
    \newif\ifGPcolor
    \GPcolortrue
  }{}%
  \@ifundefined{ifGPblacktext}{%
    \newif\ifGPblacktext
    \GPblacktexttrue
  }{}%
  % define a \g@addto@macro without @ in the name:
  \let\gplgaddtomacro\g@addto@macro
  % define empty templates for all commands taking text:
  \gdef\gplbacktext{}%
  \gdef\gplfronttext{}%
  \makeatother
  \ifGPblacktext
    % no textcolor at all
    \def\colorrgb#1{}%
    \def\colorgray#1{}%
  \else
    % gray or color?
    \ifGPcolor
      \def\colorrgb#1{\color[rgb]{#1}}%
      \def\colorgray#1{\color[gray]{#1}}%
      \expandafter\def\csname LTw\endcsname{\color{white}}%
      \expandafter\def\csname LTb\endcsname{\color{black}}%
      \expandafter\def\csname LTa\endcsname{\color{black}}%
      \expandafter\def\csname LT0\endcsname{\color[rgb]{1,0,0}}%
      \expandafter\def\csname LT1\endcsname{\color[rgb]{0,1,0}}%
      \expandafter\def\csname LT2\endcsname{\color[rgb]{0,0,1}}%
      \expandafter\def\csname LT3\endcsname{\color[rgb]{1,0,1}}%
      \expandafter\def\csname LT4\endcsname{\color[rgb]{0,1,1}}%
      \expandafter\def\csname LT5\endcsname{\color[rgb]{1,1,0}}%
      \expandafter\def\csname LT6\endcsname{\color[rgb]{0,0,0}}%
      \expandafter\def\csname LT7\endcsname{\color[rgb]{1,0.3,0}}%
      \expandafter\def\csname LT8\endcsname{\color[rgb]{0.5,0.5,0.5}}%
    \else
      % gray
      \def\colorrgb#1{\color{black}}%
      \def\colorgray#1{\color[gray]{#1}}%
      \expandafter\def\csname LTw\endcsname{\color{white}}%
      \expandafter\def\csname LTb\endcsname{\color{black}}%
      \expandafter\def\csname LTa\endcsname{\color{black}}%
      \expandafter\def\csname LT0\endcsname{\color{black}}%
      \expandafter\def\csname LT1\endcsname{\color{black}}%
      \expandafter\def\csname LT2\endcsname{\color{black}}%
      \expandafter\def\csname LT3\endcsname{\color{black}}%
      \expandafter\def\csname LT4\endcsname{\color{black}}%
      \expandafter\def\csname LT5\endcsname{\color{black}}%
      \expandafter\def\csname LT6\endcsname{\color{black}}%
      \expandafter\def\csname LT7\endcsname{\color{black}}%
      \expandafter\def\csname LT8\endcsname{\color{black}}%
    \fi
  \fi
    \setlength{\unitlength}{0.0500bp}%
    \ifx\gptboxheight\undefined%
      \newlength{\gptboxheight}%
      \newlength{\gptboxwidth}%
      \newsavebox{\gptboxtext}%
    \fi%
    \setlength{\fboxrule}{0.5pt}%
    \setlength{\fboxsep}{1pt}%
\begin{picture}(7200.00,5040.00)%
    \gplgaddtomacro\gplbacktext{%
      \csname LTb\endcsname%%
      \put(3600,4336){\makebox(0,0){\strut{}(b)}}%
    }%
    \gplgaddtomacro\gplfronttext{%
      \csname LTb\endcsname%%
      \put(5955,4316){\makebox(0,0)[r]{\strut{}${\mathrm{Re}}{\mathit \Delta}_{xx}({\mathbf k},0)$}}%
      \csname LTb\endcsname%%
      \put(5955,4096){\makebox(0,0)[r]{\strut{}${\mathrm{Re}}{\mathit \Delta}_{xy}({\mathbf k},0)$}}%
      \csname LTb\endcsname%%
      \put(4463,1459){\makebox(0,0){\strut{}$(2\pi,0)$}}%
      \put(2259,1822){\makebox(0,0){\strut{}$k_x$}}%
      \csname LTb\endcsname%%
      \put(6301,2683){\makebox(0,0)[l]{\strut{}$(2\pi,2\pi)$}}%
      \put(5922,2117){\makebox(0,0){\strut{}$k_y$}}%
      \put(920,1340){\makebox(0,0)[r]{\strut{}$-20$}}%
      \put(920,1827){\makebox(0,0)[r]{\strut{}$-10$}}%
      \put(920,2315){\makebox(0,0)[r]{\strut{}$0$}}%
      \put(920,2801){\makebox(0,0)[r]{\strut{}$10$}}%
      \put(920,3288){\makebox(0,0)[r]{\strut{}$20$}}%
      \put(122,2315){\rotatebox{90}{\makebox(0,0){\strut{}Energy (meV)}}}%
    }%
    \gplbacktext
    \put(0,0){\includegraphics{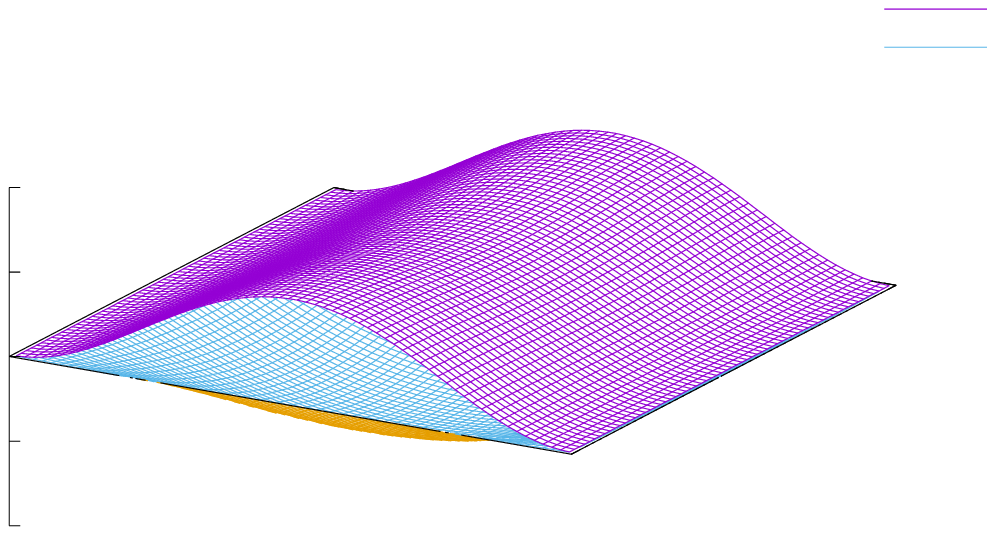}}%
    \gplfronttext
  \end{picture}%
\endgroup

%% file: fig7c.tex
% GNUPLOT: LaTeX picture with Postscript
\begingroup
  \makeatletter
  \providecommand\color[2][]{%
    \GenericError{(gnuplot) \space\space\space\@spaces}{%
      Package color not loaded in conjunction with
      terminal option `colourtext'%
    }{See the gnuplot documentation for explanation.%
    }{Either use 'blacktext' in gnuplot or load the package
      color.sty in LaTeX.}%
    \renewcommand\color[2][]{}%
  }%
  \providecommand\includegraphics[2][]{%
    \GenericError{(gnuplot) \space\space\space\@spaces}{%
      Package graphicx or graphics not loaded%
    }{See the gnuplot documentation for explanation.%
    }{The gnuplot epslatex terminal needs graphicx.sty or graphics.sty.}%
    \renewcommand\includegraphics[2][]{}%
  }%
  \providecommand\rotatebox[2]{#2}%
  \@ifundefined{ifGPcolor}{%
    \newif\ifGPcolor
    \GPcolortrue
  }{}%
  \@ifundefined{ifGPblacktext}{%
    \newif\ifGPblacktext
    \GPblacktexttrue
  }{}%
  % define a \g@addto@macro without @ in the name:
  \let\gplgaddtomacro\g@addto@macro
  % define empty templates for all commands taking text:
  \gdef\gplbacktext{}%
  \gdef\gplfronttext{}%
  \makeatother
  \ifGPblacktext
    % no textcolor at all
    \def\colorrgb#1{}%
    \def\colorgray#1{}%
  \else
    % gray or color?
    \ifGPcolor
      \def\colorrgb#1{\color[rgb]{#1}}%
      \def\colorgray#1{\color[gray]{#1}}%
      \expandafter\def\csname LTw\endcsname{\color{white}}%
      \expandafter\def\csname LTb\endcsname{\color{black}}%
      \expandafter\def\csname LTa\endcsname{\color{black}}%
      \expandafter\def\csname LT0\endcsname{\color[rgb]{1,0,0}}%
      \expandafter\def\csname LT1\endcsname{\color[rgb]{0,1,0}}%
      \expandafter\def\csname LT2\endcsname{\color[rgb]{0,0,1}}%
      \expandafter\def\csname LT3\endcsname{\color[rgb]{1,0,1}}%
      \expandafter\def\csname LT4\endcsname{\color[rgb]{0,1,1}}%
      \expandafter\def\csname LT5\endcsname{\color[rgb]{1,1,0}}%
      \expandafter\def\csname LT6\endcsname{\color[rgb]{0,0,0}}%
      \expandafter\def\csname LT7\endcsname{\color[rgb]{1,0.3,0}}%
      \expandafter\def\csname LT8\endcsname{\color[rgb]{0.5,0.5,0.5}}%
    \else
      % gray
      \def\colorrgb#1{\color{black}}%
      \def\colorgray#1{\color[gray]{#1}}%
      \expandafter\def\csname LTw\endcsname{\color{white}}%
      \expandafter\def\csname LTb\endcsname{\color{black}}%
      \expandafter\def\csname LTa\endcsname{\color{black}}%
      \expandafter\def\csname LT0\endcsname{\color{black}}%
      \expandafter\def\csname LT1\endcsname{\color{black}}%
      \expandafter\def\csname LT2\endcsname{\color{black}}%
      \expandafter\def\csname LT3\endcsname{\color{black}}%
      \expandafter\def\csname LT4\endcsname{\color{black}}%
      \expandafter\def\csname LT5\endcsname{\color{black}}%
      \expandafter\def\csname LT6\endcsname{\color{black}}%
      \expandafter\def\csname LT7\endcsname{\color{black}}%
      \expandafter\def\csname LT8\endcsname{\color{black}}%
    \fi
  \fi
    \setlength{\unitlength}{0.0500bp}%
    \ifx\gptboxheight\undefined%
      \newlength{\gptboxheight}%
      \newlength{\gptboxwidth}%
      \newsavebox{\gptboxtext}%
    \fi%
    \setlength{\fboxrule}{0.5pt}%
    \setlength{\fboxsep}{1pt}%
\begin{picture}(7200.00,5040.00)%
    \gplgaddtomacro\gplbacktext{%
      \csname LTb\endcsname%%
      \put(3600,4336){\makebox(0,0){\strut{}(c)}}%
    }%
    \gplgaddtomacro\gplfronttext{%
      \csname LTb\endcsname%%
      \put(5955,4316){\makebox(0,0)[r]{\strut{}${\mathrm{Re}}{\mathit \Delta}_{yx}({\mathbf k},0)$}}%
      \csname LTb\endcsname%%
      \put(5955,4096){\makebox(0,0)[r]{\strut{}${\mathrm{Re}}{\mathit \Delta}_{yy}({\mathbf k},0)$}}%
      \csname LTb\endcsname%%
      \put(4463,1459){\makebox(0,0){\strut{}$(2\pi,0)$}}%
      \put(2259,1822){\makebox(0,0){\strut{}$k_x$}}%
      \csname LTb\endcsname%%
      \put(6301,2683){\makebox(0,0)[l]{\strut{}$(2\pi,2\pi)$}}%
      \put(5922,2117){\makebox(0,0){\strut{}$k_y$}}%
      \put(920,1340){\makebox(0,0)[r]{\strut{}$-20$}}%
      \put(920,1827){\makebox(0,0)[r]{\strut{}$-10$}}%
      \put(920,2315){\makebox(0,0)[r]{\strut{}$0$}}%
      \put(920,2801){\makebox(0,0)[r]{\strut{}$10$}}%
      \put(920,3288){\makebox(0,0)[r]{\strut{}$20$}}%
      \put(122,2315){\rotatebox{90}{\makebox(0,0){\strut{}Energy (meV)}}}%
    }%
    \gplbacktext
    \put(0,0){\includegraphics{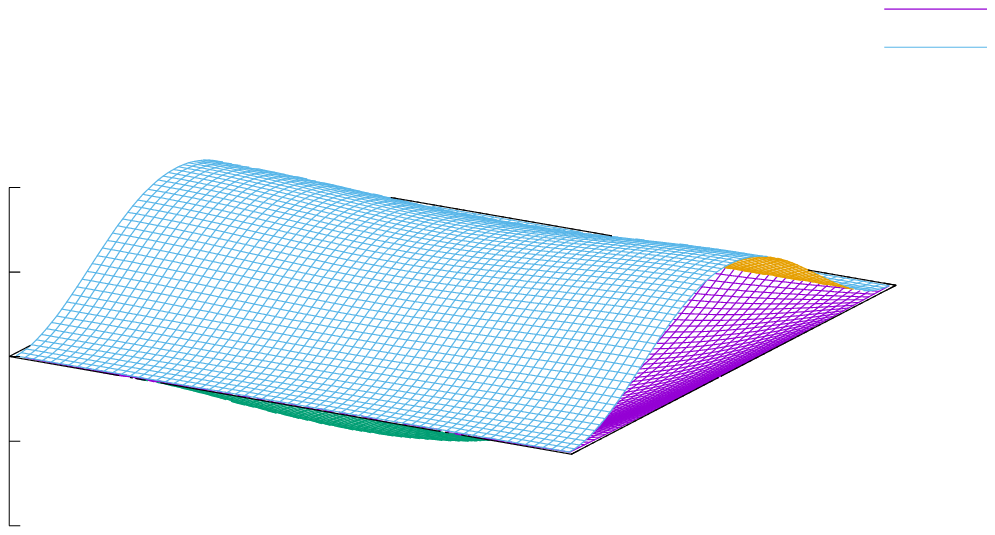}}%
    \gplfronttext
  \end{picture}%
\endgroup

%% file: fig7d.tex
% GNUPLOT: LaTeX picture with Postscript
\begingroup
  \makeatletter
  \providecommand\color[2][]{%
    \GenericError{(gnuplot) \space\space\space\@spaces}{%
      Package color not loaded in conjunction with
      terminal option `colourtext'%
    }{See the gnuplot documentation for explanation.%
    }{Either use 'blacktext' in gnuplot or load the package
      color.sty in LaTeX.}%
    \renewcommand\color[2][]{}%
  }%
  \providecommand\includegraphics[2][]{%
    \GenericError{(gnuplot) \space\space\space\@spaces}{%
      Package graphicx or graphics not loaded%
    }{See the gnuplot documentation for explanation.%
    }{The gnuplot epslatex terminal needs graphicx.sty or graphics.sty.}%
    \renewcommand\includegraphics[2][]{}%
  }%
  \providecommand\rotatebox[2]{#2}%
  \@ifundefined{ifGPcolor}{%
    \newif\ifGPcolor
    \GPcolortrue
  }{}%
  \@ifundefined{ifGPblacktext}{%
    \newif\ifGPblacktext
    \GPblacktexttrue
  }{}%
  % define a \g@addto@macro without @ in the name:
  \let\gplgaddtomacro\g@addto@macro
  % define empty templates for all commands taking text:
  \gdef\gplbacktext{}%
  \gdef\gplfronttext{}%
  \makeatother
  \ifGPblacktext
    % no textcolor at all
    \def\colorrgb#1{}%
    \def\colorgray#1{}%
  \else
    % gray or color?
    \ifGPcolor
      \def\colorrgb#1{\color[rgb]{#1}}%
      \def\colorgray#1{\color[gray]{#1}}%
      \expandafter\def\csname LTw\endcsname{\color{white}}%
      \expandafter\def\csname LTb\endcsname{\color{black}}%
      \expandafter\def\csname LTa\endcsname{\color{black}}%
      \expandafter\def\csname LT0\endcsname{\color[rgb]{1,0,0}}%
      \expandafter\def\csname LT1\endcsname{\color[rgb]{0,1,0}}%
      \expandafter\def\csname LT2\endcsname{\color[rgb]{0,0,1}}%
      \expandafter\def\csname LT3\endcsname{\color[rgb]{1,0,1}}%
      \expandafter\def\csname LT4\endcsname{\color[rgb]{0,1,1}}%
      \expandafter\def\csname LT5\endcsname{\color[rgb]{1,1,0}}%
      \expandafter\def\csname LT6\endcsname{\color[rgb]{0,0,0}}%
      \expandafter\def\csname LT7\endcsname{\color[rgb]{1,0.3,0}}%
      \expandafter\def\csname LT8\endcsname{\color[rgb]{0.5,0.5,0.5}}%
    \else
      % gray
      \def\colorrgb#1{\color{black}}%
      \def\colorgray#1{\color[gray]{#1}}%
      \expandafter\def\csname LTw\endcsname{\color{white}}%
      \expandafter\def\csname LTb\endcsname{\color{black}}%
      \expandafter\def\csname LTa\endcsname{\color{black}}%
      \expandafter\def\csname LT0\endcsname{\color{black}}%
      \expandafter\def\csname LT1\endcsname{\color{black}}%
      \expandafter\def\csname LT2\endcsname{\color{black}}%
      \expandafter\def\csname LT3\endcsname{\color{black}}%
      \expandafter\def\csname LT4\endcsname{\color{black}}%
      \expandafter\def\csname LT5\endcsname{\color{black}}%
      \expandafter\def\csname LT6\endcsname{\color{black}}%
      \expandafter\def\csname LT7\endcsname{\color{black}}%
      \expandafter\def\csname LT8\endcsname{\color{black}}%
    \fi
  \fi
    \setlength{\unitlength}{0.0500bp}%
    \ifx\gptboxheight\undefined%
      \newlength{\gptboxheight}%
      \newlength{\gptboxwidth}%
      \newsavebox{\gptboxtext}%
    \fi%
    \setlength{\fboxrule}{0.5pt}%
    \setlength{\fboxsep}{1pt}%
\begin{picture}(7200.00,5040.00)%
    \gplgaddtomacro\gplbacktext{%
      \csname LTb\endcsname%%
      \put(3600,4336){\makebox(0,0){\strut{}(d)}}%
    }%
    \gplgaddtomacro\gplfronttext{%
      \csname LTb\endcsname%%
      \put(5955,4316){\makebox(0,0)[r]{\strut{}     -20}}%
      \csname LTb\endcsname%%
      \put(5955,4096){\makebox(0,0)[r]{\strut{}     -25}}%
      \csname LTb\endcsname%%
      \put(5955,3876){\makebox(0,0)[r]{\strut{}     -30}}%
      \csname LTb\endcsname%%
      \put(4463,2433){\makebox(0,0){\strut{}$(2\pi,0)$}}%
      \put(2259,2795){\makebox(0,0){\strut{}$k_x$}}%
      \csname LTb\endcsname%%
      \put(6301,3657){\makebox(0,0)[l]{\strut{}$(2\pi,2\pi)$}}%
      \put(5922,3090){\makebox(0,0){\strut{}$k_y$}}%
      \put(920,1340){\makebox(0,0)[r]{\strut{}$-40$}}%
      \put(920,2315){\makebox(0,0)[r]{\strut{}$-20$}}%
      \put(920,3288){\makebox(0,0)[r]{\strut{}$0$}}%
      \put(122,2315){\rotatebox{90}{\makebox(0,0){\strut{}Energy (meV)}}}%
    }%
    \gplbacktext
    \put(0,0){\includegraphics{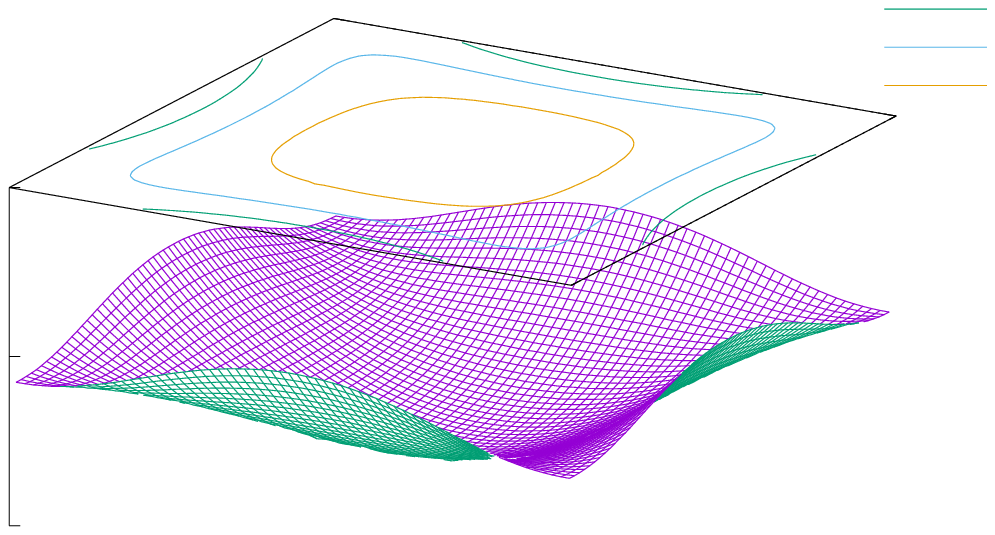}}%
    \gplfronttext
  \end{picture}%
\endgroup

%% file: JPCM-118505.R2.bbl
\begin{thebibliography}{100}
\bibitem{Chao77}
Chao K A, Spa\l ek J and Ole\'s A M 1977 \JPC {\bf 10} L271

\bibitem{Chao78}
Chao K A, Spa\l ek J and Ole\'s A M 1978 \PR B {\bf 18} 3453

\bibitem{FCZhang88}
Zhang F C and Rice T M 1988 \PR B {\bf 37} 3759

\bibitem{Zaanen88}
Zaanen J and Ole\'s A M 1988 \PR B {\bf 37} 9423

\bibitem{Matsukawa89}
Matsukawa H and Fukuyama H 1989 \JPSJ {\bf 58} 2845

\bibitem{Mori02} % t-J
Mori M, Tohyama T and Maekawa S 2002 \PR B {\bf 66} 064502

\bibitem{Plakida02} % t-J
Plakida N M 2002 {\it Cond. Matter Phys.} {\bf 5} 707

\bibitem{Rosch04} % t-J
R\"osch O and Gunnarsson O 2004 \PRL {\bf 92} 146403

\bibitem{Ishihara04} % t-J
Ishihara S and Nagaosa N 2004 \PR B {\bf 69} 144520

\bibitem{Tohyama04} % t-J
Tohyama T 2004 \PR B {\bf 70} 174517

\bibitem{Mishchenko04} % t-J
Mishchenko A S and Nagaosa N 2004 \PRL {\bf 93} 036402

\bibitem{Mishchenko06} % t-J
Mishchenko A S and Nagaosa N 2006 \PR B {\bf 73} 092502

\bibitem{Mallik20} % t-J
Mallik Aabhaas V, Gupta Gaurav K, Shenoy Vijay B and Krishnamurthy H R 2020 \PRL {\bf 124} 147002

\bibitem{FCZhang03} % t-J-U
Zhang F C 2003 \PRL {\bf 90} 207002

\bibitem{Yuan05} % t-J-U
Yuan F, Yuan Q and Ting C S 2005 \PR B {\bf 71} 104505

\bibitem{Wang06} % t-J-U
Wang Q H, Wang Z D, Chen Y and Zhang F C 2006 \PR B {\bf 73} 092507

\bibitem{Abram13} % t-J-U
Abram M, Kaczmarczyk J, J\k{e}drak J and Spa\l ek J 2013 \PR B {\bf 88} 094502

\bibitem{Spalek17} % t-J-U
Spa\l ek J, Zegrodnik M and Kaczmarczyk J 2017 \PR B {\bf 95} 024506

\bibitem{Zegrodnik17} % t-J-U
Zegrodnik M and Spa\l ek J 2017 \PR B {\bf 96} 054511

\bibitem{Fidrysiak20} % t-J-U
Fidrysiak M and Spa\l ek J 2020 \PR B {\bf 102} 014505

\bibitem{Schrieffer66}
Schrieffer J R and Wolff P A 1966 \PR {\bf 149} 491

\bibitem{Suhl59}
Suhl H, Matthias B T and Walker L R 1959 \PRL {\bf 3} 552

\bibitem{Kondo63}
Kondo J 1963 {\it Prog. Theor. Phys.} {\bf 29} 1

\bibitem{Kondo02}
Kondo J 2002 \JPSJ {\bf 71} 1353

\bibitem{Emery87}
Emery V J 1987 \PRL {\bf 58} 2794

% added, 19-Jan-2020
\bibitem{Penson86}
Penson K A and Kolb M 1986 \PR B {\bf 33} 1663

\bibitem{Belkasri96}
Belkasri A and Buzatu F D 1996 \PR B {\bf 53} 7171

\bibitem{Robaszkiewicz99}
Robaszkiewicz S and Bu\l ka B D 1999 \PR B {\bf 59} 6430

\bibitem{Kapcia14}
Kapcia K J 2014 {\it Acta Phys. Pol. A} {\bf 126} 53

\bibitem{KYamada92}
Yamada K, Yosida K and Hanzawa K 1992 {\it  Prog. Theor. Phys. Suppl.} {No. 108} 141

% IPT
\bibitem{KYamada75}
Yamada K 1975 {\it  Prog. Theor. Phys.} {\bf 53} 970

% IPT
\bibitem{KYosida75}
Yosida K and Yamada K 1975 {\it  Prog. Theor. Phys.} {\bf 53} 1286

% IPT
\bibitem{Georges92}
Georges A and Kotliar G 1992 \PR B {\bf 45} 6479

% IPT
\bibitem{XYZhang93}
Zhang X Y,  Rozenberg M J and Kotliar G 1993 \PRL {\bf 70} 1666

\bibitem{Ino1997} % phase separation
Ino A, Mizokawa T, Fujimori A, Tamasaku K, Uchida S, Kimura T, Sasagawa T and Kishio K 1997 \PRL {\bf 79} 2101

\bibitem{Fujimori1998} % phase separation
Fujimori A, Ino A, Mizokawa T, Kim C, Shen Z X, Sasagawa T, Kimura T, Kishio K, Takaba M, Tamasaku K, Eisaki H and Uchida S 1998 {\it J. Phys. Chem. Solids} {\bf 59} 1892

\bibitem{Fujimori2001} % phase separation
Fujimori A, Ino A, Yoshida T, Mizokawa T, Shen Z X, Kim C, Kakeshita T, Eisaki H and Uchida S 2001, {\it Open Problems in Strongly Correlated Electron Systems} 
ed J. Bonca {\it et al.} (Dordrecht: KluIr Academic Pub.) p 119

\bibitem{Tsukada06} % doping evolution
Tsukada A, Yamamoto H and Naito M 2006 \PR B {\bf 74} 174515

\bibitem{TYoshida06} % doping evolution
Yoshida T, Zhou X J, Tanaka K, Yang W L, Hussain Z, Shen Z X, Fujimori A, Sahrakorpi S, Lindroos M, Markiewicz R S, Bansil A, Komiya Seiki, Ando Yoichi, Eisaki H, Kakeshita T and Uchida S 2006 \PR B {\bf 74} 224510

\bibitem{Wen05} % low temperature specific heat
Wen H-H, Shan L, Wen X G, Wang Y, Gao H, Liu Z Y, Zhou F, Xiong J W and Ti W X 2005 \PR B {\bf 72} 134507 

\bibitem{Wang07} % low temperature specific heat
Wang Y, Yan J, Shan L, Wen H-H, Tanabe Y, Adachi T and Koike Y 2007 \PR B {\bf 76} 064512

\bibitem{Vidberg77} % Pade
Vidberg H J and Serene J W 1977 {\it J. Low. Temp. Phys.} {\bf 29} 179

\bibitem{Muller95} % two-gap
M\"uller K A 1995 {\it Nature} {\bf 377} 133

%%%
\bibitem{Li99} % c-axis twist Josephson experiments on Bi2Sr2CaCu2O8+? (Bi2212)
Li Qiang, Tsay Y N, Suenaga M, Klemm R A, Gu G D and Koshizuka N 1999 \PRL {\bf 83} 4160

\bibitem{Takano02} % artificial cross-whisker experiments on Bi2212
Takano Y, Hatano T, Fukuyo A, Ishii A, Ohmori M, Arisawa S, Togano K and Tachiki M 2002 \PR B {\bf 65} 140513(R) 

\bibitem{Takano03} % artificial cross-whisker experiments on Bi2212
Takano Y, Hatano T, Ohmori M, Kawakami S, Ishii A, Arisawa S, Kim S-J, Yamashita T, Togano K and Tachiki M 2003 {\it J. Low. Temp. Phys.} {\bf 131} 533

\bibitem{Takano04} % artificial cross-whisker experiments on Bi2212
Takano Y, Hatano T, Kawakami S, Ohmori M, Ikeda S, Nagao M, Inomata K, Yun K S, Ishii A, Tanaka A, Yamashita T and Tachiki M 2004 {\it Physica} C {\bf 408-410} 296

\bibitem{Latyshev04} % natural cross-whisker experiments on Bi2212
Latyshev Y I, Orlov A P, Nikitina A M, Monceau P and Klemm R A 2004 \PR B {\bf 70} 094517

\bibitem{Klemm05} % The phase-sensitive c-axis twist experiments on Bi2Sr2CaCu2O8+? and their implications
Klemm R A 2005 {\it Phil. Mag.} {\bf 85} 801

\bibitem{Zhu21} % Josephson junctions made of twisted ultrathin Bi2Sr2CaCu2O8+? flakes
Zhu Y, Liao M, Zhang Q, Xe H-Y, Meng F, Liu Y, Bai Z, Ji S, Zhang J, Jiang K, Zhong R, Schneeloch J, Gu G, Gu L, Ma X, Zhang D and Xue Q-K 2021 arXiv:1903.07965v2

\bibitem{Misra02} % c-axis STM experiments
Misra S, Oh S, Hornbaker D J, DiLuccio T, Eckstein J N and Yazdani A 2002 \PRL {\bf 89} 087002

\bibitem{Hoogenboom03} % c-axis STM experiments
Hoogenboom B W, Kadowaki K, Revaz B and Fischer {\O} 2003 {\it Physica} C {\bf 391} 376

\bibitem{Zhong16} % c-axis STM experiments
Zhong Y, Wang Y, Han S, Lv Y-F, Wang W-L, Zhang D, Ding H, Zhang Y-M, Wang L, He K, Zhong R, Schneeloch J, Gu G-D, Song C-L, Ma X-C and Xue Q-K 
2016 {\it Sci. Bull.} {\bf 61} 1239

\bibitem{Kashiwagi15_1} % A high-Tc intrinsic Josephson junction emitter tunable from 0.5 to 2.4 terahertz
Kashiwagi T, Sakamoto K, Kubo H, Shibano Y, Enomoto T, Kitamura T, Asanuma K, Yasui T, Watanabe C, Nakade K, Saiwai Y, Katsuragawa T, Tsujimoto M, Yoshizaki R, Yamamoto T, Minami H, Klemm R A and Kadowaki K 2015 {\it Appl. Phys. Lett.} {\bf 107} 082601

\bibitem{Kashiwagi15_2} % Efficient Fabrication of Intrinsic-Josephson-Junction Terahertz Oscillators with Greatly Reduced Self-Heating Effects
Kashiwagi T, Yamamoto T, Minami H, Tsujimoto M, Yoshizaki R, Delfanazari K, Kitamura T, Watanabe C, Nakade K, Yasui T, Asanuma K, Saiwai Y, Shibano Y, Enomoto T, Kubo H, Sakamoto K, Katsuragawa T, Markovi{\'c} B, Mirkovi{\'c} J, Klemm R A and Kadowaki K 2015 \PR {\it Appl.} {\bf 4} 054018

\bibitem{Wang11} % low-temperature specific heat in near optimally doped Bi2Sr2?x Lax CuO6+? (x \UTF{223C} 0.4)
Wang Y, Liu Z-Y, Lin C T and Wen H-H 2011 \PR B {\bf 83} 054509

\bibitem{Chen19} % ARPES
Chen S-D, Hashimoto M, He Y, Song D, Xu K-J, He J-F, Devereaux T P, Eisaki H, Lu D-H, Zaanen J, Shen Z-X 2019 {\it Science} {\bf 366} 1099
%%%

\bibitem{Masui03} % Raman, YBCO237 and BSCCO
Masui T, Limonov M, Uchiyama H, Lee S, Tajima S and Yamanaka A 2003 \PR B {\bf 68} 060506(R)

\bibitem{Khasanov07_01} % magnetic field penetration depth by muon-spin rotation, LSCO
Khasanov R, Shengelaya A, Maisuradze A, La Mattina F, Bussmann-Holder A, Keller H and M\"uller K A 2007 \PRL {\bf 98} 057007

\bibitem{Khasanov07_02} % magnetic field penetration depth by muon-spin rotation, YBCO237
Khasanov R, Str\"assle S, Di Castro D, Masui T, Miyasaka S, Tajima S, Bussmann-Holder A and Keller H 2007 \PRL {\bf 99} 237601

\bibitem{Khasanov08} % magnetic field penetration depth by muon-spin rotation, YBCO248
Khasanov R, Shengelaya A, Bussmann-Holder A, Karpinski J, Keller H and M\"uller K A 2008 {\it J. Supercond. Nov. Magn.} {\bf 21} 81

\bibitem{Moriya00}
Moriya T and Ueda K 2000 {\it Adv. Phys.} {\bf 49} 555 and references therein

%%%
\bibitem{Que87}
Que W-M and Kirczenow G 1987 \SSC {\bf 64} 1052

%%% Kondo-lattice model
\bibitem{Prelovsek88}
Prelov{\v{s}}ek P 1988 \PL A {\bf 126} 287

\bibitem{Ramsak89}
Ram{\v{s}}ak A and Prelov{\v{s}}ek P 1989 \PR B {\bf 40} 2239

\bibitem{Castellani88_1}
Castellani C, Di Castro C and Grilli M 1988 {\it Physica} C {\bf 153-155} 1659

\bibitem{Castellani88_2}
Castellani C, Di Castro C and Grilli M 1988 {\it Int. J. Mod. Phys.} B {\bf 2} 659

\bibitem{Cancrini91}
Cancrini N, Caprara S, Castellani C, Di Castro C, Grilli M and Raimondi R 1991 {\it Europhys. Lett.} {\bf 14} 597

\bibitem{Kamimura88}
Kamimura H 1988 {\it Int. J. Mod. Phys.} B {\bf 2} 699

\bibitem{Andrei89}
Andrei N and Coleman P 1989 \PRL {\bf 62} 595

\bibitem{Hatsugai89}
Hatsugai Y, Imada M and Nagaosa N 1989 \JPSJ {\bf 58} 1347
%%%

\bibitem{Hirsch92} % kinetic
Hirsch J E 1992 {\it Physica} C {\bf 201} 347

\bibitem{Tsunetsugu98} % kinetic
Tsunetsugu H and Imada M 1998 \JPSJ {\bf 67} 1864

\bibitem{Tsunetsugu99} % kinetic
Tsunetsugu H and Imada M 1999 \JPSJ {\bf 68} 3162

\bibitem{Imada01} % kinetic
Imada M and Onoda S 2001, {\it Open Problems in Strongly Correlated Electron Systems} 
ed J. Bonca {\it et al.} (Dordrecht: KluIr Academic Pub.) p 69

\bibitem{Sarker09} % charge pair-hopping
Sarker S K and Lovorn T 2010 \PR B {\bf 82} 014504

\bibitem{Sarker12} % charge pair-hopping
Sarker S K and Lovorn T 2012 \PR B {\bf 85} 144502

\bibitem{Feng12} % kinetic
Feng S, Zhao H and Huang Z 2012 \PR B {\bf 85} 054509

\bibitem{Feng15} % kinetic
Feng S, Kuang L and Zhao H 2015 {\it Physica} C {\bf 517} 5

\bibitem{Gao18} % kinetic
Gao D, Mou Y and Feng S 2018 {\it J. Low Temp. Phys.} {\bf 192} 19

\end{thebibliography}
